\documentclass[aoas]{imsart}

\RequirePackage{amsthm,amsmath,amsfonts,amssymb}
\RequirePackage[authoryear]{natbib}
\RequirePackage[colorlinks,citecolor=blue,urlcolor=blue]{hyperref}
\RequirePackage{graphicx}

\usepackage{bm}
\usepackage{booktabs}
\usepackage{tabularx}
\usepackage{xcolor}
\usepackage{color}
\usepackage{graphicx}
\usepackage{bbm}
\usepackage{subcaption}
\usepackage[section]{placeins}
\usepackage{mathtools}

\usepackage{adjustbox}

\usepackage{multirow}

\usepackage{url}
\usepackage{ulem}

\usepackage{hyperref}
\hypersetup{
    colorlinks,
    urlcolor=blue,
    citecolor=blue
}

\usepackage{etoolbox}
\usepackage{graphicx}

\makeatletter

\newcommand\bigcdot[1][.5]{\mathbin{\vcenter{\hbox{\scalebox{#1}{$\bullet$}}}}}
\makeatother

\DeclareTextFontCommand{\emph}{}

\def\ie{{{i.e.}, }}
\def\eg{{{e.g.}, }}

\def\FIGPATH{Plots/}

\DeclareMathOperator{\logit}{logit}

\startlocaldefs

\theoremstyle{remark}

\endlocaldefs

\begin{document}

\begin{frontmatter}
\title{Spatiotemporal wildfire modeling through point processes with moderate and extreme marks}
\runtitle{Wildfire modelling}

\begin{aug}
\author[A]{\fnms{Jonathan} \snm{Koh}\ead[label=e1]{jonathan.koh@epfl.ch}},
\author[B]{\fnms{Fran\c cois} \snm{Pimont}\ead[label=e2]{francois.pimont@inrae.fr}}
\author[B]{\fnms{Jean-Luc} \snm{Dupuy}\ead[label=e3]{jean-luc.dupuy@inrae.fr}}
\and
\author[C]{\fnms{Thomas} \snm{Opitz}\ead[label=e4]{thomas.opitz@inrae.fr}},

\address[A]{Institute of Mathematics, EPFL,
\printead{e1}}

\address[B]{URFM UR629, INRAE, 
\printead{e2,e3}}

\address[C]{BioSP, INRAE,
\printead{e4}}
\end{aug}

\begin{abstract}
Accurate spatiotemporal modeling of conditions leading to moderate and large wildfires provides better understanding of  mechanisms driving fire-prone ecosystems and improves risk management. We here develop a joint model for the occurrence intensity and the wildfire size distribution by combining extreme-value theory and point processes within a novel Bayesian hierarchical model, and use it to study daily summer wildfire data for the French Mediterranean basin during 1995--2018. The occurrence component models wildfire ignitions as a spatiotemporal log-Gaussian Cox process. Burnt areas are numerical marks attached to points and are considered as extreme if they exceed a high threshold. The size component is a two-component mixture varying in space and time that jointly models moderate and extreme fires. We capture non-linear influence of covariates (Fire Weather Index, forest cover) through component-specific smooth functions, which may vary with season. We propose estimating shared random effects between model components  to reveal and interpret common drivers of different aspects of wildfire activity. This leads to increased parsimony and reduced estimation uncertainty with better predictions.   Specific stratified subsampling of zero counts is implemented to cope with large observation vectors. We compare and validate models through predictive scores and visual diagnostics. Our methodology provides a holistic approach to explaining and predicting the drivers of wildfire activity and associated uncertainties.  

\end{abstract}

\begin{keyword}
\kwd{Bayesian~hierarchical~model}
\kwd{Cox~process}
\kwd{Extreme-value~theory}
\kwd{Forest~fires}
\kwd{Shared~random~effects}
\end{keyword}

\end{frontmatter}

\section{Introduction}
\label{sec:introduction}



Wildfires are defined as uncontrolled fires of combustible natural vegetation such as trees in a forest. Their activity usually shows seasonal cycles, as several conditions must coincide for their occurrence: the presence of combustible material as fuel, its easy flammability resulting from weather conditions such as droughts, and a trigger. Triggers  include natural causes such as lightning, but the majority of occurrences in Europe  are caused by human activity, either intentional (arson), neglectful (cigarette stubs) or accidental (agriculture).    

 Wildfires represent major environmental and ecological risks worldwide. They provoke many human casualties and substantial economic costs, and can trigger extreme air pollution episodes and entail important losses of biomass and biodiversity. While climate change is expected to exacerbate their frequency and extent \citep{Jones.2020}, wildfires themselves contribute an important fraction of global greenhouse gases that can accelerate climate change. To aid in wildfire prevention and risk mitigation, one must identify the factors contributing to wildfires and predict their spatiotemporal distribution. Prediction maps of various components of wildfire risk are relevant for the study of historical periods, for short-term forecasting and for long-term projections. 

The study of wildfire activity has led to a large body of statistical and machine learning literature on  methods for identifying risk factors and producing risk maps \citep[][]{Preisler.2004,Xi2019,Pereira2019}. Most studies focus on modeling either occurrence counts or sizes, the latter usually represented by the burnt areas of spatially and temporally contiguous wildfire events. In occurrence modeling, the spatial or spatiotemporal pattern of ignition points (or other representative points of separate wildfire events) can be analyzed with point process tools \citep[][]{Peng2005,Genton2006,Xu2011,Serra2013,Tonini2017,Pereira2019,Opitz2020}.
Often, data are available as presence/absence or counts over dense spatial or spatiotemporal grids, 
or have been transformed to such representations to facilitate modeling and to harmonize different spatial-temporal scales of wildfire and predictor data such as weather conditions, land cover and land use. 

Burnt area, a key measure of wildfire impact, usually provides a good proxy for biomass loss and greenhouse gas emissions, and it allows interpretation of  impacts on ecosystem services such as biodiversity or clean air. Many univariate probability distributions have been explored for modeling fire sizes \citep[\eg][]{Cumming2001,Schoenberg2003,Cui2008,Pereira2019}.
Empirical distributions are usually heavy-tailed, which is also the case with the wildfire data we consider in Mediterranean France. This lead to a very small number of the most extreme wildfires accounting for a very large fraction of total burnt area. 
There is no consensus on which parametric distribution family provides the best fit \citep{Pereira2019}. Distributions suggested by extreme-value theory, such as the generalized Pareto distribution (GPD) arising for threshold exceedances, have been studied \citep[\eg approaches by][]{DeZeaBermudez2009,Mendes2010,Turkman2010,Pereira2019}. 

Joint statistical analyses of wildfire occurrence and sizes have been proposed and often use tools for marked point processes, where numerical marks represent burnt areas.  Descriptive approaches  \citep[\eg][]{Tonini2017}
characterize different regimes of wildfire activity (\ie numbers, sizes, spatialtemporal autocorrelation) by taking into account weather, land cover, fire management and environmental factors. For explanatory and predictive modeling, Bayesian hierarchical models are useful, where we include latent Gaussian components to allow for observation and estimation uncertainty, and to capture nonlinear influences of observed covariates.  One may consider only categorical information (\eg small and large wildfires) without attempting to model the continuous distribution of values; for example, \citet{Serra2014} construct a Bayesian spatiotemporal ``hurdle" model to focus on occurrences of large wildfires. As to continuous distributions, \citet{Rios2018} implement MCMC inference for zero-inflated Beta-regression to model the occurrence of wildfires in spatial units, with absence corresponding to zero-inflation, while positive area fraction covered by wildfires is captured through the Beta distribution. \citet{Joseph2019} estimate separate regression models with random effects for occurrence numbers in areal units and for sizes, and they study 
posterior predictive distributions for block maxima of wildfire sizes. \citet{Pimont2020} developed a marked spatiotemporal log-Gaussian Cox process model, called \emph{Firelihood}, for daily data 
by applying the integrated nested Laplace approximation \citep[INLA,][]{Rue.al.2009} for Bayesian inference of most components of the model.  
Their distribution of wildfire sizes over positive values is based on estimating exceedance probabilities and excess distributions over a range of severity thresholds. 
Weather information is included through a nonlinear effect of the Fire Weather Index \citep[FWI,][]{vanWagner1977},  constructed to yield high correlation with wildfire activity.

In this work, we develop the following novelties to address critical shortcomings of the works cited above. 
Since large wildfires play a dominant and critical role for fire activity due to the heavy tails of burnt areas, 
we focus on accurate modeling of their distribution, and in particular its spatiotemporal variation. However, models constructed using only extreme wildfires would lead to high estimation uncertainty when inferring complex spatiotemporal structures. We therefore propose the novel joint estimation of extreme and non extreme wildfires where the model borrows strength from the latter to help estimate the former; the large number of observations available for moderate fires improves the prediction of larger fires, so changes in extreme fire activity are better accounted for.

Complex models such as \emph{Firelihood} require separate estimation of the occurrence and size model components, thus hampering inferences exploiting stochastic interactions between them. 
Temporal stochastic structures are often restricted to the spatiotemporal variability in covariates. 
In \citet{Pimont2020}, simulated predictive distributions of wildfire activity for various divisions of the space-time domain  failed to capture some very extreme events, specifically the year 2003. Here we increase the flexibility of the spatiotemporal structure, especially for extremes. 

Our new approach leverages a combination of marked point processes defined over continuous space and time and extreme-value theory to represent the mechanisms leading to wildfires 
exceeding a high severity threshold for burnt areas. The point pattern of extreme fires is viewed as a \emph{thinning} of the full pattern, and we select a suitable threshold before using the theoretically justified GPD model for threshold excesses. 

We also advocate  \emph{sharing} spatial random effects  that affect several model components simultaneously: these effects are estimated for one response variable (\eg wildfire counts) but we also include them with scaling coefficients in other response variables (\eg wildfire size exceedances). This approach decreases uncertainty in the estimation of those regression equations whose vector of observed responses carries too little information to estimate complex predictive structures. We will highlight the improved inferences through sharing in our wildfire application. 
Besides being a tool to increase model parsimony, it also provides new scientific insight by highlighting joint drivers of different wildfire components. 

The FWI 
quantifies the influence of weather drivers on wildfire activity and is often mapped as an index for fire danger, for instance by the French weather service M\'et\'eo France. Model diagnostics of \citet{Pimont2020} showed  that the predictive power of FWI in France may diminish depending on season, such that the danger rating of fire activity using FWI should not be constant throughout the fire season.  Therefore, we here develop estimation of a more sophisticated seasonal nonlinear FWI effect to assess and interpret differences of wildfire response to FWI across months. 


Predictive model validation is intricate 
because of heavy tails and high prediction uncertainty for individual wildfires. Customary validation  scores, such as means of squared or absolute errors, are not useful. In addition to visual diagnostics, we tackle this difficulty through joint assessment of several numerical criteria, either through scores for binary data \citep[\eg Area under the Curve,][]{Fawcett.2006} to assess the  exceedance behavior over a relevant severity threshold, or through comparison of the distribution of probabilistic scores for continuous predictions, such as the  scaled Continuous Ranked Probability Score \citep{Bolin-Wallin.2020}.

We estimate our marked log-Gaussian Cox process in a Bayesian setting using INLA \citep[][]{Illian.al.2012}
by adopting Penalized Complexity (PC) priors for hyperparameters \citep{Simpson.al.2017}. 
Gaussian process priors follow the Mat\'ern covariance function, and we use the Stochastic Partial Differential Equation (SPDE) approach of \citet{Lindgren.al.2011} for numerically efficient Gauss--Markov approximation. Fully Bayesian inference is out of reach with several millions of observations of wildfire counts for pixel-days as given here. 
Therefore, we devise a specific subsampling scheme for zero counts that keeps a relatively larger proportion of observations with high FWI, for which most wildfires occur. This allows for joint Bayesian inference of all components, and we ensure that our subsample sizes allow fitting models on standard personal computers, in contrast to other highly computer-intensive approaches  in the  recent literature \citep[\eg][]{Joseph2019,Pimont2020,Opitz2020} requiring  high memory resources. 

In the remainder of the paper, we first 
explore available data on wildfires and predictors in \S\ref{sec:data}.
 We provide general background on extreme-value theory and point processes, and on how to combine them in a Bayesian hierarchical model using the INLA-SPDE method, in \S\ref{sec:methods-general}.
The specific hierarchical structure for the joint analysis of extreme and non-extreme wildfires is developed in \S\ref{sec:model}. Estimation with subsampling  of pixel-days without wildfire occurrences is detailed in \S\ref{sec:likelihood_COX}. After a comparative analysis of models in \S\ref{sec:model:selection}, we highlight key findings and prediction of wildfire activity components in \S\ref{sec:mode:results} and \S\ref{sec:application}, and we conclude 
in \S\ref{sec:conclusion}. 



\section{Wildfire data}
\label{sec:data}


Since 1973, wildfires occurring in the fire-prone French Mediterranean region have been recorded in the Prom\'eth\'ee database (\url{www.promethee.com}). Each wildfire occurrence is reported with its fire ignition cell in a $2\times 2$km$^2$ grid,
day of detection and burnt area in hectare (ha).
Inconsistent reporting was found for small wildfires, especially smaller than $1$~ha, and 
we keep only data with reported burnt area larger than $1$~ha; \ie of \emph{escaped} wildfires 
that could not be extinguished at an early stage. 
We use the observation period 1995--2018, for which gridded weather reanalysis data (\emph{SAFRAN} model of \emph{M\'et\'eo France}) and information on forested area are available. 

\begin{figure}[!t]
    \centering
    \includegraphics[width=0.78\textwidth]{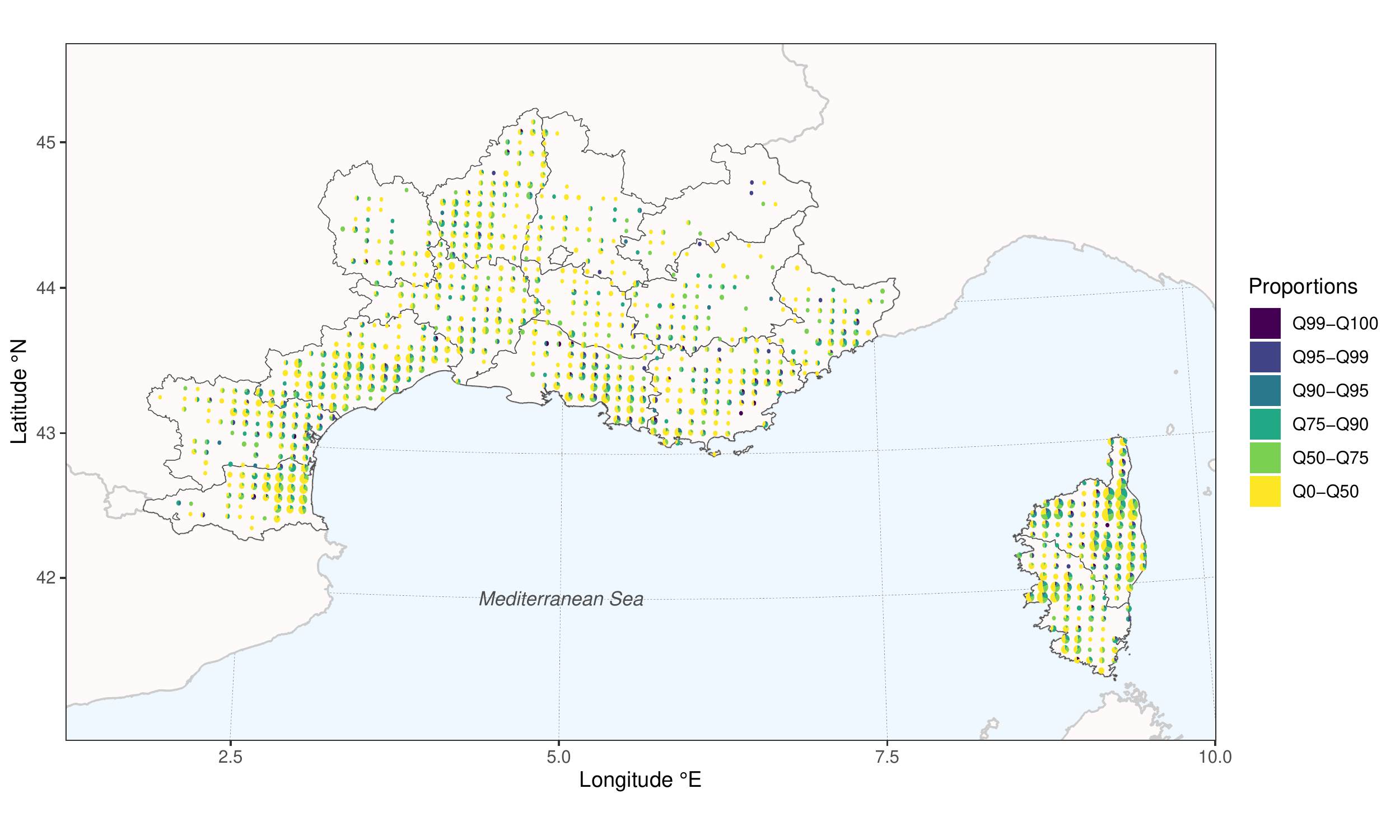}\vspace{.25cm}\\
    \includegraphics[width=0.78\textwidth]{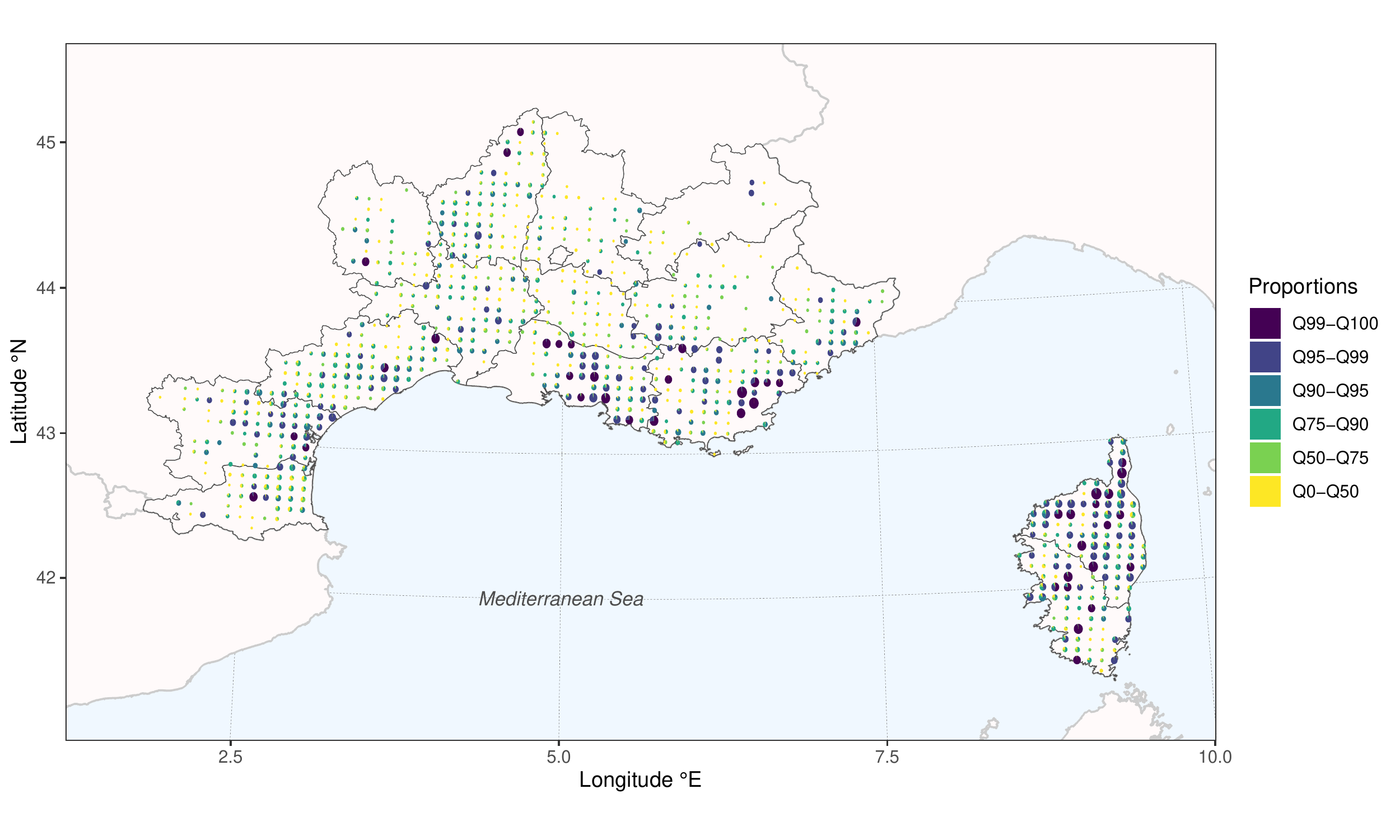}
    \caption{
    Maps of Prom\'eth\'ee data aggregated to the SAFRAN  grid  at $8$km resolution. The pie charts in the grid cells are based on  $6$ wildfire size classes with boundaries given by empirical quantile levels $0,0.5,0.75,0.9,0.95,0.99,1$ of all burnt areas (June--October). Top display: pie charts show relative count proportions over the six classes and have size increasing with increasing counts. Bottom display: pie charts show relative burnt area proportions and have size increasing with increasing aggregated burnt area. }
    \label{fig:maps-safran-grid}
\end{figure}

Figure~\ref{fig:maps-safran-grid}  illustrates the  heavy tails in the distribution of burnt areas and strong spatial variability in numbers and sizes of wildfires. It also shows the contours of  administrative areas (``\emph{d\'epartements}") in the study region. 
Small to moderately large wildfires strongly dominate the pie charts for wildfire counts, while large wildfires strongly dominate the pie charts of aggregated burnt area. Certain spatial patterns are similar in the distribution of numbers and sizes of wildfires (top and bottom display of Figure~\ref{fig:maps-safran-grid}, respectively), but we also discern notable differences. For example, large wildfire numbers do not always entail large aggregated burnt areas, as we see for the Pyr\'enees-Orientales d\'epartment in the southwest.  The disparities among the two displays show the need to model spatiotemporal structures in both wildfire numbers and sizes, as well as their interaction. Figure~\ref{fig:burnt-surfaces} (left panel) shows a histogram of burnt area values. 
Incidentally, the sum of burnt areas  exceeding the empirical $99\%$-quantile is larger than the corresponding sum of the remaining wildfires.  

The SAFRAN model provides gridded weather reanalyses at $8$km resolution.  The joint influence of weather variables such as temperature, precipitation and wind speed on fire activity patterns is highly complex. Meteorological indices of fire danger have been constructed, such as the widely used unitless Fire Weather Index (FWI) that was originally defined for Canadian forests. Its values are often used for direct interpretation and fire danger mapping. Instead, we here study its relationship to components of fire risk, such as  occurrence frequency and wildfire sizes. 
For our models, we preprocess SAFRAN data to daily FWI and use the SAFRAN grid by aggregating daily wildfire counts to its cells; \citet{Pimont2020} provide arguments to use this spatial-temporal resolution.
Forest cover is another crucial explanatory variable.
The study area hosts approximately $60\%$ of forested areas or  vegetation types that ignite easily (shrubland; other natural herbaceous vegetation).  Wildfires do not propagate easily through the other available land cover types.
We consider  relevant fuel material through proportion covered by this vegetation in each SAFRAN grid cell (and day) based on  CORINE Land Cover  data (CLC). 
CLC dynamics are captured by linear temporal interpolation of several inventories.
We refer to the resulting pixel-day predictor as forested area (FA), in $\%$. 


\section{Methods for point patterns with extreme marks}
\label{sec:methods-general}

\subsection{Extreme-value theory}\label{sec:explore:evt}

Given a random variable $X\sim F$ with distribution $F$ satisfying mild regularity conditions, the generalized Pareto distribution (GPD) arises asymptotically for the positive excesses of $X$ above a threshold increasing to $x^\star=\sup\{x: F(x)<1\}$ \citep{Coles.2001}. Therefore, given a large threshold $u<x^\star$, the tail behavior of a wide class of random variables $X$ can be approximated as
\begin{equation}\label{eq:gpd}
    \mathrm{Pr}(X>x+u \mid X>u) \approx \mathrm{GPD}_{\sigma,\xi}(x) = \left \{
\begin{array}{ll}
(1+\xi x/\sigma)_{+}^{-1/\xi}& \quad \xi \neq 0, \\
\exp (-x/\sigma)& \quad \xi =0, 
\end{array}
\quad x >0,
\right.
\end{equation}
with shape parameter $\xi\in\mathbb{R}$ and  scale parameter $\sigma=\sigma(u)>0$, where $a_{+}=\max(a,0)$. 
The shape parameter determines the rate of tail decay, with slow power-law decay for $\xi>0$, exponential decay for $\xi=0$, and polynomial decay towards a finite upper bound for $\xi<0$.  
Writing $p_{\mathrm{exc}}=1-F(x)$  for the exceedance probability of $X$ above  $u$, we use \eqref{eq:gpd} to approximate the cumulative distribution function $F$ of $X$ above the threshold $u$ \citep{Davison-Smith.1990} as 
\begin{equation}\label{eq:excd}
F(x) \approx 1 - p_{\mathrm{exc}} \mathrm{GPD}_{\sigma,\xi}(x-u), \quad x>u, 
\end{equation}
where $\xi,\sigma$ and $p_{\mathrm{exc}}$ are parameters to be estimated. We account for dependence and non-stationarity among observations by including auxiliary variables and Gaussian random effects into  $\sigma$ and $p_{\mathrm{exc}}$. Nonstationarity in $\xi$ is often hard to identify, and we therefore keep $\xi$ stationary.

\begin{figure}[!t]
    \centering
    \includegraphics[width=.3\textwidth]{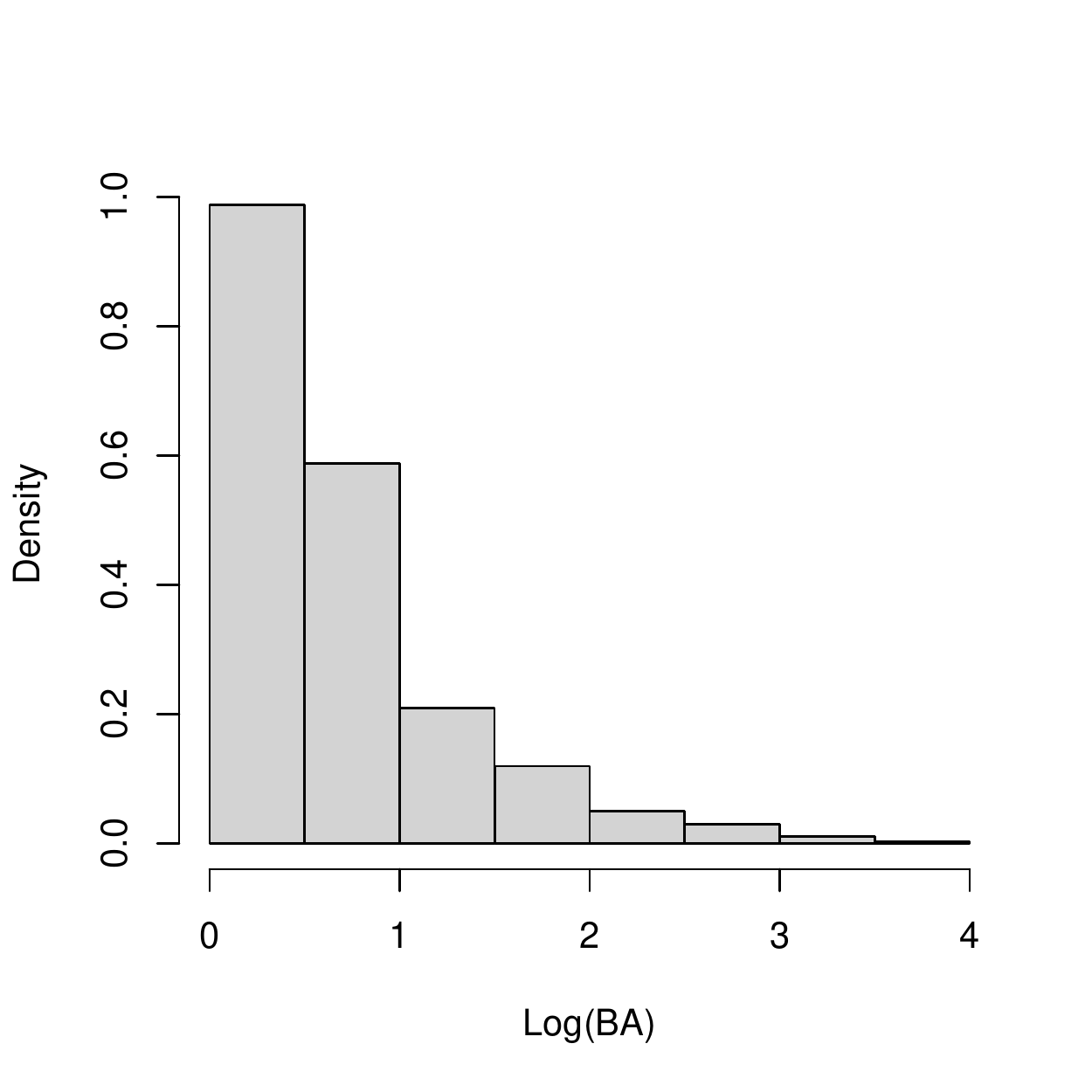}
    \includegraphics[width=.3\textwidth]{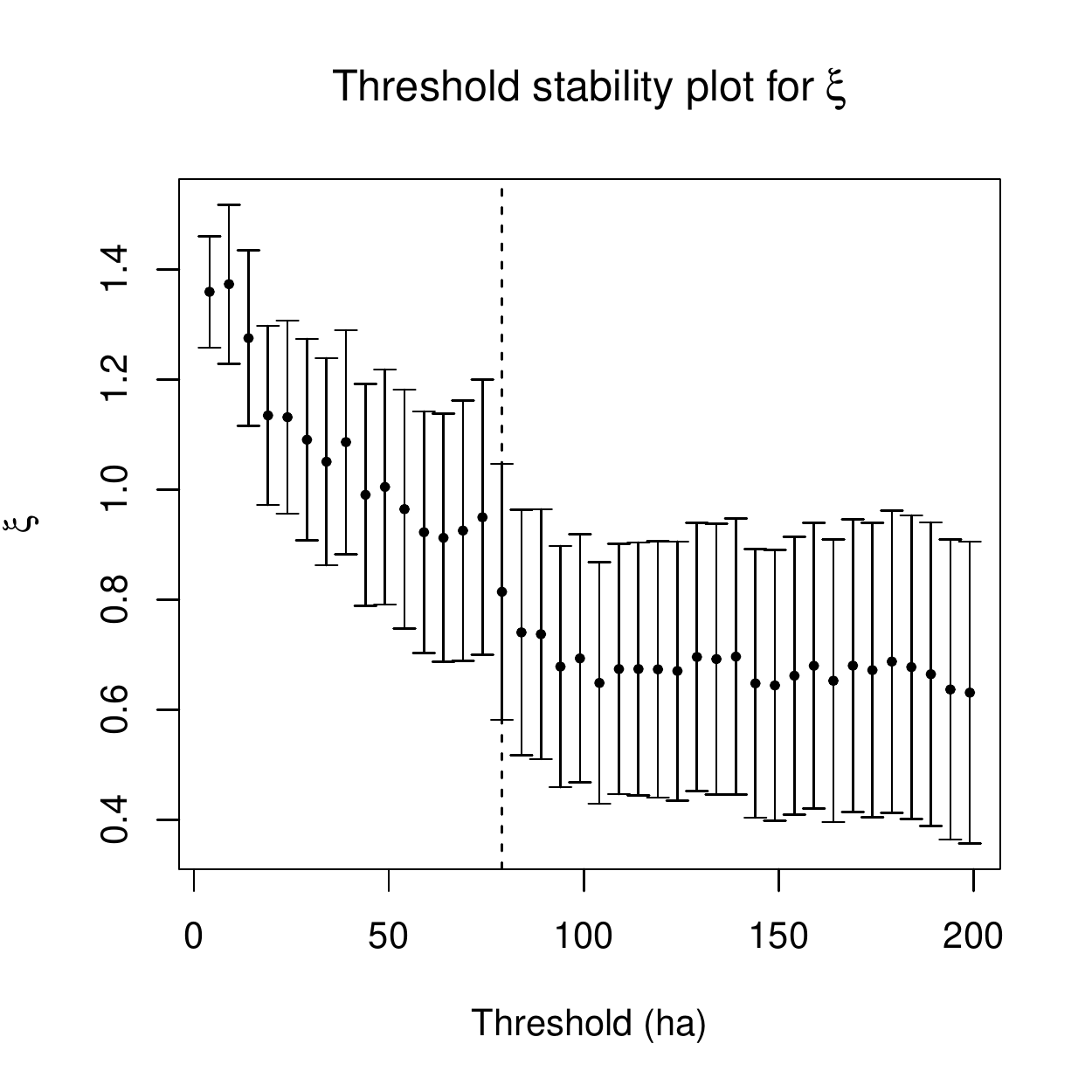}
    \includegraphics[width=.3\textwidth]{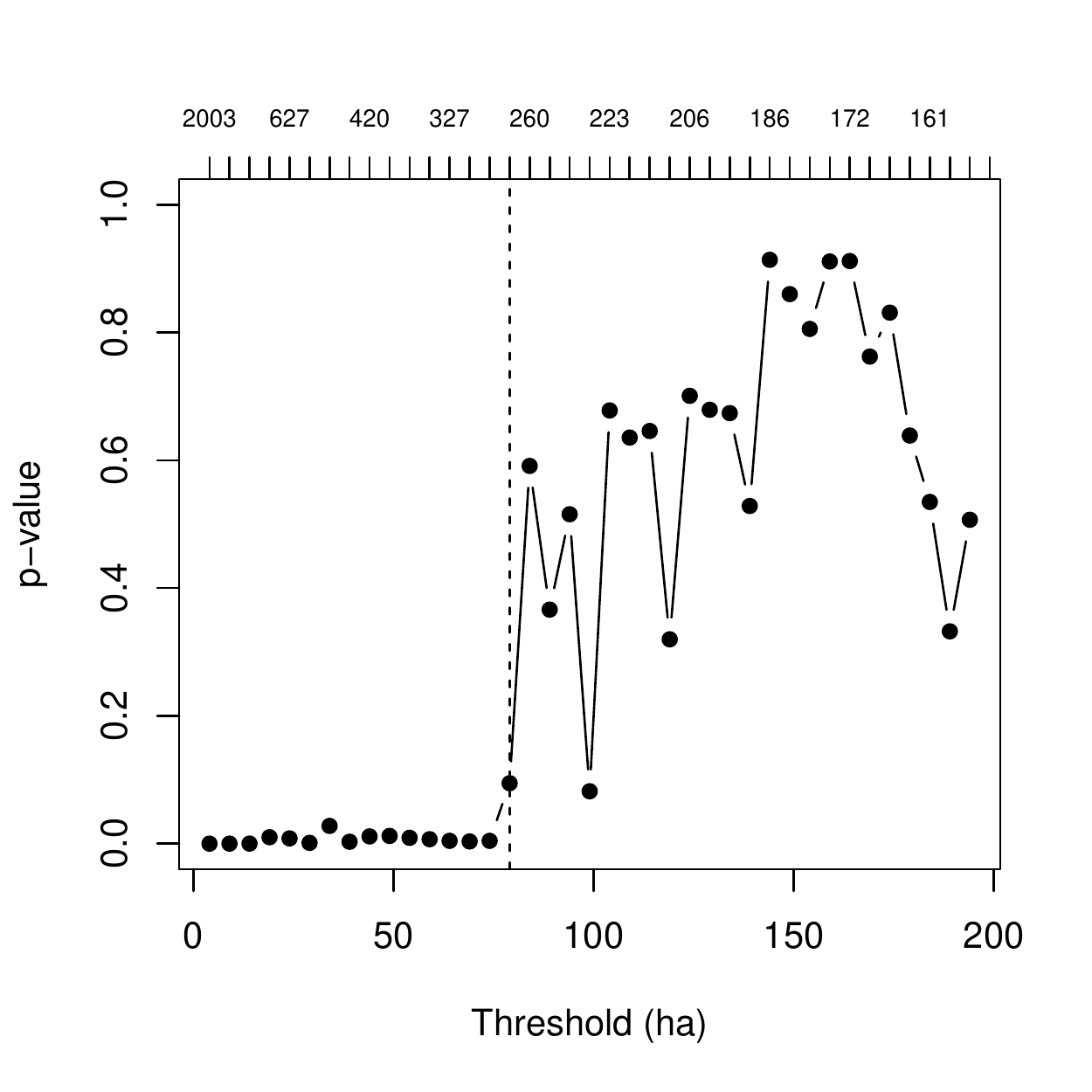}
    \caption{Burnt area distribution. Left: Histogram of burnt areas (ha) in base-10-logarithm. Middle: Parameter stability of the tail index. Right: p-values for the null hypothesis of a GPD distribution above the threshold; tick labels on top indicate the number of fires above the thresholds.}
    \label{fig:burnt-surfaces}
\end{figure}

Based on \eqref{eq:excd}, we model the conditional GPD of fire size excesses and $p_{\mathrm{exc}}$. To explore the tail behavior of all fire sizes pooled together and choose an appropriate threshold $u$, we can use tools such as mean excess plots (see Supplement) 
or the following threshold stability plot of parameters, here considered for the GPD shape  $\xi$, estimated by maximum likelihood for a range of  increasingly high thresholds $v_1<\ldots<v_m$. We use  multiple statistical tests  \citep{Northrop-Coleman.2014} to
test the null hypotheses that the data come from a  common truncated GPD on all intervals $(v_k, v_{k+1})$, $k=1,\ldots,m$, where $v_{m+1}=\infty$. Using $m=40$ equidistant intervals of length $5$ha for fire sizes,  Figure~\ref{fig:burnt-surfaces} provides evidence that stability is reached above approximately the $95\%$ quantile ($79$ha), with  failure to reject the null  hypothesis of $\xi_k=\cdots=\xi_m$ for intervals with $v_k > 79$ha and estimated shape $\hat{\xi}_k\approx 0.7$. 

\citet{Joseph2019} modeled fire sizes in the contiguous United States and concluded that the GPD leads to overestimation of extreme fire sizes. However, they fitted the GPD to the full distribution;  
Figure~\ref{fig:burnt-surfaces} shows that we would have obtained a very different value $\hat{\xi}\approx 1.4$ for $u=1$, which entails an extremely slow tail decay. 

\subsection{Mark-dependent thinning of point processes}
\label{sec:markdep_cox}

We consider the point pattern of fire ignitions and burnt areas as a realization of a spatiotemporal marked point process; \ie of a random count measure $N$ that attributes value $N(B)\in\{0,1,2,\ldots\}$ to  Borel sets $B\subset \mathbb{R}^2 \times \mathbb{R}$. 
We model the intensity function $\lambda(x)$ of the point process in the observation window $\mathcal{D}\subset \mathbb{R}^2 \times \mathbb{R}$. It defines the expected number of points $\mu(B)$ for any $B\subset \mathcal{D}$ as 
$$
\mu(B) = \mathbb{E} N(B)= \mathbb{E}\sum_{i=1}^N 1(x_i\in B) = \int_B \lambda(x)\,\mathrm{d}x.
$$
We focus on Poisson point processes characterized by the counts   $N(B)\sim \mathrm{Pois}\{\mu(B)\}$. 
With two types of points, such as non-extreme and extreme points, the point pattern is a superposition of the two single-type patterns: $\lambda = \lambda_1+\lambda_2$.  
The points of a specific type, say type $2$, are  obtained by \emph{thinning} the full point pattern; \ie by removing the points of other types (here type $1$) using the thinning probability 
$p(x)= \lambda_2(x)/\lambda(x)$, $x \in \mathcal{D}$. 
Extreme events, characterized as points $x_i$ whose magnitude mark $y_i$ exceeds a fixed high value $u(x_i)$ are obtained by thinning the full point pattern. 
Given a point pattern $\{x_1,\ldots,x_N\}$, $N\geq 1$, we define variables $E_i=\mathbb{I}\{y_i>u(x_i)\}\sim \mathrm{Bernoulli}\{p(x_i)\}$, $i=1,\ldots,N$. 
An independently thinned Poisson process (\ie $E_i$ are independent) is again a Poisson process.

\subsection{Spatiotemporal Log-Gaussian Cox processes}\label{sec:likelihood_COX}

Log-Gaussian Cox processes (LGCPs) are Poisson processes with log-Gaussian intensity function $\lambda(x)$. This random specification of the  intensity function allows us to explain spatiotemporal variability not captured by other deterministic parameters, and provides a natural framework for the Bayesian modeling of point processes with Gaussian process priors. 
Two major challenges arise for likelihood-based inference in LGCPs: (i) intensity functions are conceptually defined over continuous space; (ii) the Gaussian random effects lead to an intractable likelihood with no general closed-form expression. Challenge (ii) requires estimation techniques to handle latent variables; see \S\ref{sec:methods-inlaspde}. 
As to (i), without considering  the marks, LGCPs have no general closed-form expression for their probability densities
\begin{equation}\label{eq:lgcp-density}
(x_1,\ldots,x_N) \mapsto  \mathbb{E}_{{\lambda}}  \exp\left(-\int_{\mathcal{D}} \lambda(x)\, \mathrm{d}x \right) \prod_{i=1}^n \lambda(x_i), 
\end{equation}
where $x= x(s,t)$ is a point in the space-time observation window $\mathcal{D}$. Different approximation strategies allow numerical computation of the integral $\int_{\mathcal{D}} \lambda(x)\, \mathrm{d}x$ for a given intensity function. We opt for discretizing the observation window using the SAFRAN grid, and assume that the  intensity function  does not vary within pixel-day grid cells. 
Conditional on $\lambda$, the number of points observed in a cell $C_k$, $k=1,\ldots,K$, is Poisson distributed.  
Therefore,  estimating the LGCP corresponds to performing a (mixed) Poisson regression with log-link:
\begin{equation}\label{eq:lgcp-discretized}
 N_k \overset{\text{ind}}{\sim} \mathrm{Pois}(| C_k |\, \lambda_k), \quad \mathbb{E}\left[N_k\mid \lambda_k\right]=\lambda_k, \quad \log(\lambda_k)= \mu_k, \quad k=1,\ldots, K,
\end{equation}
where $|C_k|$ is the Lebesgue volume, $\bigcup_{k}^K C_k = \mathcal{D}$ and  $C_{k_1}\cap C_{k_2}=\emptyset$ if $k_1\not=k_2$. The \emph{linear predictor} $\mu_k$ is additively composed of fixed and random effects.
For space-varying random effects, we use the value at the center of the grid cell.  Likelihood-based inference for latent Gaussian processes is often based on Laplace approximation \citep{Tierney.Kadane.1986}. In particular, the INLA framework assumes conditional independence of the observations given the latent Gaussian predictor and is thus well suited for LGCPs, where the Poisson observations $N_k$ are conditionally independent given $\mu_k$ \citep{Illian.al.2012,Opitz2020b}.
  Other approaches for numerically approximating the integral in \eqref{eq:lgcp-density} exist. Typically, they use appropriately weighted sums $\sum_{k}\omega_k\lambda(\tilde{x}_k)$ with discretization points $\tilde{x}_k$ and weights $\omega_k>0$, which lead to variants of Poisson and logistic regression (\eg the Berman--Turner \citeyear{Berman1992} device); 
see \citet{Baddeley2010}. 
 
 \subsection{Data aggregation and subsampling schemes}
 \label{ssec:ss}
 
Spatiotemporal hierarchical modeling is notoriously computer-intensive due to large datasets and numerical challenges with covariances. With the \texttt{R-INLA} implementation \citep{Rue.al.2017}, up to  several hundred thousand observations can be handled.
Stable inferences may require compromises with respect to the complexity of the latent model and the number of observations, which jointly determine the size and sparsity of the Gaussian precision matrices, which in turn influence computation times, memory requirements and well-conditioned numerical behavior.
Even stronger restrictions arise with methods such as Markov Chain Monte Carlo (MCMC) to achieve  approximation quality comparable to INLA \citep{Taylor.Diggle.2014,Niekerk2019}. \citet[][\S $8.4$]{Krainski.al.2018} develop strategies for LGCPs by aggregating the events to larger mapping units and lowering spatial-temporal resolution of random effects to decrease computation times, which, however, would impede the modeling of structures arising at small spatiotemporal scales.  

Another way to cope with this issue is subsampling \citep{Baddeley.Turner.2000,Rathbun2007,Baddeley2010,Rathbun2013,Baddeley.al.2014}, 
where the model is estimated  using an appropriately reweighted subsample of data points, which keeps the loss of information small. 
Since maximum likelihood is equivalent to maximizing the empirical expectation of the log-density of observations, a subsampling scheme is appropriate if it ensures a faithful approximation of this expectation. Subsampling  in likelihood-based estimation can be interpreted as importance sampling \citep{Tokdar2010}: the original sample with observation weight unity is replaced by a subsample with typically larger observation weights.
Weighted subsampling theory goes back to  \citet{Horvitz1952a}.

The Poisson intensities $\lambda_k=\exp(\mu_k)$ ($k=1,\ldots,K$) in \eqref{eq:lgcp-discretized} are the parameters to be estimated, and we need a subsample $N_{k_j}$ with weights $\omega_j$ ($j=1,\ldots,J$) such that the subsample likelihood is close  to the full density \eqref{eq:lgcp-density}. 
The sample size $K$ exceeds $5$ million due to over $1000$ daily-replicated spatial pixels. To enable \texttt{R-INLA}-based estimation, we devise a stratified subsampling scheme to reduce the number of observations  by hundredfold.
Observations $N_k>0$ are not subsampled since they are rare and highly informative; we keep them with each weight unity. 
For the zero wildfire occurrence counts, we link subsampling to Poisson additivity. The likelihood contribution $\exp(-\lambda_k)^{\omega_k}=\exp(-\omega_k\lambda_k)$ with weight $\omega_k\in\mathbb{N}$ is equal to the likelihood of the sum 
of $\omega_k$ observations with count $0$; the size of the initial sample is divided by the factor $\omega_k$. 
The predictors (covariates, random effects), and therefore of intensities $\lambda_k$, 
differ between different pixel-days $k$ in our  models, so Poisson additivity cannot be applied without additional approximations. However, the values of such predictors may often be very similar for cells located close in space and time, so we control the loss of information due to subsampling that preserves a representative coverage of space and time.

We partition our data by years and pixels and then apply subsampling within each partition.  
The subsample contains a fixed number of observations (here set to two) for each year-pixel combination. 
We thus obtain approximately $50,000$ observations in the subsample, in line with the rule of thumb of \citet{Baddeley.al.2014,Baddeley2015} that the subsample  should be at least a factor four larger than the number of event points. The resulting models can be run on standard desktop computers (16Gb of memory).  
Within pixel-year combinations, we use non-uniform  random sampling to overweight specific parts of the predictor space. For inference on the FWI-month interaction, we set different sampling probabilities for FWI values above and below the empirical FWI-quantile at $p_{\text{FWI}}$ for each pixel-year. Values above the threshold are expected to correspond to more fire-prone conditions, and we over-represent them,
\eg  by fixing sampling probabilities $p_{\text{SS}}=0.9$ for FWI values below the threshold. To appropriately identify seasonal effects, 
we choose the month among June--October at random. For instance, high FWI values tend to be less frequent in October, but uniform subsampling of months gives them more weight. With this scheme, we obtain a positive sampling probability $p_k>0$ for each observation $N_k$ in \eqref{eq:lgcp-discretized}, and likelihood weights  are $\omega_k=1/p_k$ for the selected observations. Simulation experiments (see Supplement) 
motivated taking $(p_{\text{FWI}},p_{\text{SS}})=(0.7,0.9)$.

\subsection{Fully Bayesian inference using INLA-SPDE}
\label{sec:methods-inlaspde}

The integrated nested Laplace approximation \citep[INLA][]{Rue.al.2009,Lindgren.al.2015,Opitz.2017b} is a Bayesian technique for generalized additive models with Gaussian random effects.  It uses astutely designed deterministic approximations for accurate posterior inference on model parameters, random effects and predictions conditional on data. 
INLA enables transfer of information across components, appropriate uncertainty assessment and estimation of shared effects. 
We implement \emph{Penalized Complexity priors} \citep[PC priors,][]{Simpson.al.2017} in our models to control the complexity of model components.
Such priors penalize the distance of the prior of a model component towards a simpler baseline at a constant rate. 

\begin{figure}[t]
\centering
\begin{subfigure}{.35\linewidth}
    \includegraphics[width=.99\textwidth]{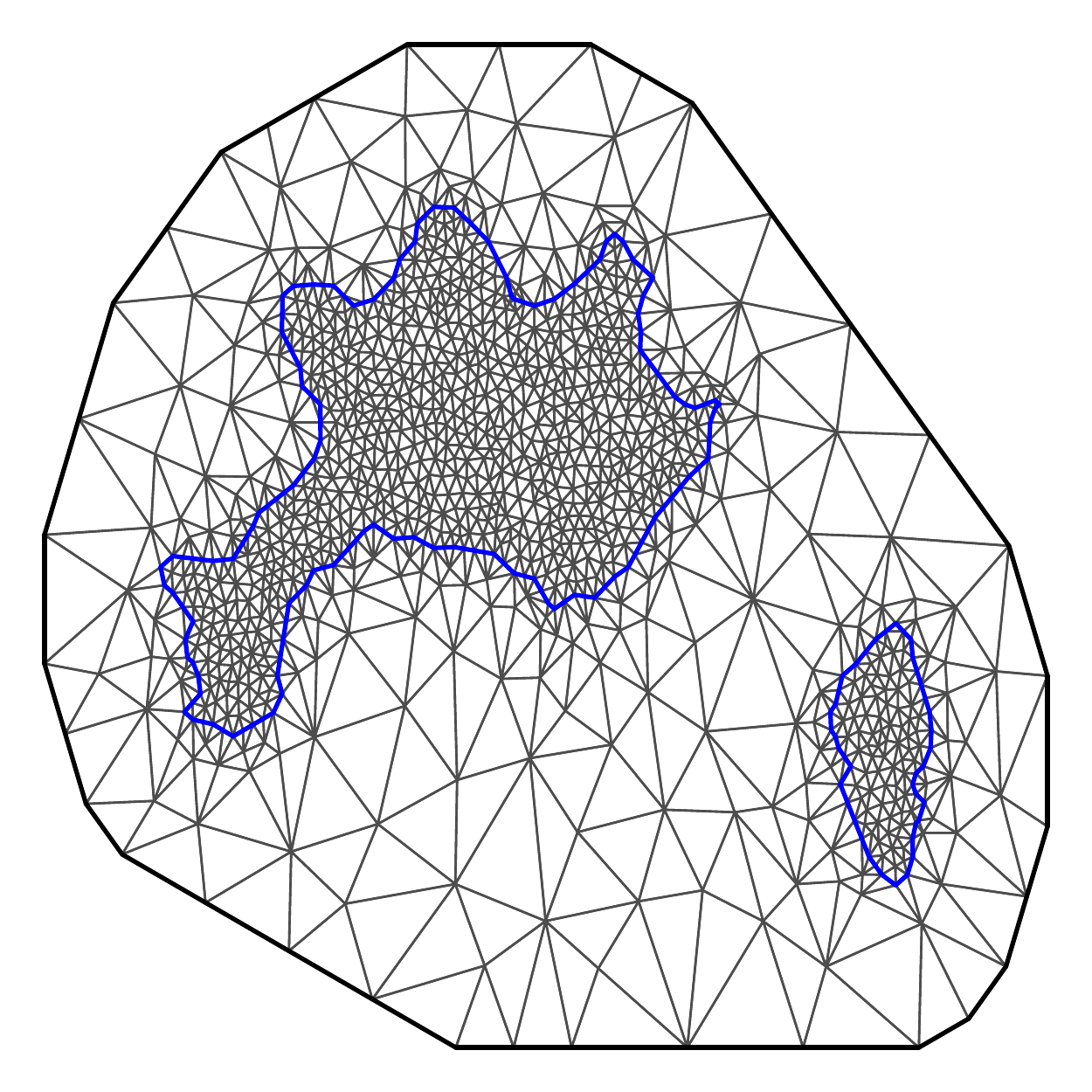}
\end{subfigure}
\begin{subfigure}{.3\linewidth}
    \includegraphics[width=.99\textwidth]{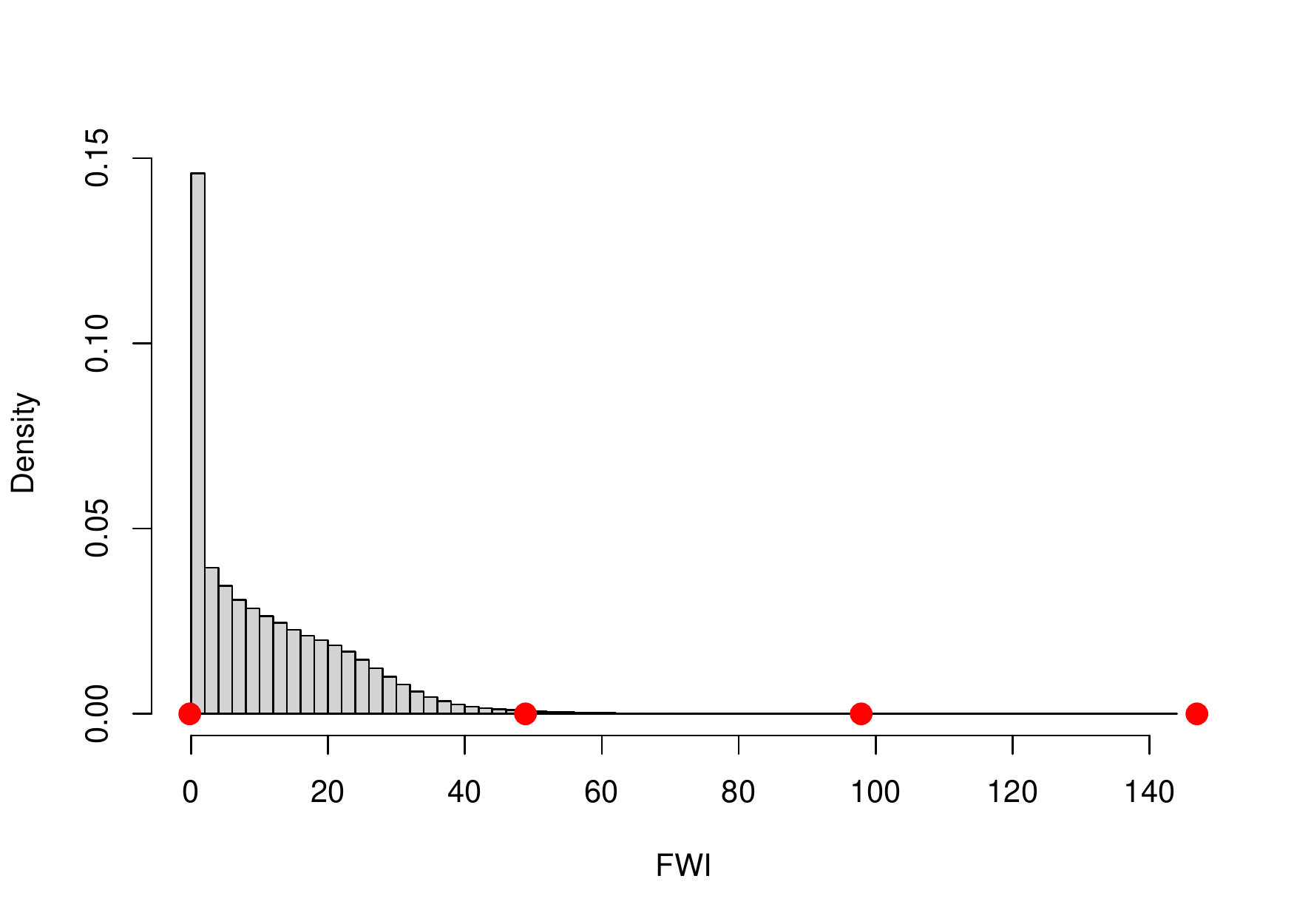}\\
    \includegraphics[width=.99\textwidth]{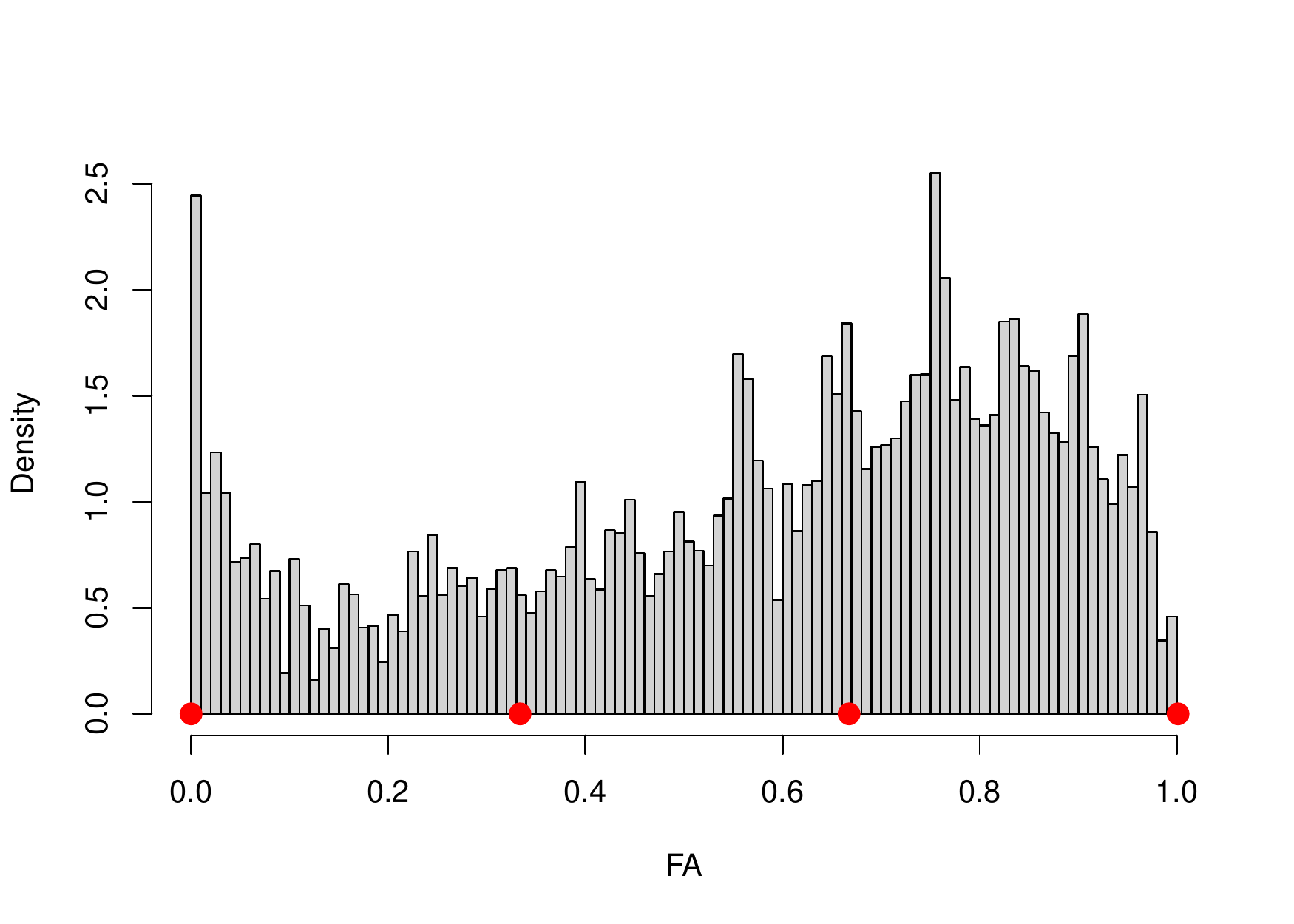}
\end{subfigure}
  \caption{Discretization of random effects with SPDE-based Gaussian prior processes. Left: Triangulation mesh of the study area (blue contours) for the SPDE approach. Neumann boundary conditions are set on the exterior (black) boundary to obtain a unique solution. The finite element solution defines a Gauss--Markov random vector with one variable in each node. Right: Histograms of FWI and FA values. The red points indicate where the spline knots are placed.}
    \label{fig:mesh}
\end{figure}

Owing to the large number of pixels in our problem, spatial Gaussian random effects and their  conditional distributions must be tractable in this setting. We use the Mat\'ern covariance function for random effects (denoted $g$), given  as follows for two points $s_1$ and $s_2$:
$$
\mathrm{Cov}\{g( s_1), g( s_2)\} = \sigma^2 2^{1-\nu} (\kappa||  s_1 -  s_2||)^{\nu} K_{\nu}(\kappa ||  s_1 -  s_2||) /\Gamma(\nu), \quad \sigma, \nu >0,
$$
with Euclidean distance $||\cdot||$, gamma function $\Gamma$, the modified Bessel function of the second kind $K_{\nu}$, and the standard deviation and smoothness parameters $\sigma$ and $\nu$. The \emph{empirical range} at which the correlation drops to approximately $0.1$ is $r=\sqrt{8\nu}/\kappa$.
Numerically convenient representations through approximating Gauss-Markov random fields (GMRF, characterized by sparse precision, \ie inverse covariance, matrices) are constructed by solving a stochastic partial differential equation 
\citep[SPDE,][]{Lindgren.al.2011,Krainski.al.2018}, where we fix the smoothness $\nu$ at unity.   
The discretization points  
are chosen as the nodes of a finite element representation (\eg the triangulation of space for $d=2$, or spline nodes for $d=1$), which enables efficient inference for random effects representing spatial variation ($d=2$) or nonlinear functions ($d=1$ for the FWI and FA effects). Our spatial triangulation mesh in Figure~\ref{fig:mesh} has $1114$ nodes. It is less dense in the extended zone around the study area to ensure that SPDE boundary conditions have negligible influence on the study area. The four splines knots for FWI and FA are evenly spaced throughout the feature space.

\section{Point processes  with moderate and extreme marks}\label{sec:model}

\emph{Point processes} govern the  space-time point patterns of occurrences; \emph{size processes} govern the moderate-level and extreme quantitative marks. 
We write $N_{it}$ for the number of wildfire occurrences on day $t\in\{1,\dots,n\}$ and over the $8\text{km} \times 8\text{km}$ grid cell $i \in \{1,\dots, 1143\} $ with centroid  $s_i$, 
and $\mathcal{A}_{i,t}\subset \mathcal{D}$ for the space-time cell with volume $|\mathcal{A}_{i,t}|=64$ ($\text{km}^2 \times \text{day}$). If $N_{it}>0$, we let $\bm Y_{it}=(Y_{it,1},\ldots,Y_{it,N_{it}}) \in (1,\infty)^{N_{it}}$ denote the corresponding quantitative marks. We write $z_k(s,t)$ ($k=1,\ldots,K$) for known deterministic covariates.

We model data of escaped fires ($>1$~ha), whose occurrence structure is captured by a regression component \textsc{COX} defining a LGCP. A logistic regression component \textsc{BIN} is used to classify fires into moderate ($0$) and large ($1$) according to exceedance above a fixed threshold $u$, \ie to provide the thinning of the point pattern and leave only extreme wildfires. Based on Figure~\ref{fig:burnt-surfaces}, we consider a fire size $Y_{it,k}$ to be extreme if $Y_{it,k}>79\text{ha}$ ($k=1,\dots,N_{it}$); \ie $u=79$.  We write $\bm R_{it}=(R_{it,1},\ldots,R_{it,N_{it}}) \in \{0,1\}^{N_{it}}$ for the vector of binary exceedance indicators $R_{it,k}=\mathbb{I}(Y_{it,k}>u)$.  
Moderate wildfire sizes $Y_{it,k}\in(1,u]$  are modeled through a Beta regression component \textsc{BETA} applied to pretransformed values $(Y_{it,k}-1)/(u-1)$.
 The Beta distribution, usually parametrized by two shape parameters $a,b>0$, 
 is here parametrized through a precision parameter $\phi=a+b>0$ and the mean $\mu_{it}^\mathrm{BETA}=a/(a+b)\in(0,1)$ with $\mathrm{logit}$-link function, such that $a=\mu_{it}^\mathrm{BETA} \phi$ and $b=\phi(1-\mu_{it}^\mathrm{BETA})$; it is a flexible location-shape family for interval-valued data, which can be used with INLA. For large wildfires, we build on the  extreme-value framework in~\S\ref{sec:explore:evt} and model excesses $Y_{it}-u>0$ above $u$
through a GPD regression component \textsc{GPD} to characterize extreme wildfires. Following \citet{Opitz.al.2018}, we use a log-link function for the median $\mu_{it}^\mathrm{GPD}$ of the GPD.  

 Some hyperparameters (\eg precision parameters of priors for fixed effects) are fixed a priori, but those that may strongly influence the posterior model structure are estimated.  Priors are fully detailed in the Supplement. 



\subsection{Bayesian hierarchical multi-response regression}
\label{sec:regression-equation}


Our modeling assumptions in \S\ref{sec:likelihood_COX} entail  the following structure for the linear COX predictor:
$$
\mu^{\mathrm{COX}}_{it} = \log\int_{\mathcal{A}_{it}} \lambda(s,t) \mathrm{d}(s,t) = \log \lambda(s_i,t) +\log |\mathcal{A}_{it}|.$$ 
We construct the system of regression equations in a Bayesian generalized additive mixed model (GAMM) as follows: 
\begin{align*}
N_{it}\mid \mu^{\mathrm{COX}}_{it} \sim\  &\mathrm{Poisson}\{\exp(\mu^{\mathrm{COX}}_{it}) \}, \\
R_{it,k}\mid \mu^{\mathrm{BIN}}_{it} \sim\  &\mathrm{Bernoulli}\{ \logit^{-1}( \mu^{\mathrm{BIN}}_{it} )\}, \quad k=1,\dots,N_{it},\\
\{ Y_{it,k} - u \mid R_{it,k} =1,\, \mu^{\mathrm{GPD}}_{it} \} \sim\  &\mathrm{GPD}\{ \exp( \mu^{\mathrm{GPD}}_{it} ), \xi\}, \\ 
 \{ (Y_{it,k}-1) / (u-1)  \mid R_{it,k} =0,\, \mu^{\mathrm{BETA}}_{it}\} \sim\  & \mathrm{Beta}\{\logit^{-1}(\mu^{\mathrm{BETA}}_{it}), \phi\};
\end{align*}
\begin{align*}
\mu_{it}^\mathrm{COMP} &= \sum_{k=1}^K g_k^\mathrm{COMP}\{z_k(s_i, t); \bm\theta^\mathrm{COMP},  \bm\theta^\mathrm{SHR}  \}, \quad \mathrm{COMP}=\{\mathrm{COX}, \mathrm{BIN}, \mathrm{GPD}, \mathrm{BETA}\}; 
\end{align*}
\begin{align*}
   \bm \theta = (\xi,\phi,\bm \theta^\text{COX}, \bm\theta^\text{BIN}, 
   \bm \theta^\text{GPD}, 
   \bm \theta^\text{BETA},
   \bm\theta^\mathrm{SHR})   \sim \text{Hyperpriors},
\end{align*}
where terms $g_k^\mathrm{COMP}$
capture linear or nonlinear influence of the covariates in the corresponding model component.
The specifics of $\bm\theta$ are discussed below. 


By construction, the intensity function $\lambda_{\mathrm{exc}}$ of the point process of large fires satisfies $\lambda_{\mathrm{exc}}(s_i,t)\leq \lambda(s_i,t)$. 
The exceedance probability $\mathrm{logit}^{-1}{\mu^{\mathrm{BIN}}_{it}}=\lambda_{\mathrm{exc}}(s_i,t)/ \lambda(s_i,t)$ defines the  independent Bernoulli probability of the full point pattern in COX. Since $
\lambda_{\mathrm{exc}}(s_i,t) = \exp(\mu^{\mathrm{BIN}}_{it}) \exp(\mu^{\mathrm{COX}}_{it}) / \{ 1+\exp(\mu^{\mathrm{BIN}}_{it}) \}$ and typically $\exp(\mu^{\mathrm{BIN}}_{it})\approx 0$, 
we obtain  $\log \lambda_{\mathrm{exc}}(s_i,t) \approx \mu^{\mathrm{BIN}}_{it}+\mu^{\mathrm{COX}}_{it}$.




\subsection{Sharing latent effects} \label{sec:model:sharing}

For maximal flexibility, we could incorporate mutually independent spatial effects into all model components. However, models would become overly complex, with too many spatial effects and hyperparameters to estimate,  and with high posterior uncertainties in the spatial effects of the BIN and GPD components due to the relatively small number of large wildfires. We strike a balance by sharing spatial random effects between model components of the point and size processes, though with a preliminary model selection procedure (see \S\ref{sec:model:selection}) that avoids compromising the quality of model fit and predictions. We set SPDE-based spatial GMRF priors $g^{\text{COX-BETA}}$, $g^\text{COX-BIN}$ and $g^{\text{BIN-GPD}}$ (recall \S\ref{sec:methods-inlaspde}) for the shared spatial effects. We use superscripts to indicate  the two components into which we jointly incorporate an effect, and we use $n$ to indicate the number of latent random variables for the corresponding effect (in superscript):
\begin{align*}
{ g^{\text{COX-BETA}} ( s_i )} & \sim  \mathcal{GP}_{\mathrm{2D\text{-}SPDE}}(\bm \omega_1), & \quad n^{\text{COX-BETA}} &= 1114 , \\ 
{ g^\text{COX-BIN} ( s_i ) } & \sim  \mathcal{GP}_{\mathrm{2D\text{-}SPDE}}(\bm \omega_2), & \quad n^{\text{COX-BIN}} &= 1114 , \\ 
{ g^{\text{BIN-GPD}} ( s_i )} & \sim  \mathcal{GP}_{\mathrm{2D\text{-}SPDE}}(\bm \omega_3), & \quad n^{\text{BIN-GPD}} &= 1114 ,
\end{align*}
where $\bm \omega_1$, $\bm \omega_2$ and $\bm \omega_3$ consist of separate Mat\'ern range $r$ and standard deviation $\sigma$ parameters with PC priors \citep{Fuglstad.al.2018}. Each shared effect is additively included in the linear predictor of the second component and then shared towards the first component with a scaling factor $\beta \in \mathbb{R}$, with superscripts to denote the two components.
We denote the vector of sharing-related hyperparameters by $\bm \theta^\mathrm{SHR}= (\bm \omega_1, \bm \omega_2, \bm \omega_3, \beta^\text{COX-BETA}, \beta^\text{COX-BIN}, \beta^\text{BIN-GPD} )$, and use   flat, independent zero-centered Gaussian hyperpriors for  the scaling factors.

Sharing allows modeling of residual spatial effect components  that jointly affect multiple model responses, such as land-use features at the Wildland-to-Urban interface \citep{Stewart.2007}, where human activities intermingle with wildland vegetation. Accurate sharing improves parsimony of the model and borrows estimation strength for random effects across model components by simultaneously using data from several response types.
Expert knowledge should guide the choice of which spatial effects are shared between specific components; sharing coefficients different from zero provide novel insight into the interplay of spatial structures across these components.

\subsection{Prior structure of linear predictors} \label{sec:model:point}

We let $z_\text{FWI}(s_i,t)$ and $z_\text{FA}(s_i,t)$ denote the average FWI and FA on day $t$ in grid cell $i$, and by $a(t)$ and $m(t)$ the corresponding year and month of day $t$. Using notation $\alpha$ for the intercept and $g$ for the other GAMM components, the prior structure of the model component COX for escaped fire occurrences is
\begin{align*}
\mu^\mathrm{COX}_{it} = &\alpha^\text{COX} + g_{1}^\text{COX} ( s_i ) + \beta^{\text{COX-BETA}} { g^{\text{COX-BETA}} ( s_i )} + \beta^\text{COX-BIN} { g^\text{COX-BIN} ( s_i ) } \\
 &+ g^\text{COX}_\text{2}\{z_\text{FA}(s_i,t)\} + g^\text{COX}_\text{3}\{z_\text{FWI}(s_i,t); m(t)\}  \\
 &+ g_{4}^\text{COX}\{a(t)\} + g_{5}^\text{COX}\{m(t)\};
\end{align*}
\begin{align*}
   g_{1}^\text{COX}(s_i) 
   & \stackrel{\mathrm{iid}}{\sim}   
   \mathcal{N}\{0, 1/\tau_1\}, &\quad n_1^{\text{COX}} &= 1143 , \\   
   g_{2}^\text{COX}(\bigcdot)   & \sim  \mathcal{GP}_{\mathrm{1D\text{-}SPDE}}(\bm \phi_1),  &\quad  n_2^{\text{COX}} &= 4 , \\
   g^\text{COX}_\text{3}(\,\bigcdot\, ; m)  & \sim  \mathcal{GP}_{\mathrm{1D\text{-}SPDE}}(\bm \phi_2),  \\
  g^\text{COX}_\text{3}(z_\text{FWI}; \,\bigcdot\,)   & \sim  \mathcal{GP}_{\mathrm{RW1}}(1/\tau_2 ), &\quad n_3^{\text{COX}} &= 4\times 5 = 20,\\
  g^\text{COX}_\text{4}(\bigcdot\,) & \sim  \mathcal{GP}_{\mathrm{RW1}}(1/\tau_3), &\quad n_4^{\text{COX}} &= 20, \\
   g^\text{COX}_\text{5}(\bigcdot\,)    &\sim  \mathcal{GP}_{\mathrm{RW1}}(1/\tau_4), &\quad n_5^{\text{COX}} &= 5; 
\end{align*}
\begin{align*}
   \bm \theta^\text{COX} = \{\alpha^\text{COX}, \bm \phi_1, \bm \phi_2, \tau_1, \tau_2, \tau_3, \tau_4\} \sim \text{Hyperpriors}.
\end{align*}
Spatial occurrence hot-spots (see Supplement), 
may arise due to time-invariant land-use features. Moreover, spatial variation may be shared from patterns in the BETA and BIN components through the components ${ g^{\text{COX-BETA}} ( s_i )}$ and ${ g^{\text{COX-BIN}} ( s_i )}$, respectively. 
 The  month and year effects, $g^\text{COX}_\text{4}$ and $g^\text{COX}_\text{5}$, capture spatially homogeneous temporal variations in occurrence intensities. They are endowed with first-order random-walk priors $\mathcal{GP}_{\mathrm{RW1}}$ with a sum-to-zero constraint for identifiability; \eg for the yearly effect and for $a=1995,\dots,2013$,
 $$
 g_4^\mathrm{COX}(a+1)-g_4^\mathrm{COX}(a)\sim \mathcal{N}(0, 1/\tau_3), \quad
 \sum_{i=1995}^{2014}g_4^\mathrm{COX}(i)=0.
 $$
 The quadratic B-spline functions of FWI and FA are endowed with priors $\mathcal{GP}_{\mathrm{1D-SPDE}}$,  constrained to zero at the left boundary $0$ and constrained to sum to zero, respectively. Most wildfires in the region are caused by human activity, possibly leading to a nonlinear relationship between FA and occurrence intensity, as dense forest areas are often  exposed to low human activity. 
We allow for monthly variation of the nonlinear FWI effect  through separate $\mathcal{GP}_{\mathrm{1D-SPDE}}$-terms in $g_3$ for each month, linked across successive months with a $\mathcal{GP}_{\mathrm{RW1}}$-structure in the prior model.

The regression equation used for the Bernoulli process is 
\begin{align*}
\mu^\mathrm{BIN}_{it} =  &\alpha^\text{BIN} + { g^\text{COX-BIN} ( s_i ) } + \beta^{\text{BIN-GPD}} { g^{\text{BIN-GPD}} ( s_i )} + g^\text{BIN}_\text{1}\{z_\text{FWI}(s_i,t)\}  \\ &+ g^\text{BIN}_\text{2}\{z_\text{FA}(s_i,t)\} + g_3^\text{BIN}\{a(t)\}; 
\end{align*}
\begin{align*}
   g_{k}^\text{BIN}(\bigcdot) \sim &\mathcal{GP}_{\mathrm{1D\text{-}SPDE}}(\bm \zeta_k), \quad k=1,2, &\quad n_1^{\text{BIN}}, n_2^{\text{BIN}} &=5,\\
g_3^\text{BIN}(\bigcdot) \sim & \mathcal{GP}_{\mathrm{RW1}}( 1/\tau_5), &\quad n_3^{\text{BIN}} &= 5;
\end{align*}
\begin{align*}
   \bm \theta^\text{BIN} = \{\alpha^\text{BIN}, \bm \zeta_1, \bm \zeta_2, \tau_5 \} \sim \text{Hyperpriors}.
\end{align*}
The linear predictor of the Bernoulli probability has a simpler form than that of the occurrence component but still allows the capture of specific nonlinear effects of FWI and FA.  
In Figure~\ref{fig:maps-safran-grid}, we discern hot-spot areas of large fire occurrences that differ substantially from the  overall occurrence structure, and we aim to capture these residual effects through the shared spatial effects. 



The prior structure for the two mixture components of quantitative marks is
\begin{align*}
\mu^\mathrm{BETA}_{it} =  &\alpha^\text{BETA} + { g^\text{COX-BETA}( s_i ) }+  g^\text{BETA}_\text{1}\{z_\text{FWI}(s_i,t)\} + g^\text{BETA}_\text{2}\{z_\text{FA}(s_i,t)\}, \\
\mu^\mathrm{GPD}_{it}=  &\alpha^\text{GPD} + { g^{\text{BIN-GPD}} ( s_i )} + g^\text{GPD}_\text{1}\{z_\text{FWI}(s_i,t)\}  + g^\text{GPD}_\text{2}\{z_\text{FA}(s_i,t)\} \\ & + g_3^\text{GPD}\{a(t)\};
\end{align*}
\begin{align*}
 g_{k}^\text{BETA}(\bigcdot), g_{k}^\text{GPD}(\bigcdot) \sim &\mathcal{GP}_{\mathrm{1D\text{-}SPDE}}(\bm \kappa_k), \quad k=1,2, &\quad n_1^{\text{GPD}},n_2^{\text{GPD}},n_1^{\text{BETA}},n_2^{\text{BETA}} &=5,\\
g_3^\text{GPD}(\bigcdot) \sim & \mathcal{GP}_{\mathrm{RW1}}( 1/\tau_6), &\quad n_3^{\text{GPD}} &= 5;
\end{align*}
\begin{align*}
   \bm \theta^\text{MARK} = \{\alpha^\text{GPD}, \alpha^\text{BETA}, \bm \kappa_1, \bm \kappa_2, \tau_6 \} \sim \text{Hyperpriors}.
\end{align*}
A year effect, endowed with a random-walk prior, was included in some of the components (COX, BIN, GPD).
In all components (BETA, BIN, COX, GPD), we allow for  non-linear relationships with respect to FWI or FA. 


\subsection{Alternative model specifications}\label{sec:model:alternative}

We also consider size processes that do not model the moderate-level and extreme marks separately; \ie with no mixture representation of the size process. Similar models have been proposed in the literature \citep[\eg][]{Joseph2019}, although without the sharing of random effects. We use either the Gamma distribution for the full range of marks: $ Y_{it,k} \mid \mu^{\mathrm{SIZE}}_{it} \sim  \mathrm{Gam}\{ \exp( \mu^{\mathrm{SIZE}}_{it} ), \phi_\mathrm{Gam}\}$, or the Normal distribution for the logarithmic transformed marks: $\log Y_{it,k} \mid \mu^{\mathrm{SIZE}}_{it} \sim\  \mathcal{N}\{ \exp( \mu^{\mathrm{SIZE}}_{it} ), \phi_\mathcal{N}\} $, where the distributions are parameterized by the link function $\mu^{\mathrm{SIZE}}_{it}$ modeling the mean and precision parameters $\phi_\mathrm{Gam}=\exp(\mu^{\mathrm{SIZE}}_{it})^2/\mathrm{Var}(Y_{it,k})$ and $\phi_\mathcal{N}=1/\mathrm{Var}(\log Y_{it,k})$, respectively. In both cases 
\begin{align*}
\mu^\mathrm{SIZE}_{it}=  &\alpha^\text{SIZE} + { g^{\text{SIZE-COX}} ( s_i )} + g^\text{SIZE}_\text{1}\{z_\text{FWI}(s_i,t)\}  + g^\text{SIZE}_\text{2}\{z_\text{FA}(s_i,t)\} \\ & + g_3^\text{SIZE}\{a(t)\} + g^\text{SIZE}( s_i );
\end{align*}
\begin{align*}
g_{k}^\text{SIZE}(\bigcdot) \sim &\mathcal{GP}_{\mathrm{1D\text{-}SPDE}}(\bm \iota_k), \quad k=1,2, &\quad n_1^{\text{SIZE}}, n_2^{\text{SIZE}} &=5,\\
g_3^\text{SIZE}(\bigcdot) \sim & \mathcal{GP}_{\mathrm{RW1}}( 1/\tau_7), &\quad n_3^{\text{SIZE}} &= 5;
\end{align*}
\begin{align*}
   \bm \theta^\text{SIZE} = \{\alpha^\text{SIZE}, \bm \iota_1, \bm \iota_2, \tau_7 \} \sim \text{Hyperpriors},
\end{align*}
where the spatial effects ${ g^\text{SIZE-COX}( s_i ) }$  and ${ g^\text{SIZE}( s_i ) }$ are controlled by Mat\'ern parameters $\bm \omega_4$ and $\bm \omega_5$, similar to those in \S\ref{sec:model:sharing}.

\section{Results}
\label{sec:results}


\subsection{Model selection and comparison}\label{sec:model:selection}

Estimation was carried out using the INLA-SPDE approach described in \S \ref{sec:methods-inlaspde} by applying the subsampling scheme proposed in \S \ref{ssec:ss}. 
In a preliminary analysis of the regression models described in \S \ref{sec:model}, we used the Widely Applicable Information Criterion \citep[WAIC,][]{Watanabe.2010} in a step-wise manner to compare nested models with different components  in the regression equations (\eg linear vs nonlinear effects of explanatory variables) to choose their final forms. Due to the small number of extreme wildfires, their influence on WAIC is relatively small; we subsequently proceed with other model comparison tools that give more weight to large wildfires and their prediction. 

We label the model with prior structure detailed in \S\ref{sec:model:point} M1, and the model without spatial effects in the size and extreme occurrence components M2.
We also consider other models developed in the recent wildfire modeling literature. We refer to model M2 but without monthly variation in the FWI effect as M3, which is similar to the approach of \citet{Pimont2020}. 
We let M4 and M5 denote the models with the same point process model as M1 but with no mixture representation of the size process, for which we use a log-Normal or a Gamma response distribution with prior structure detailed in \S\ref{sec:model:alternative},  respectively. These models do not differentiate between extreme and non-extreme fires, but their response distributions have the been found to be good modeling candidates in \citet{Joseph2019}, though their approach does not use shared random effects as we do here. 

For the observed individual fires in the training (1995--2014) and validation (2015--2018) periods, we generated posterior predictive distributions of each model based on $500$ posterior simulations.
First, we evaluated the models' ability to predict exceedances above the empirical $90\%$ quantile of burnt areas using the AUC \citep{Fawcett.2006} and the Brier score \citep{Brier.1950}. The severity threshold chosen here is sufficiently high for extreme risk assessment, but not too high so as to retain enough observations to evaluate these scores with moderate uncertainty. 
Next, we also computed the scaled Continuous Ranked Probability Score (sCRPS)  suggested by \citet{Bolin-Wallin.2020} for averages of CRPS over non-identical predictive distributions, which corresponds to our setting. For these analyses, we kept the original locations of observed fires, and simulation is done from the size components only. By combining posterior simulations of the occurrence and size components, we also evaluated predictive  performance for burnt areas aggregated at the month-d\'epartement scale. 


\begin{table}[t]
\centering
\begin{adjustbox}{width=0.95\textwidth}
\centering
\begin{tabular}{lrrrrrr}
  \hline
  \\
 & Score & \multicolumn{5}{c}{Model} \\
\cline{3-7}
 & & M1 & M2 & M3 & M4 & M5\\ 
  \hline
   \multirow{4}{*}{Individual fires, $n=823$} &
sCRPS &  {2.74} &  {2.87} & 2.94 & 2.84 & 3.19 \\ 
& p-value  & - & $<5\%$ & $<1\%$& $<5\%$& $<1\%$ \\
\cline{3-7}
& $\text{Brier}_\text{q90}$ & 0.0855 & 0.0868 & 0.0866 & 0.0944 & 0.0967 \\ 
& p-value & - & $<5\%$ & $6\%$ & $<1\%$ & $<1\%$ \\  
\cline{3-7}
\cline{3-7}
  & $1-\text{AUC}_\text{q90}$ & 0.3052 & 0.3502 & 0.3516 & 0.3184 & 0.3122 \\ 
  & p-value & - & $<5\%$ & $<5\%$ & $40\%$ & $41\%$ \\ 
\cline{3-7}
  \hline
  D\'ep-month, $n=75$ & sCRPS & 3.55 & 3.62 & 3.64 & 3.62 & 3.58 \\ 
    & p-value & - & $7\%$ & $7\%$ & $9\%$ & $39\%$ \\ 
  \hline
\end{tabular}
\end{adjustbox}
\caption{Comparison of models using predictive scores (averaged over $n$ observations) calculated with data from the validation period: sCRPS, Brier and AUC scores for individual fires, and sCRPS for the spatiotemporally aggregated burnt areas at month-département scale, based on 500 simulations of the posterior models, with p-values for a permutation test comparing  to the best performing model M1. A lower score is better. }
\label{tab:benefits:spatial}
\end{table}
Table~\ref{tab:benefits:spatial} shows good relative performance of M1  for all scores when evaluating wildfire predictions on the validation period. To better grasp the uncertainty in scores, we show p-values of a permutation test assessing the significance of negative values in the differences of scores between M1 and the other models, based on 2000 permutations. 
For the sCRPS of individual fires, the score differences are all significant at the $5\%$ level. 
A general finding is that 
using sophisticated structures such as the mixture representation of size processes, sharing and monthly variation of FWI effect improves predictions; it further allows for the novel scientific insights presented in \S\ref{sec:application}. 

Comparison of M1 and M2 confirms the benefits of incorporating spatial random effects in the size model components in M1 using parsimonious structures, thanks to the sharing detailed in \S\ref{sec:model:sharing}. M1 performs better than M2, and
performances of  M2 and M3 are  similar for predicting wildfire sizes and their aggregation. 
 Model M1 performs better than  M4 and M5 especially with Brier and sCRPS scores, though in some cases improved scores have relatively lower confidence levels in light of the p-values. 
Models M4 and M5 perform better than M2 and M3 for some scores like the AUC and sCRPS at the month-département aggregation because of the additional sharing and spatial random effects in the size component. However, it performs worse for the other scores due to having no components focusing specifically on large wildfires. Despite good scores of M4 and M5 on the training set (not shown), their comparatively worse results on the validation sample suggest that the log-Normal and Gamma distribution for burnt areas do not predict the extremes in new data as well as 
M1.
In particular, M1 does not show issues of overfitting. 


\subsection{Visual inspection of posterior predictive densities}

We also assess the predictive behavior of our chosen model M1 with visual diagnostics,  especially for tail behavior. 
First, we assess whether the size component correctly predicts extreme wildfires at the regional level of specific départements.  In the Supplement 
(Figure~\ref{fig:val:exceedreg2:test}), we use simulations from the posterior model at pixel-days where fires have been observed to compare empirical and predicted excess probabilities over increasingly high thresholds, starting at $100$ha. 
Predictions are generally good since most empirical exceedance probabilities  fall within the inter-quantile range of  simulations, except for the d\'epartements of Var and Haute-Corse with small underestimation at very large thresholds. These two d\'epartements have large continuous forest areas and saw unusually many large wildfires in the summer of 2017. 
Much of their land has acidic soils that favor biomass production and are covered by tall and dense shrubland, 
so 2017 fires were harder to contain due to their higher heat release. 
Overall, the tail behavior in  fire-prone and less fire-prone regions is well discriminated by the model.

Next, we consider the occurrence component by comparing the number of simulated and observed fires  aggregated by year over the study region  (Supplement 
: Figure~\ref{fig:val:prediction:BA}, left display). Observed annual fire numbers for both test and training set fall within the inter-quantile range of simulations for more than half of the study period. M1 captures the relatively high observed numbers of 2001, 2003 and 1998 (training) and 2017 (test), while it also accurately predicts  the sharp decrease in 2018.

Lastly, we jointly evaluate the size and occurrence components of our model M1.
We aggregated simulated burnt areas   by year, over the whole spatial region in Figure~\ref{fig:val:prediction:BA} (right display), and over d\'epartements  in Figure~\ref{fig:val:reg1}.  The global time trend in observed burnt areas is well captured throughout the years in Supplement
: Figure~\ref{fig:val:prediction:BA}, with inter-quantile coverage of $42\%$. M1 captures the exceptional peak in 2003, which is poorly predicted by M4 and M5  and the Firelihood model of \citet{Pimont2020}. M1 also succeeds in accurately predicting the moderately high burnt areas in  2001 and 2017, and it generally discriminates well between fire conditions leading to small, moderate, large and very large fire numbers. 
Figure~\ref{fig:val:reg1} further shows that regional differences across d\'epartements are well captured by M1,  with most panels showing roughly $50\%$ inter-quantile range coverage. Overall, our model appropriately captures spatiotemporal variation and provides satisfactory regionalized forecasts  for operational purposes. 



\subsection{Principal results of the main model M1}\label{sec:mode:results}

\subsubsection{Covariate effects}

\begin{figure}[t]
\centering
  \begin{subfigure}[b]{.19\linewidth}
    \centering
    \includegraphics[width=.99\textwidth]{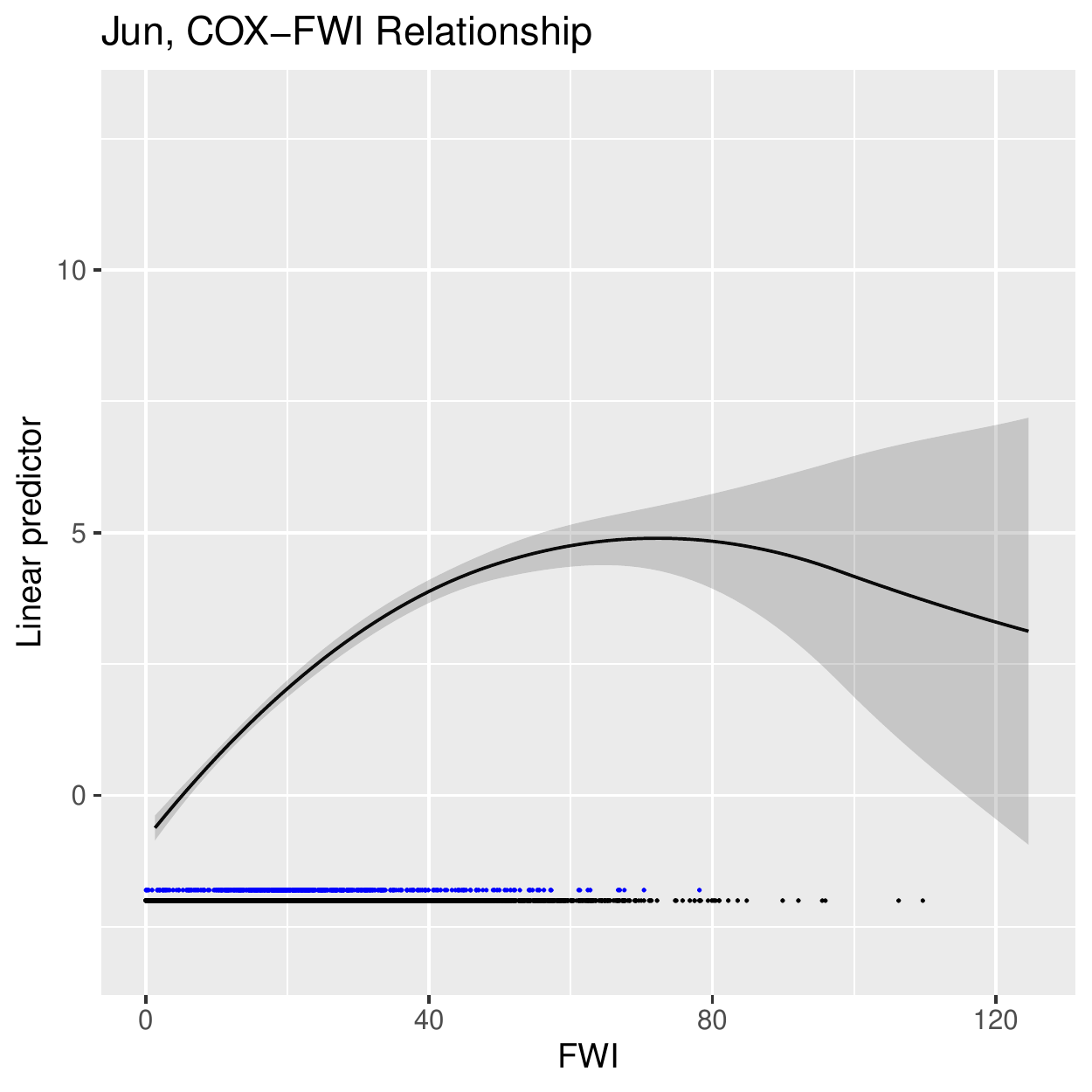}
  \end{subfigure}
   \begin{subfigure}[b]{.19\linewidth}
    \centering
    \includegraphics[width=.99\textwidth]{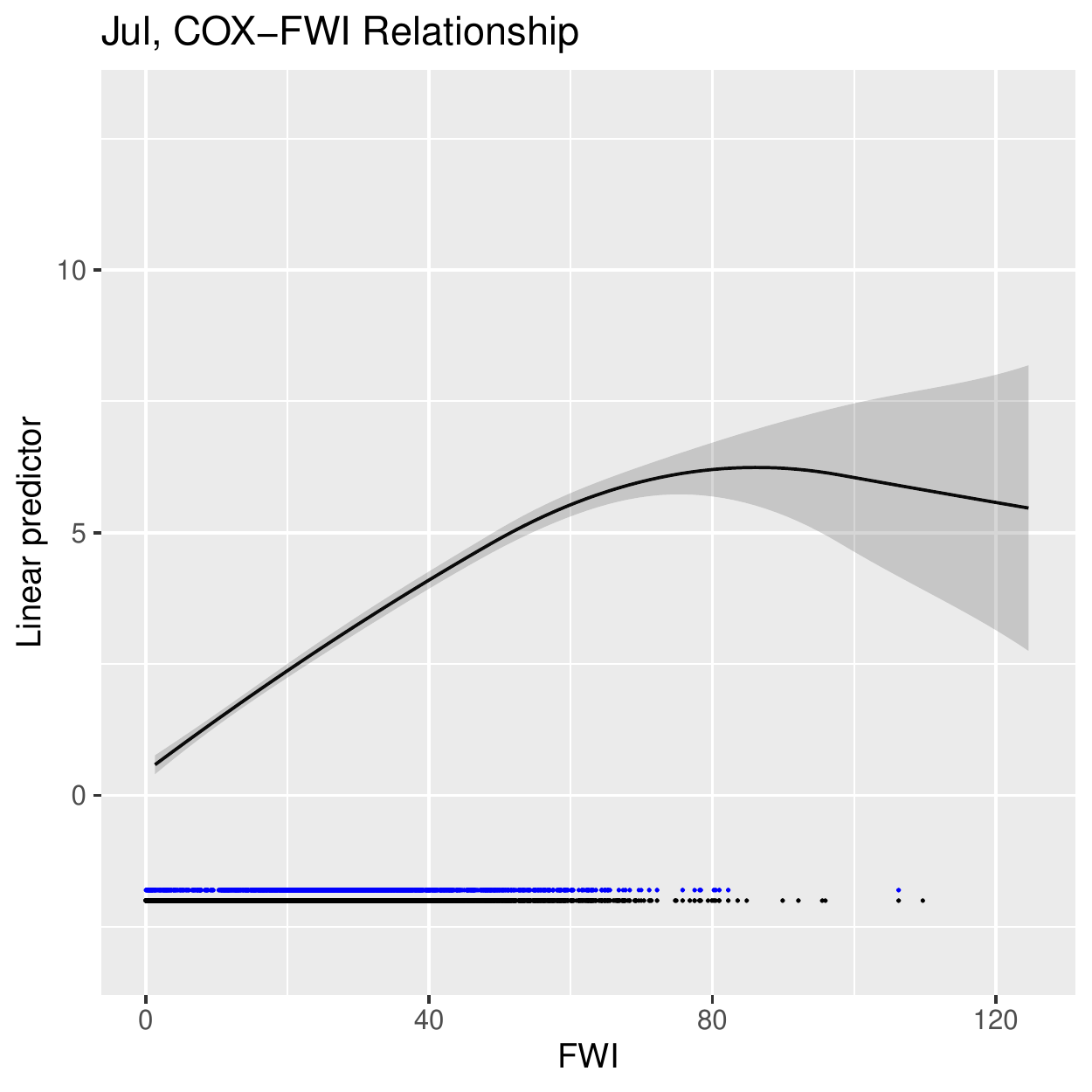}
  \end{subfigure}%
  \begin{subfigure}[b]{.19\linewidth}
    \centering
    \includegraphics[width=.99\textwidth]{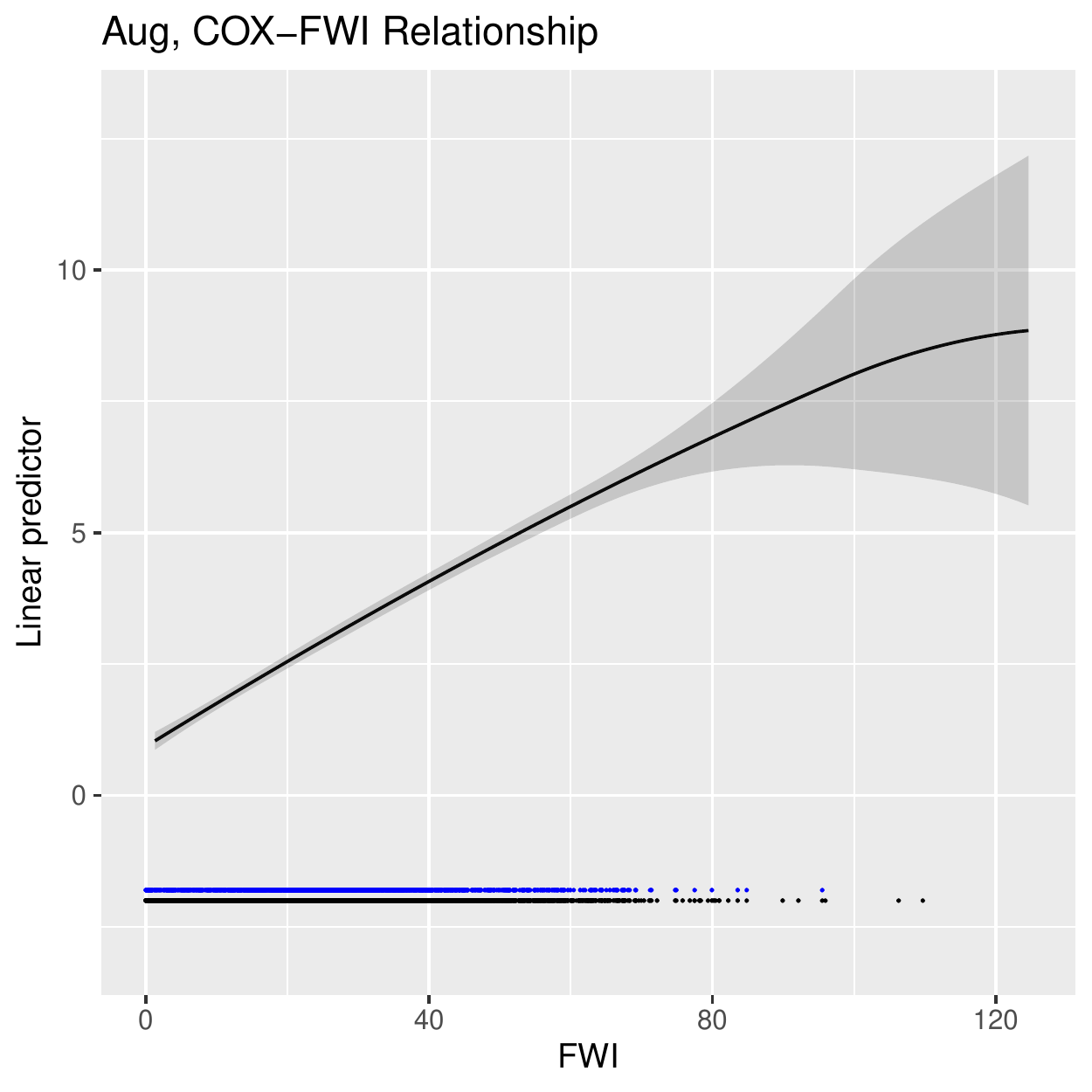}
  \end{subfigure}
  \begin{subfigure}[b]{.19\linewidth}
    \centering
    \includegraphics[width=.99\textwidth]{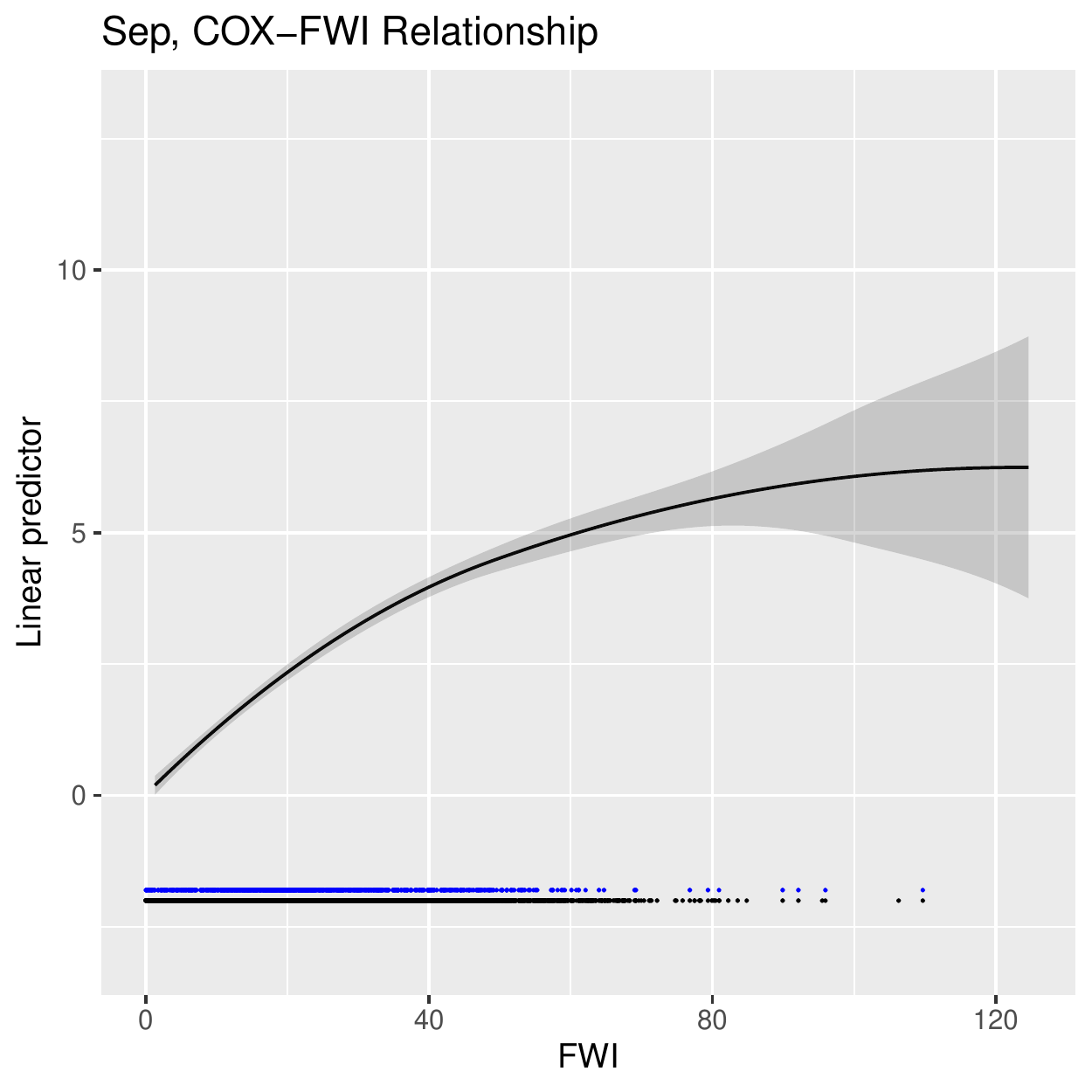}
  \end{subfigure}%
  \begin{subfigure}[b]{.19\linewidth}
    \centering
    \includegraphics[width=.99\textwidth]{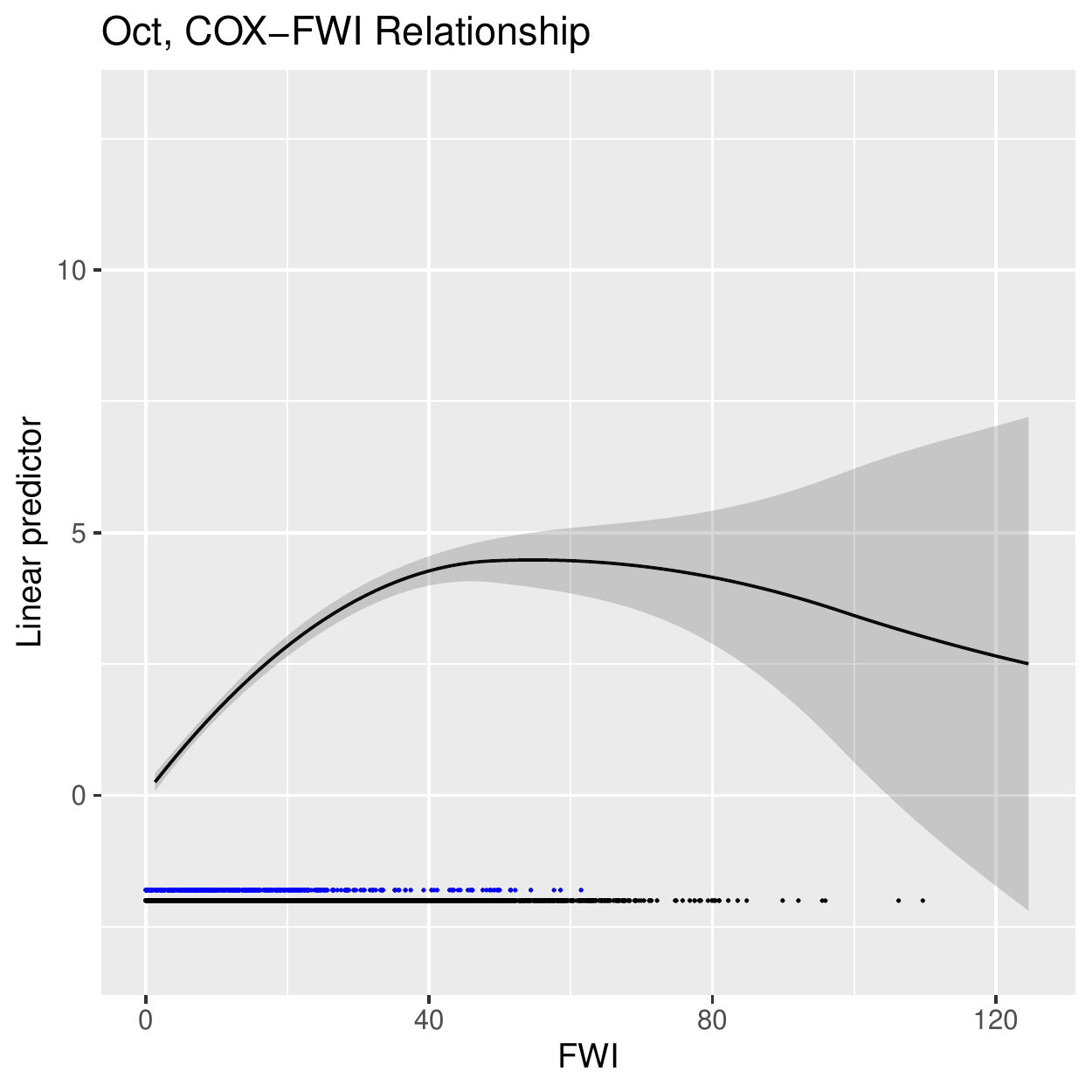}
  \end{subfigure}
  \caption{Posterior estimates of $g^\text{COX}_\text{3}(\bigcdot ; m) + g^\text{COX}_\text{5}(m)$, $m=1,\dots,5$, the joint FWI-month effect, for June--October in the linear predictor of the point process (COX) component.
  The blanket of black and blue points at the bottom of each plot shows FWI values for pixel-days with fires in any month and the specific month, respectively.}
  \label{fig:results:cox:fwimonth}
\end{figure}

For the COX component, Figure~\ref{fig:results:cox:fwimonth} shows the month-specific  FWI effect, with significant differences across months. For easier comparison, we have subtracted the same value from all curves such that the posterior mean is $0$ for FWI$=0$ in September. Throughout, the posterior means are monotonically increasing up to FWI of 75.  Curves flatten for higher values of FWI especially at the beginning and end of the wildfire season, with a slight decrease of the curve towards the highest FWI. 

\begin{figure}[t]
\centering
  \begin{subfigure}[b]{.25\linewidth}
    \centering
    \includegraphics[width=.99\textwidth]{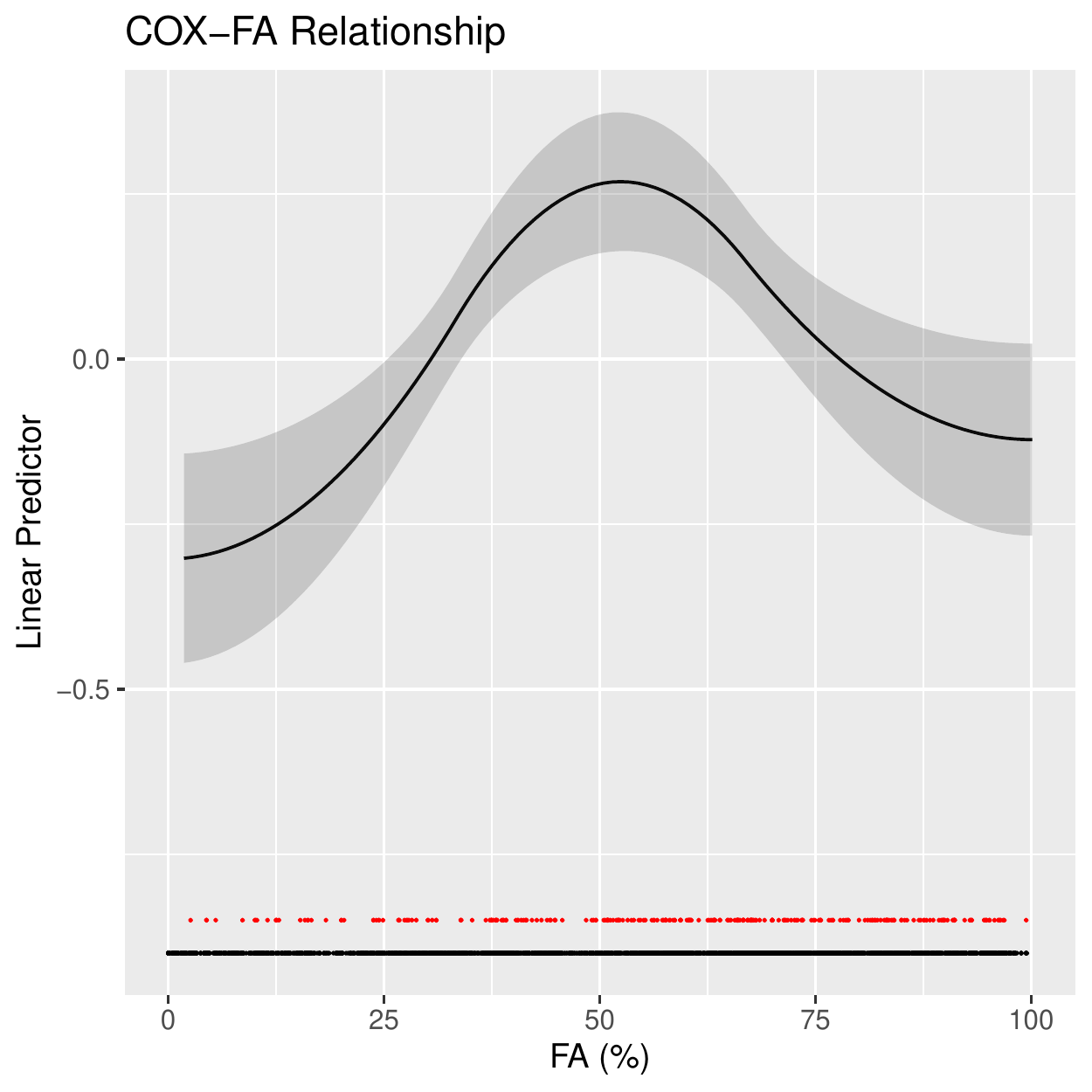}
  \end{subfigure}%
  \begin{subfigure}[b]{.25\linewidth}
    \centering
    \includegraphics[width=.99\textwidth]{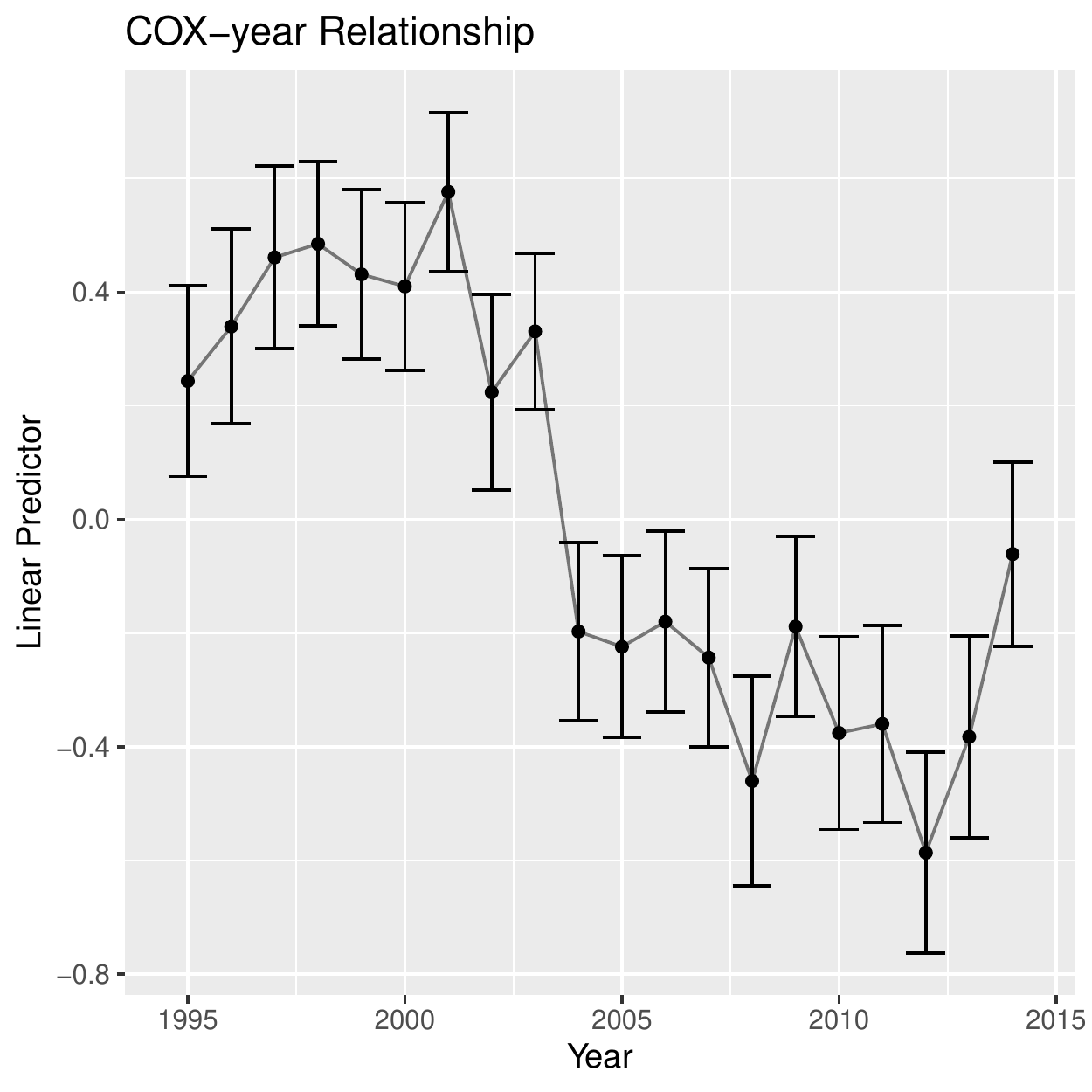}
  \end{subfigure}
  \begin{subfigure}[b]{.25\linewidth}
    \centering
    \includegraphics[width=.99\textwidth]{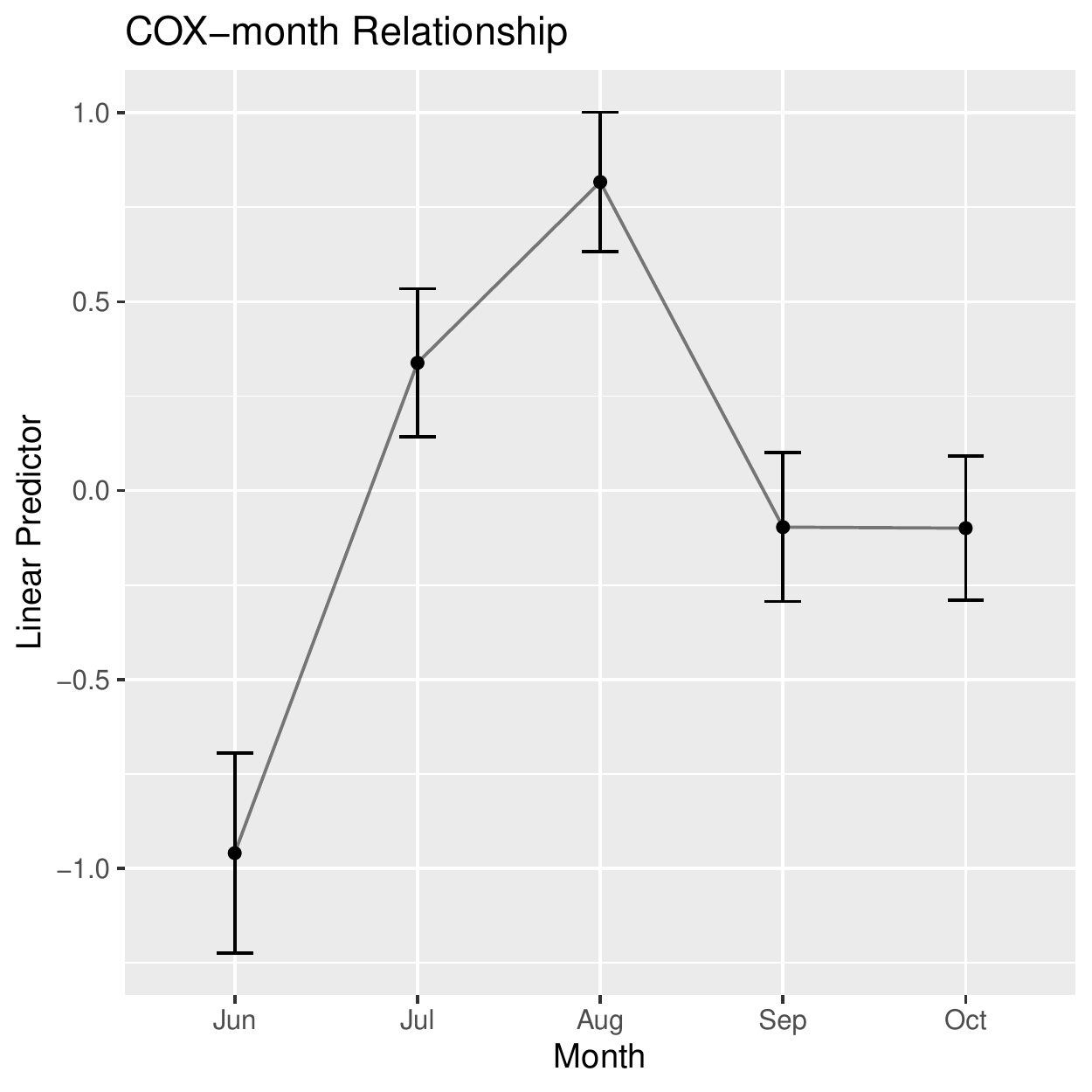}
  \end{subfigure} \\
  \begin{subfigure}[b]{.25\linewidth}
    \centering
    \includegraphics[width=.99\textwidth]{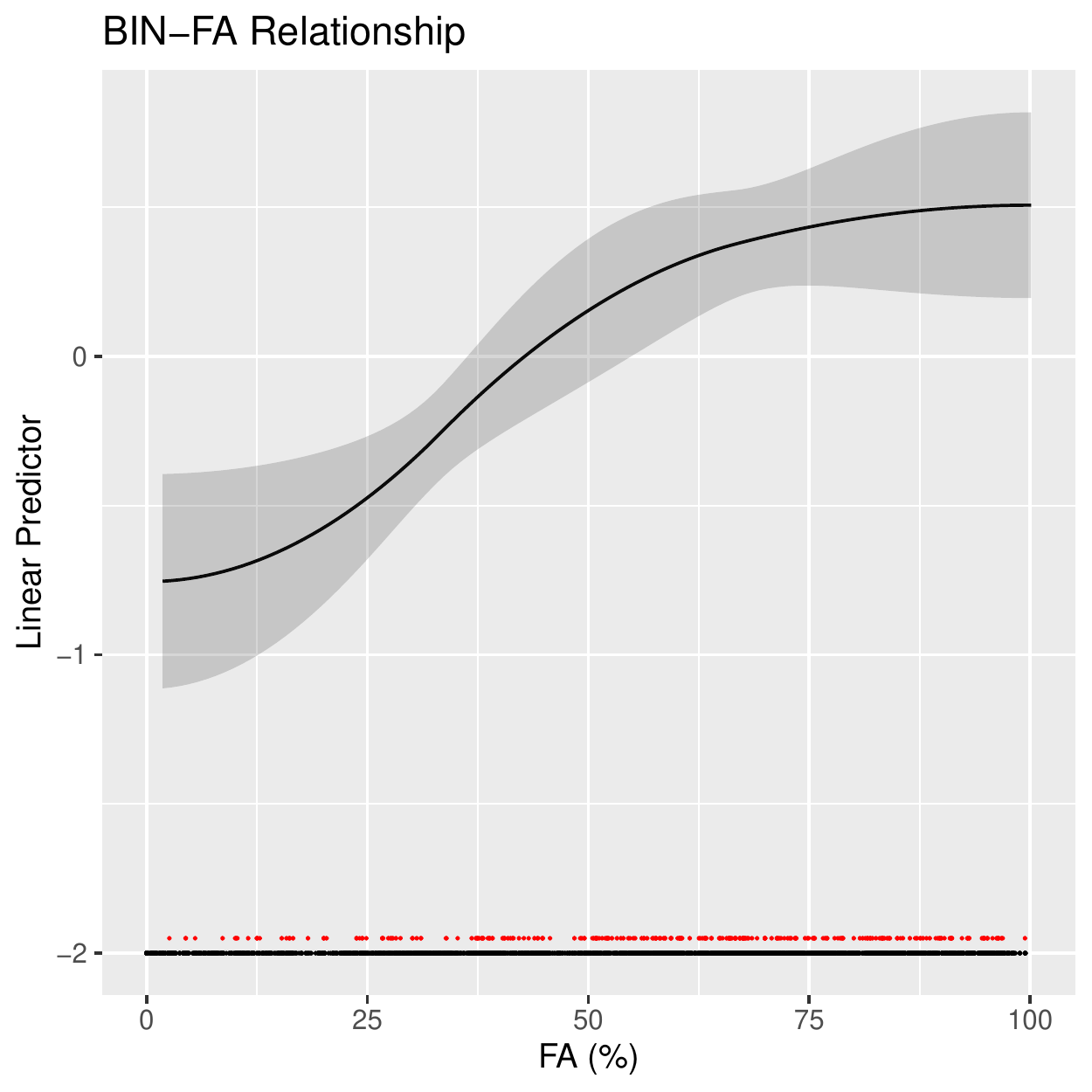}
  \end{subfigure}%
  \begin{subfigure}[b]{.25\linewidth}
    \centering
    \includegraphics[width=.99\textwidth]{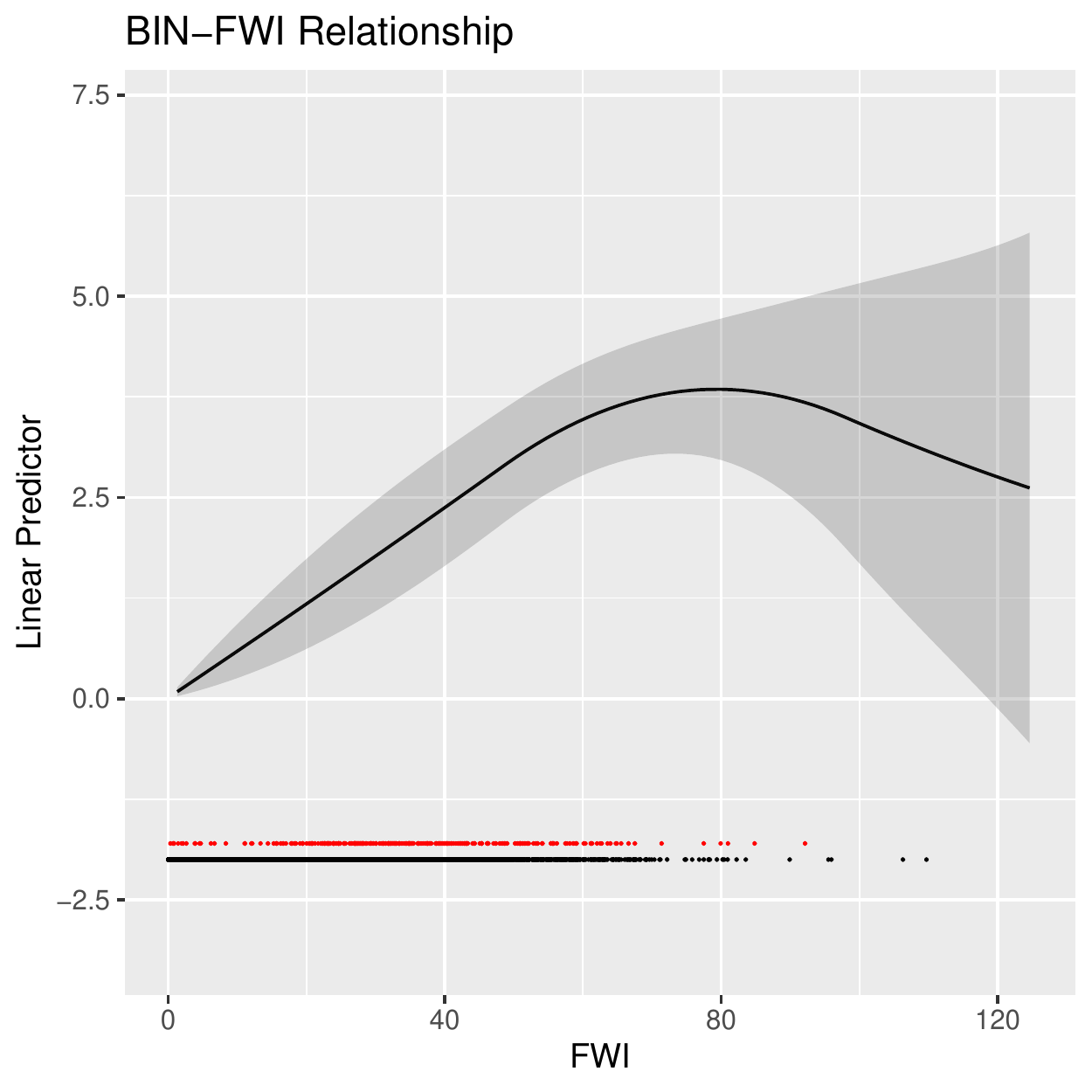}
  \end{subfigure}
  \begin{subfigure}[b]{.25\linewidth}
    \centering
    \includegraphics[width=.99\textwidth]{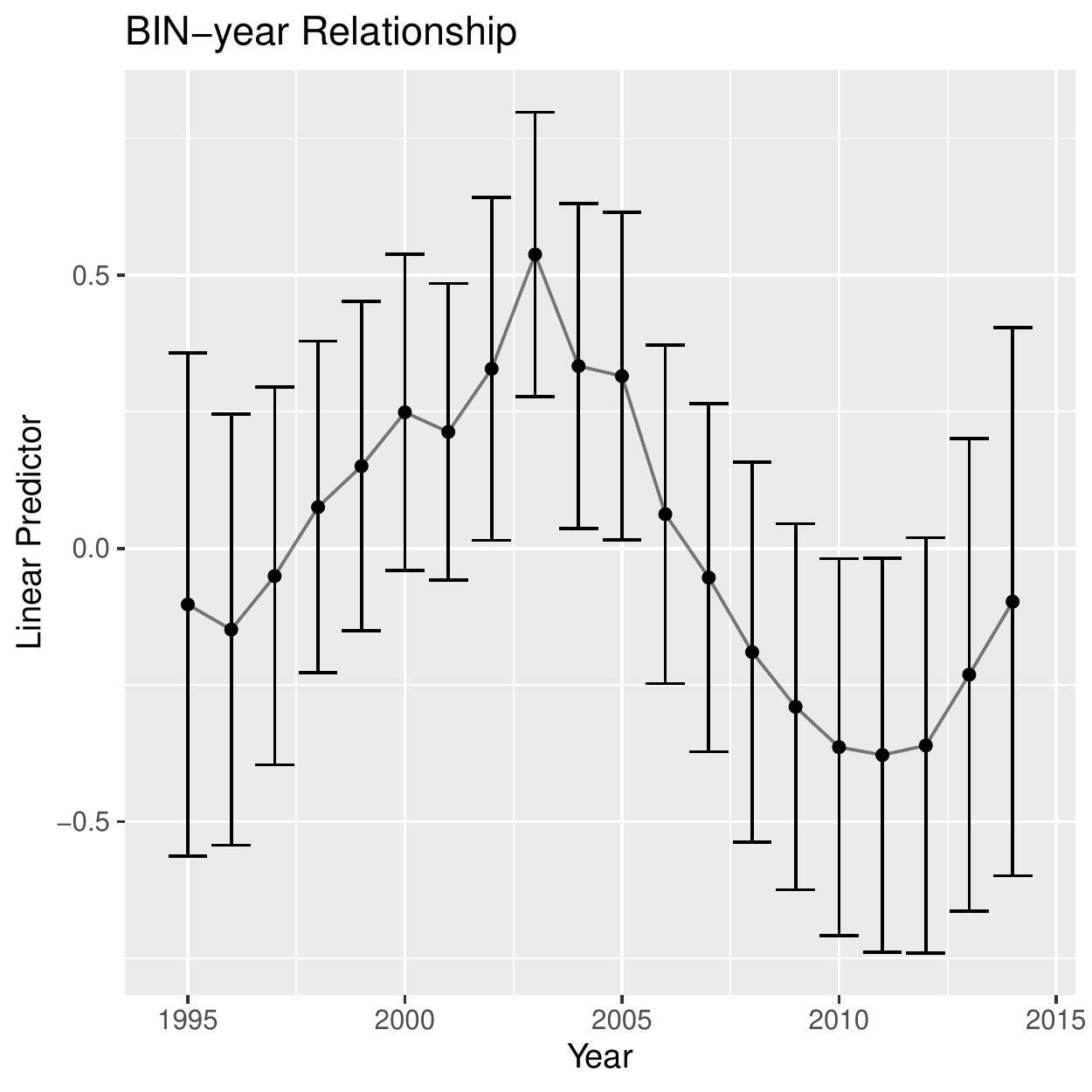}
  \end{subfigure} \\
  \caption{Posterior estimates of $g^\text{COX}_\text{2}(\bigcdot)$ (FA effect, top left panel), $g^\text{COX}_\text{4}(\bigcdot)$ (year effect,  top middle panel), $g^\text{COX}_\text{5}(\bigcdot)$ (month effect, top right panel), $g^\text{BIN}_{\text{2}}(\bigcdot{})$ (FA effect, bottom left panel), $g^\text{BIN}_{\text{1}}(\bigcdot{})$ (FWI effect, bottom middle panel) and $g^\text{BIN}_\text{3}(\bigcdot{})$ (year effect, bottom right panel)  in the linear predictor of the point process (COX) component and large wildfire probability component (BIN). At the bottom of some displays, the blanket of black and red points shows FA/FWI values for  pixel-days with moderate and large fires, respectively.}
    \label{fig:results:cox:fayear}
\end{figure}

The posterior partial effect of FA on the COX component in  Figure~\ref{fig:results:cox:fayear} indicates a  ``bump"-shaped effect of FA, which is significant based on pointwise credible intervals. Very high FA can be considered as a good proxy for relatively few human-induced wildfire ignitions, while very low FA means lack of fuel. Clearly, expected wildfire ignition numbers are not proportional to forest area. 

As to temporal partial effects without spatial variation (Figure~\ref{fig:results:cox:fayear}), the posterior year effect suggests a strong, significant drop in wildfire activity after 2003, potentially related to policy changes after the exceptional 2003 events. 
The partial month effect (top right display of Figure~\ref{fig:results:cox:fayear}, corresponding to the intercept of its combined effect with FWI in Figure~\ref{fig:results:cox:fwimonth}) is lowest at the start of the wildfire season and peaks in August. 

As to the probability of occurrence of large fires (BIN),  Figure~\ref{fig:results:cox:fayear} (bottom middle display)  highlights a strong positive posterior effect of FWI, increasing monotonically and significantly up to FWI values of around $75$, before it dampens at very large FWI values, similar to the COX component: large wildfires are relatively more frequent with moderate to high FWI values.  The probability of large wildfires tends to increase with increasing FA in a grid cell (Figure~\ref{fig:results:cox:fayear}, bottom left display), which is reasonable since larger FA fuel is available over large areas. The pointwise credible bounds of yearly effects across the study period suggest that the occurrence of large events was significantly higher around the peak in 2003.

\begin{figure}[t!]
\centering
  \begin{subfigure}[b]{.25\linewidth}
    \centering
    \includegraphics[width=.99\textwidth]{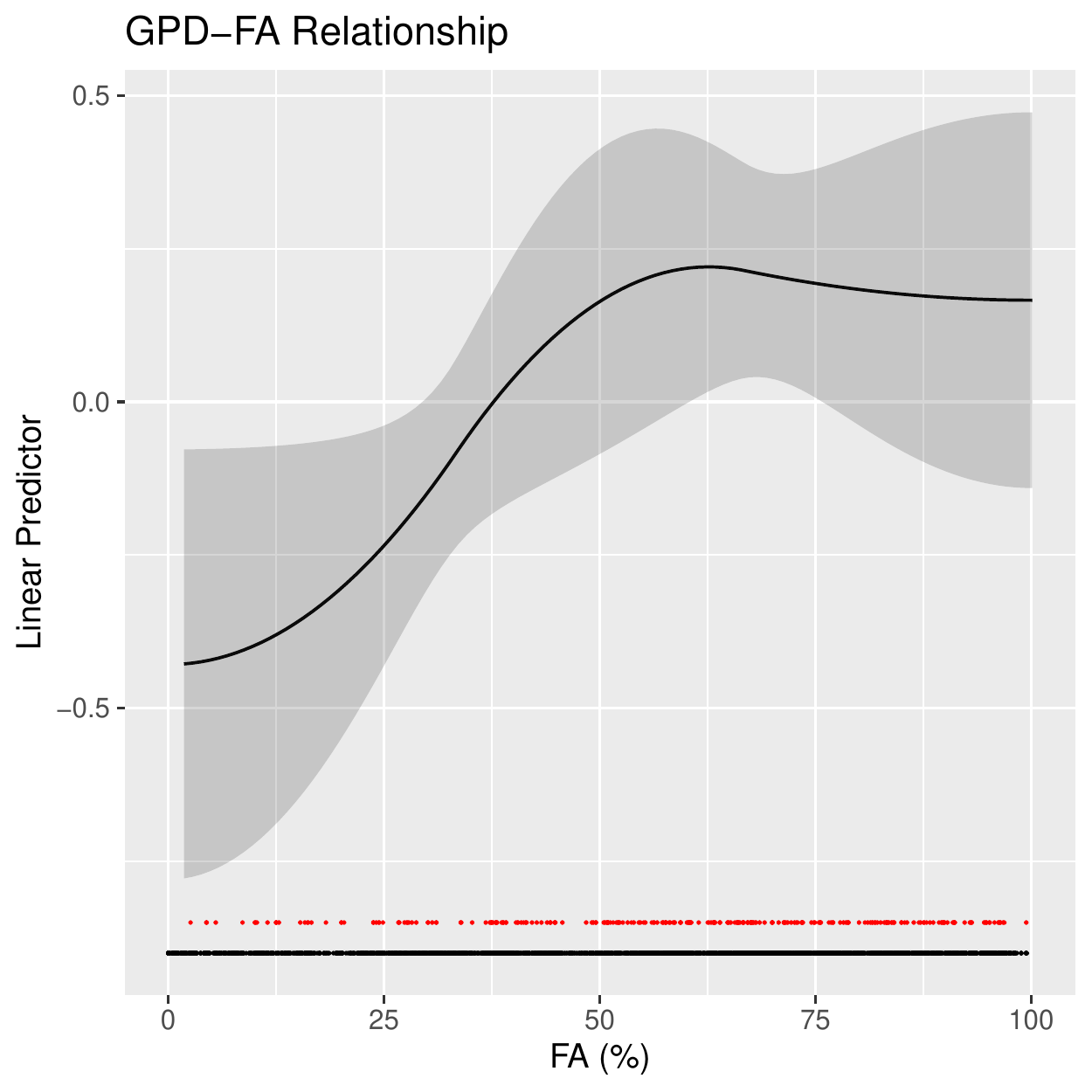}
  \end{subfigure}%
  \begin{subfigure}[b]{.25\linewidth}
    \centering
    \includegraphics[width=.99\textwidth]{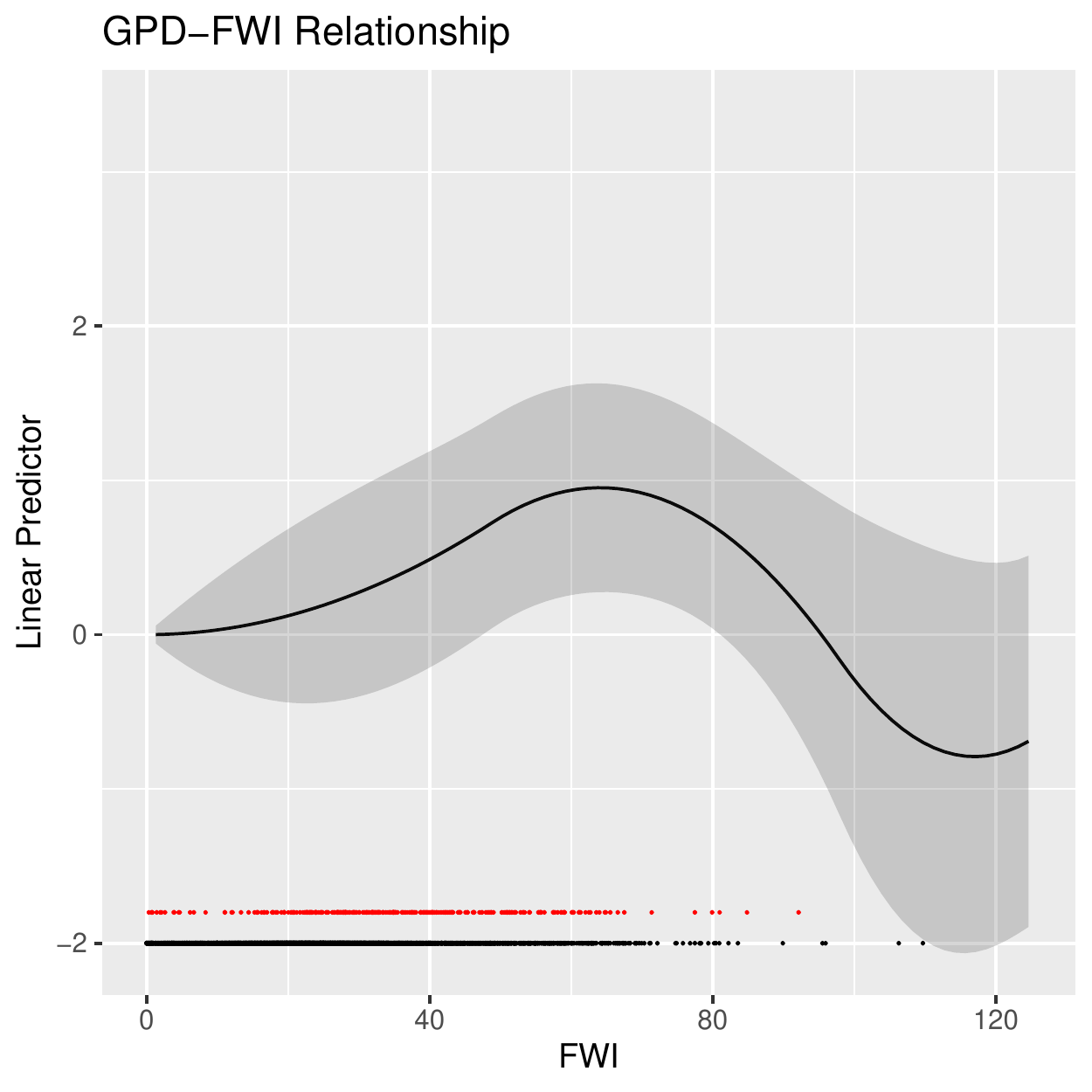}
  \end{subfigure}
  \begin{subfigure}[b]{.25\linewidth}
    \centering
    \includegraphics[width=.99\textwidth]{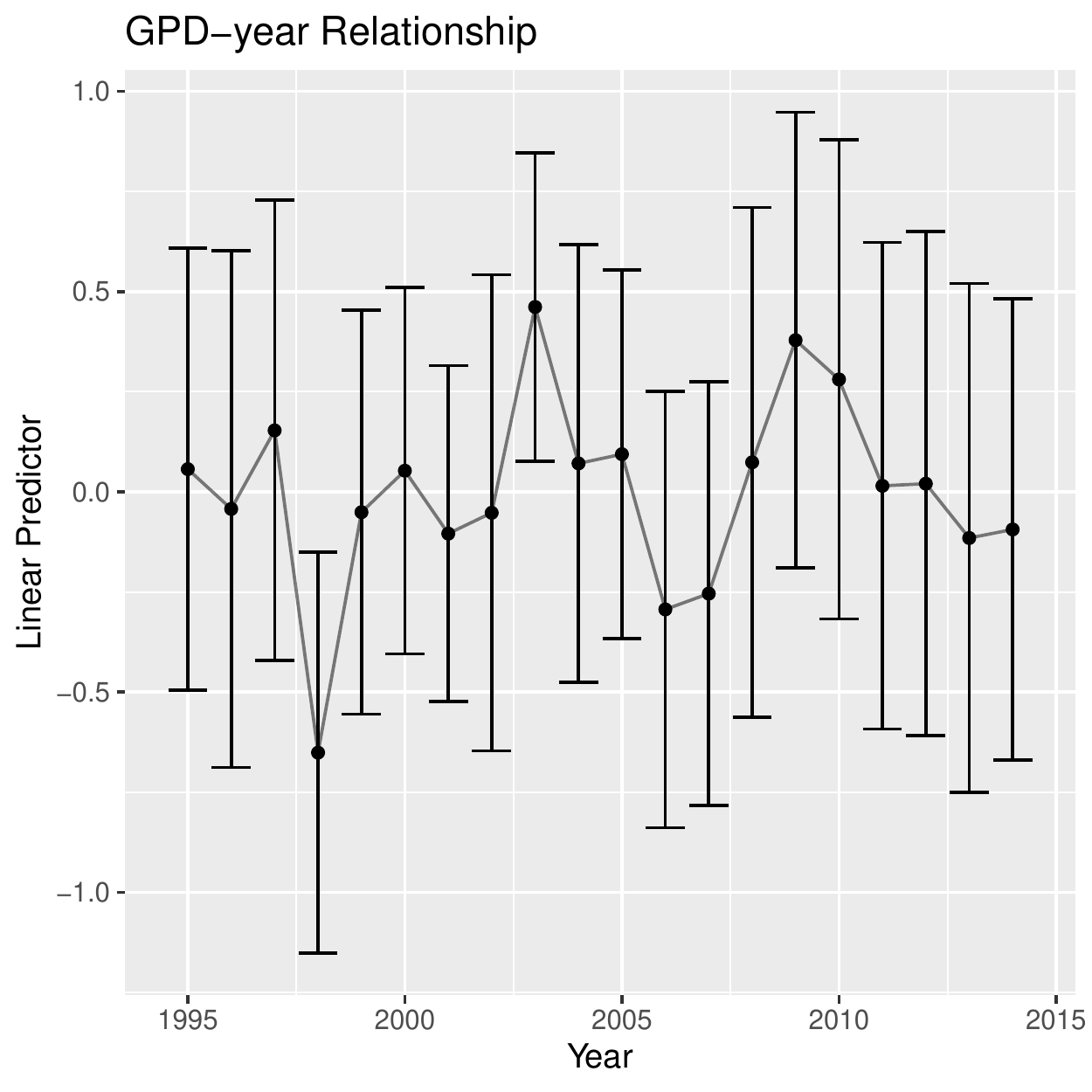}
  \end{subfigure} \\
  \begin{subfigure}[b]{.25\linewidth}
    \centering
    \includegraphics[width=.99\textwidth]{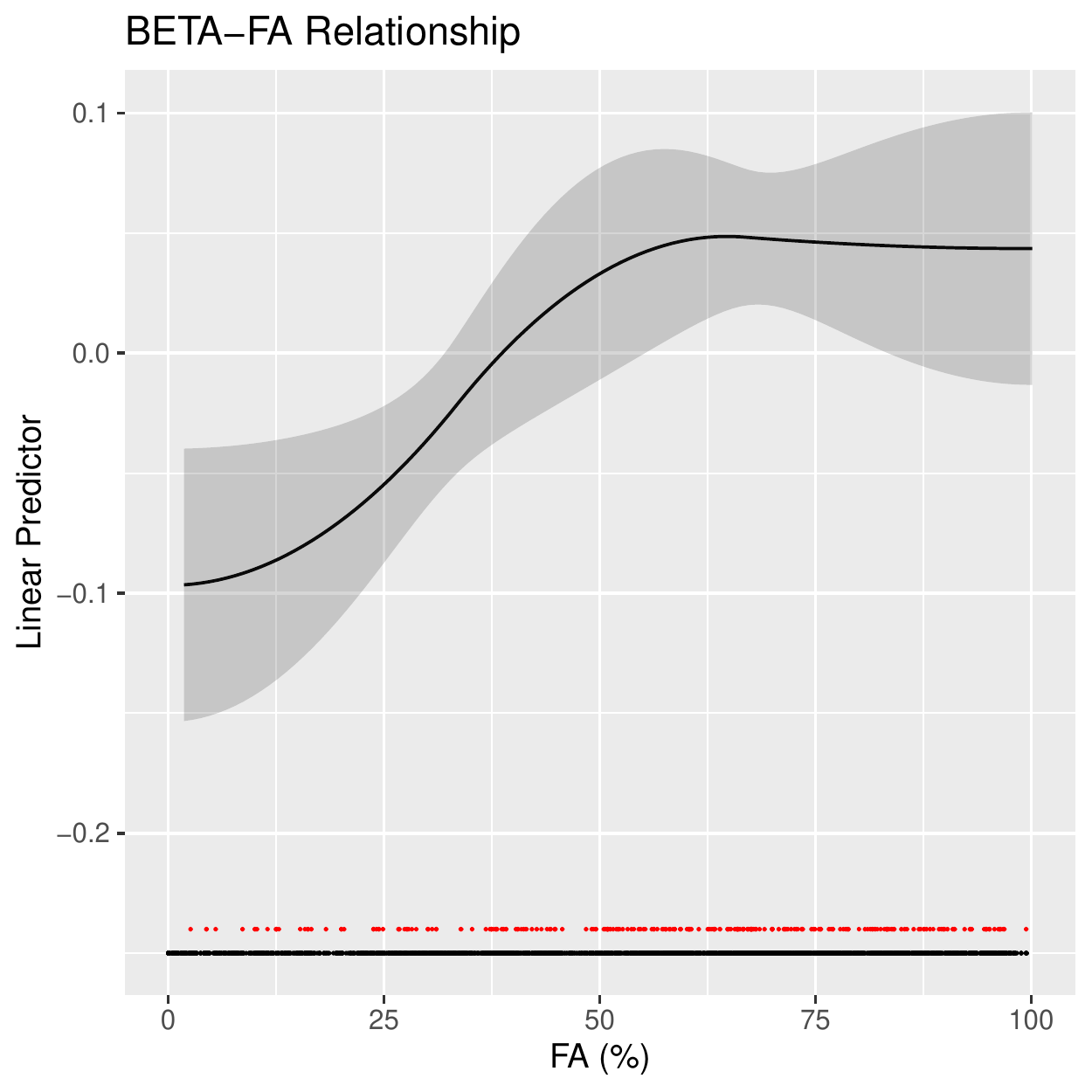}
  \end{subfigure}%
  \begin{subfigure}[b]{.25\linewidth}
    \centering
    \includegraphics[width=.99\textwidth]{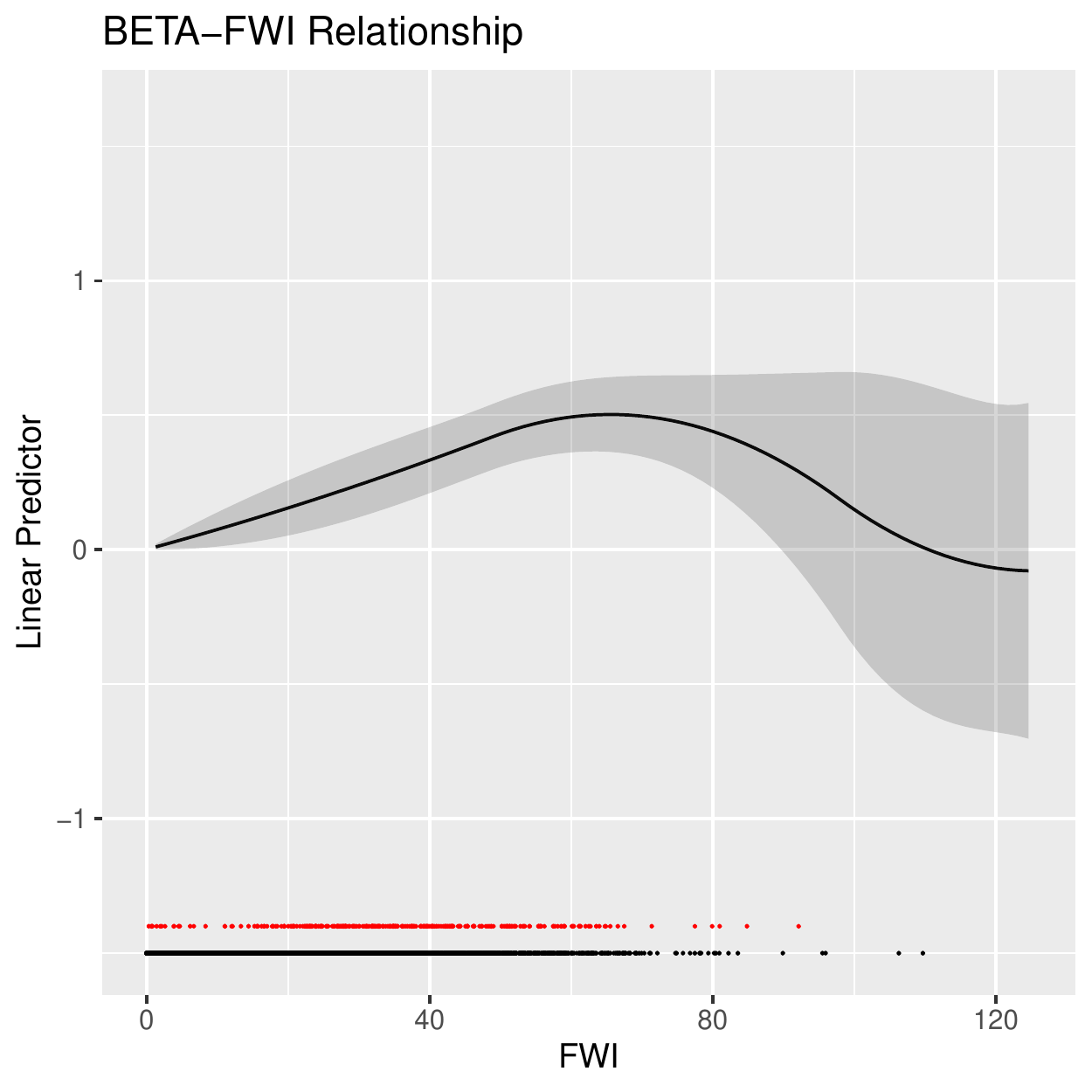}
  \end{subfigure}\\
  \caption{Panels as in Figure \ref{fig:results:cox:fayear}. Posterior estimates of $g^\text{GPD}_\text{2}(\bigcdot{})$ (FA effect, top left), $g^\text{GPD}_\text{1}(\bigcdot{})$ (FWI effect, top middle), $g^\text{GPD}_\text{3}(\bigcdot{})$ (year effect, top right), $g^\text{BETA}_\text{2}(\bigcdot{})$ (FA effect, bottom left) and $g^\text{BETA}_\text{1}(\bigcdot{})$ (FWI effect, bottom right) in the linear predictor of the large wildfire size component (GPD) and moderate wildfire size component (BETA). }
        \label{fig:results:gpd:fwifayear}
\end{figure}

In the additive effects of the two mixture components GPD and BETA of the size distribution shown in Figures~\ref{fig:results:gpd:fwifayear}  we find similar posterior effects of FWI and forest area for  extreme and moderate sizes. Posterior estimates imply that fires become larger when FWI increases up to around 60 but the effect flattens for higher FWI. 
Increasing FA leads to increasing wildfire size in both components up to $50\%$ and then reaches a plateau. For the year effect in the extreme component GPD, no clear trend arises, though 2003 has a significantly higher effect than 1998.

\subsubsection{Sharing effects induce correlated wildfire activity components}
 
We here focus only on the spatial effects that were shared between model components. 
The $95\%$ credible intervals for the scaling parameters $\beta^{\text{COX-BETA}}$, $\beta^{\text{COX-BIN}}$ and $\beta^{\text{BIN-GPD}}$ do not cover $0$; their posterior estimates for the triplet ($2.5\%$ quantile, mean, $97.5\%$ quantile) are $(6.4,{10.3},14.0)$, $(-3.1,{-1.8},-0.9)$, and $(0.5,{1.0},1.6)$, respectively. The posterior mean of $\beta^{\text{COX-BETA}}$ is positive and the one of $\beta^{\text{COX-BIN}}$ is negative, which confirms  significant positive and negative sharing between the COX and BETA, and the COX and BIN model components, respectively; these findings provide new spatial insights for fire risk management in \S\ref{sec:application}. The posterior means for the effective range parameters of the shared spatial fields, $r^{\text{COX-BETA}}$, $r^{\text{COX-BIN}}$ and $r^{\text{BIN-GPD}}$,  are 34.3km, 26.2km and 156.9km, respectively.  Posterior mean maps of their corresponding spatial random effects are shown in the Supplement. 

\begin{figure}[t!]
\centering
\begin{subfigure}{.2\linewidth}
    \includegraphics[width=.99\textwidth]{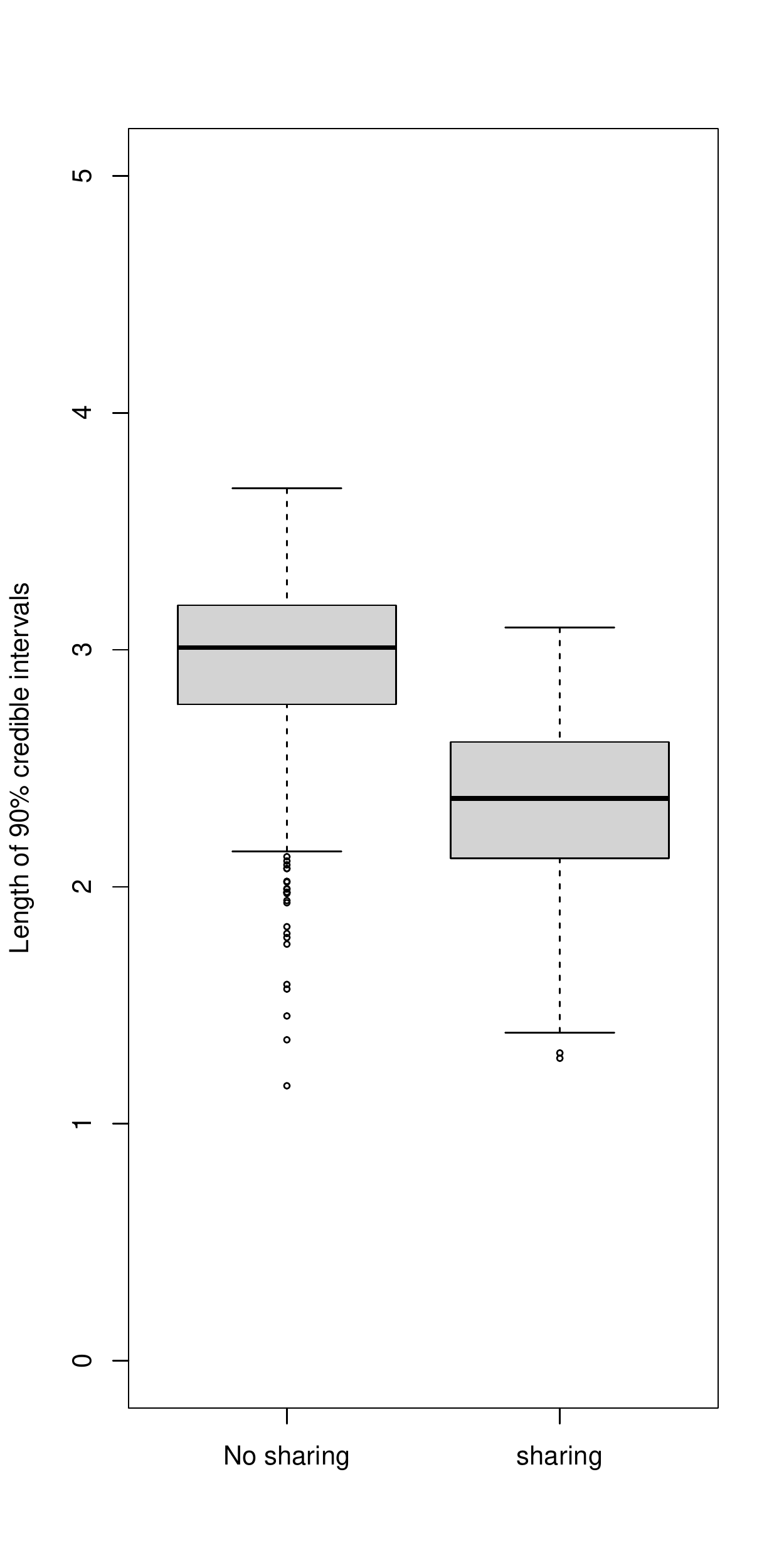}
\end{subfigure}
\begin{subfigure}{.4\linewidth}
    \includegraphics[width=.99\textwidth]{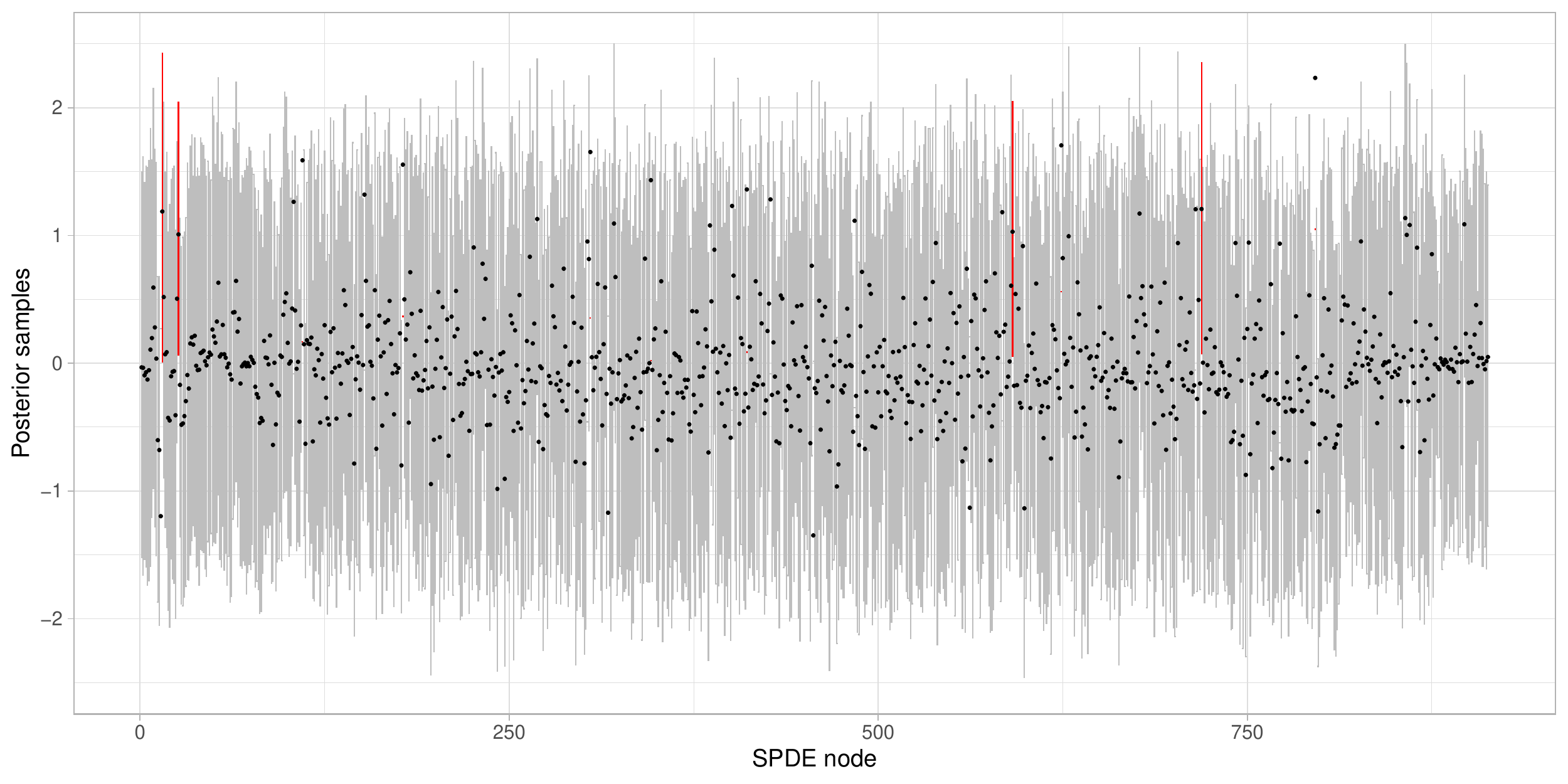}\\
    \includegraphics[width=.99\textwidth]{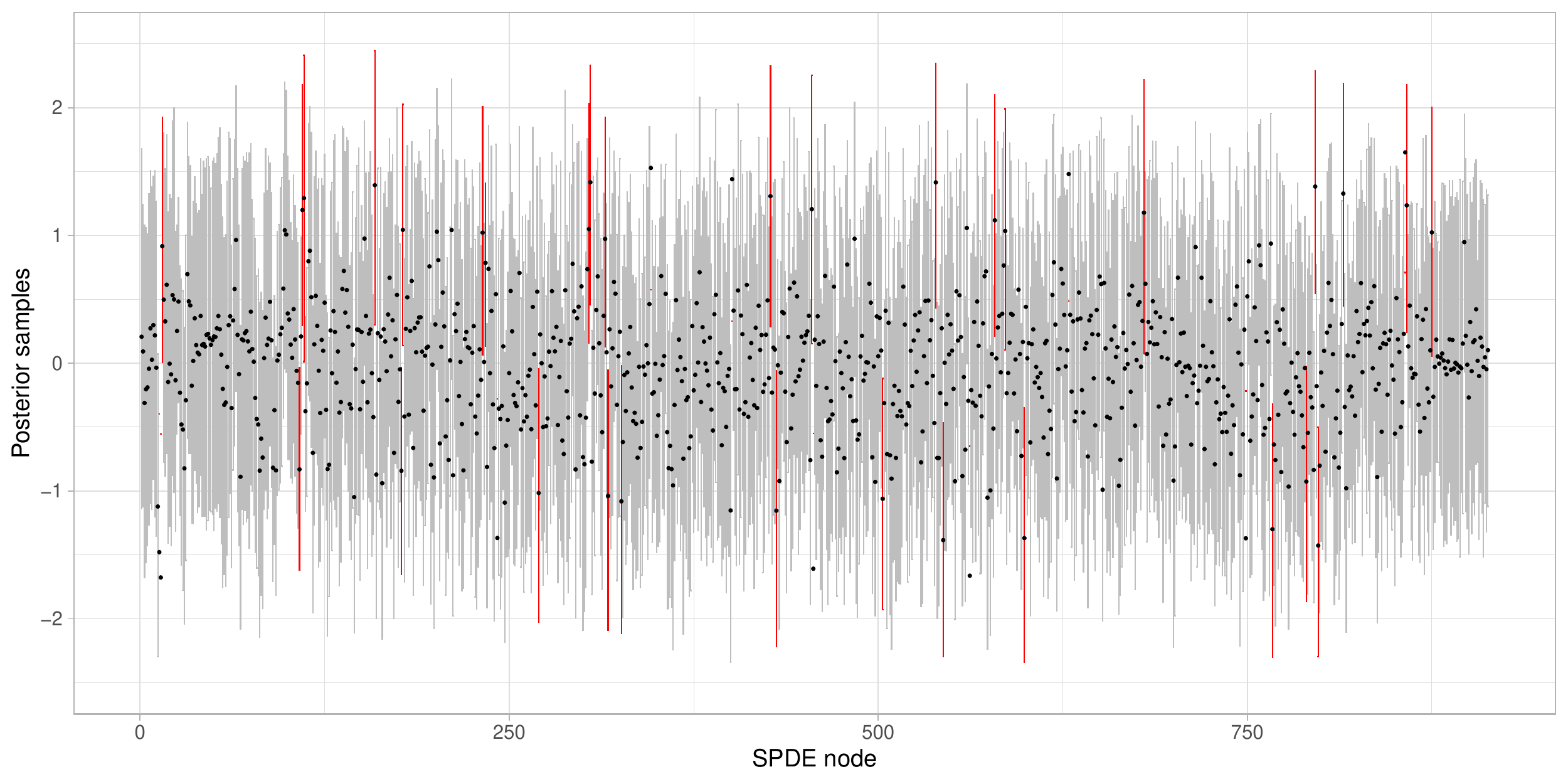}
\end{subfigure}
  \caption{Lengths of the $90\%$ credible intervals of spatial random effect variables at the SPDE triangulation nodes within the study area in the BIN component, based on 500 posterior simulations. Boxplots (left), and error bar plots for the models without (top right) and with sharing (bottom right). Red error bars indicate nodes where the intervals do not include zero.}   
  \label{fig:result:CIshare}
\end{figure}

Sharing decreases uncertainty by  borrowing estimation strength between model components. The average lengths of $95\%$ posterior credible intervals of variables constituting the random effect shrink by up to $30\%$ (Figure~\ref{fig:result:CIshare}) because of a higher  observation-to-parameter ratio that enables us to better capture relevant spatial signals.  

\begin{figure}[t!]
\centering
\begin{subfigure}[b]{.45\linewidth}
    \centering
    \includegraphics[width=.99\textwidth]{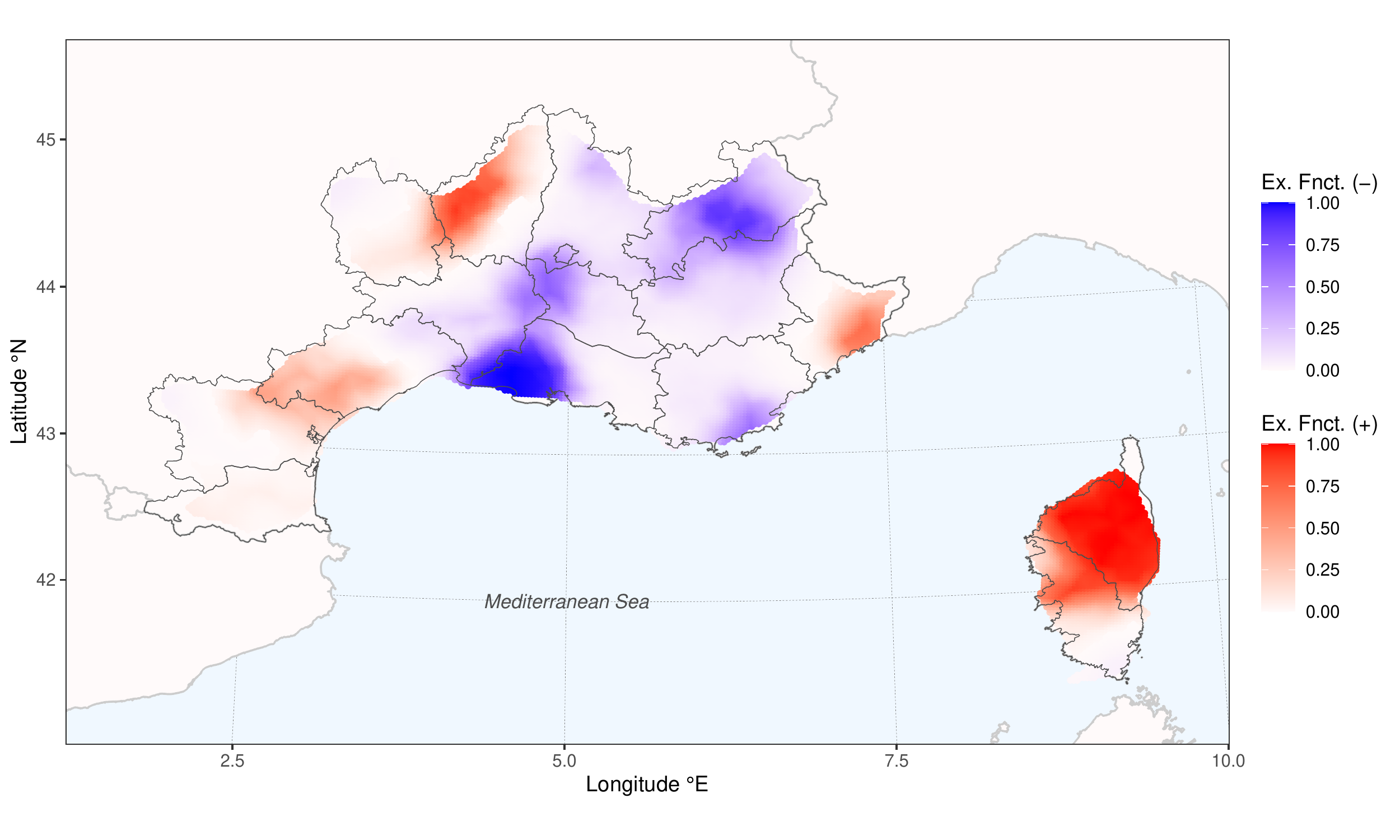}
  \end{subfigure} 
 \begin{subfigure}[b]{.45\linewidth}
    \centering
    \includegraphics[width=.99\textwidth]{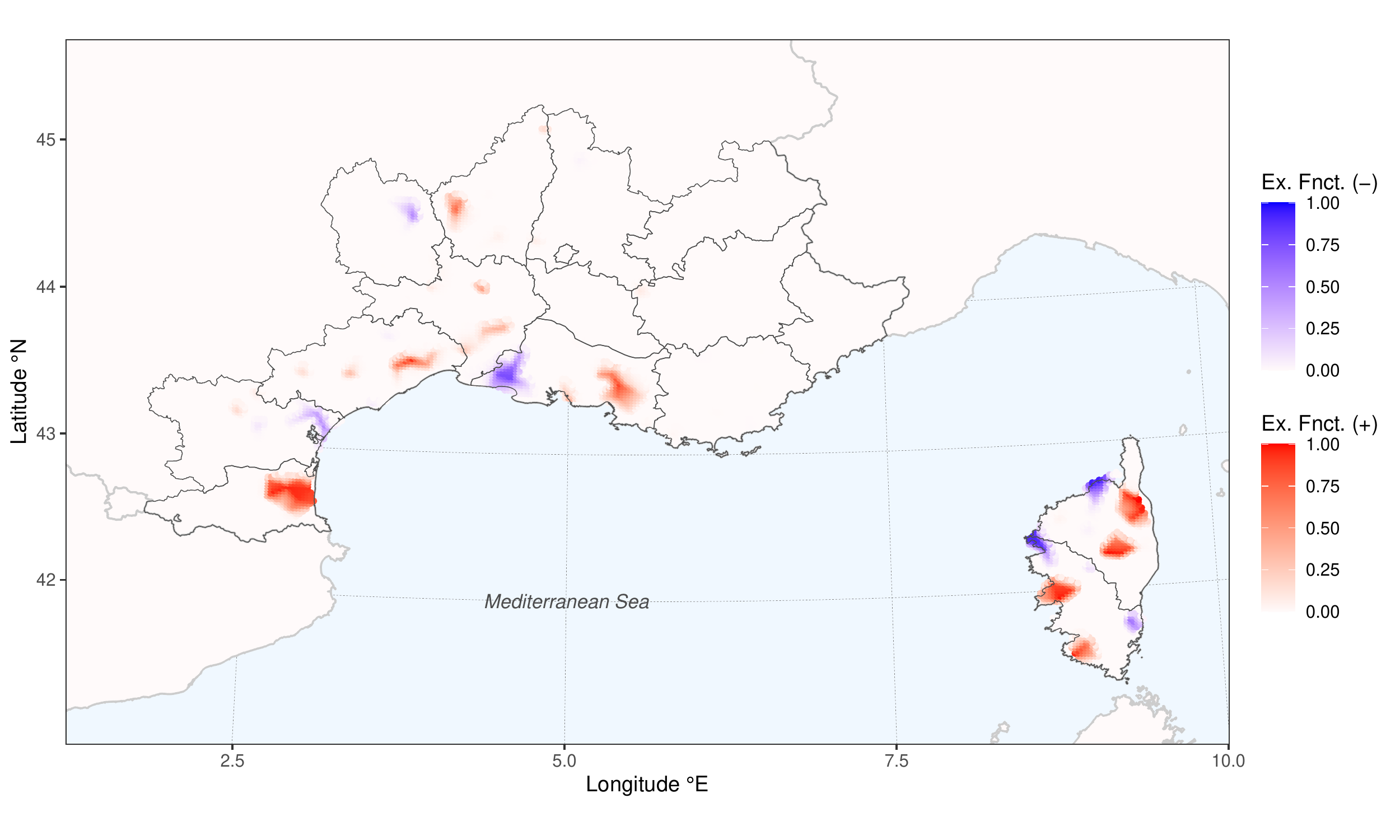}
  \end{subfigure}
  \caption{Excursion functions of posterior latent fields above $0.1$ and below $-0.1$. Plots show $\max\{ F^{+}_{0.1}(\bigcdot), F^{-}_{0.1}(\bigcdot)\}$ for the shared spatial random fields $g^{\text{COX-BETA}}$ (left panel) and $g^{\text{COX-BIN}}$ (right panel). }  
      \label{fig:results:sp:spatial}
\end{figure}

To identify the hot-spot regions of spatial random effects, we study credible sets for excursion regions \citep{Bolin-Lindgren.2015}. We evaluate where the fields exceed or fall below the thresholds  $u=0.1$ and $-u$, respectively. These thresholds approximately correspond to a $10\%$ increase and decrease, respectively,  on the scale of the response when taking into account the log or logistic link.  The $u$-excursion set with probability $\alpha$, $\mathrm{E}^{+}_{u,\alpha}(X)$, is defined as the largest set for which   the level $u$ is exceeded at all locations in the set with probability $1-\alpha$. The  negative $u$ excursion set with probability $\alpha$, $\mathrm{E}^{-}_{u,\alpha}(X)$, is defined as the largest set for which the process remains below the level $-u$ at all locations in the set with probability $1-\alpha$. This approach determines the largest set contained in the exceedance set with a minimum probability threshold, and it assumes a parametric family for the exceedance sets. 
To visualize excursion sets simultaneously for all values of $\alpha$, \citet{Bolin-Lindgren.2015} introduced the positive and negative excursion functions $F^{+}_{u}(s)  =  1-\inf\{\alpha\mid s\in \mathrm{E}^{+}_{u,\alpha} \}\in [0,1]$ and $F^{-}_{u}(s)  =  1-\inf\{\alpha \mid s\in \mathrm{E}^{-}_{u,\alpha}\} \in[0,1]$. 
Figure~\ref{fig:results:sp:spatial} highlights several hot-spot regions for the shared spatial effects, which we interpret with respect to wildfire management in \S \ref{sec:application}.




\section{New insights for wildfire science}
\label{sec:application}

\citet{Pimont2020} have pointed out several critical divergences between simulations of their model and observed wildfire activity, and they have put forward hypotheses to explain them. The novel models developed here, especially M1, do not suffer from this lack of fit by including components to estimate the sources of space-time variability conjectured by \citet{Pimont2020}. This leads to more reliable inferences and predictions, and we outline the new insights.  

\subsection{FWI and seasonal effects}
\label{ssec:interpretation-fwi-season}

The estimated FWI effect on all wildfire  components (COX, BIN, BETA, GPD) is  nonlinear with a strong increase when moving from FWI$=0$ towards FWI$\approx 60$--$80$, followed by a dampening and a slight decrease for extreme FWI values but with relatively wide credible bounds. Moreover, seasonal patterns emerge in the joint FWI-month effect in the occurrence component COX. The common practice of using FWI directly as a proxy for wildfire activity, without a nonlinear and seasonally varying transfer function as estimated here, would predict extreme wildfires badly and miss seasonally varying response of fire activity to this index.

This non-linear, even decreasing, response to high FWI and seasonal biases 
can be attributed to the excessively sharp exponential response of FWI to wind speed in its upper range and to the limited ability of the Drought Code (a subcomponent of the FWI)
to reproduce live fuel moisture dynamics in France \citep{Ruffault2018}. In spring, vegetation budburst produces new foliage with a high water content that is maintained until the onset of the summer drought, typically in early July. 
The timing of periodic events in plant life cycles (\ie plant phenology) and stomatal control under drought might also explain why dynamics of soil and vegetation water contents are unsynchronized at certain times.
In our COX model, we allow not only for a seasonal effect but also for different responses of FWI across the five months. The shapes of these monthly responses vary greatly, so seasonal variations cannot be handled solely through a separate seasonal random effect. The response in August did not exhibit any saturation in the upper part of the FWI range, suggesting that higher values in mostly dry conditions correspond to increased fire activity; the contribution of wind to FWI could be adequate in these already-dry conditions. On the contrary, a flattening and notable decrease of the COX response to FWI was observed at FWI$\approx45$--$50$ for relatively moist conditions in June and October. This supports the hypotheses that the desynchronization of soil and 
fuel moistures caused by plant phenology in Spring could be involved, and the response of the FWI to high wind would be inaccurate  in such moist conditions. July and September, with their mixture of dry and moist days, show intermediate response levels to very high FWI. These findings confirm a need to develop better wildfire danger indices in the study region. 

\subsection{Time trends during the study period}

The year 2003 was catastrophic in terms of fire sizes and burnt area. It has a pivotal role with a decrease of occurrence numbers and sizes afterwards, as highlighted by the year component of our posterior model that captures temporal trends not explained through weather and land-cover related predictors. 
In 2003, a heat wave coincided with severe drought conditions, leading to an unusually high number of escaped fires ($>1$ha), and of fires larger than $10$ha for several weeks, whose occurrence was not matched by very high values of FWI due to its weaknesses outlined in \S\ref{ssec:interpretation-fwi-season}. The drop in the estimated yearly effect after 2003 could be due to official policy measures that have slightly evolved after 2003,
and to increased awareness of fire managers to strengthen prevention or suppression policies \citep{Pimont2020}.

A finding of our model that should garner attention of wildfire managers is the yearly effect of its BIN component, as the probability of observing a large fire 
tends to increase over the most years following a decade of continuous decrease.
Our results also confirm that of \citet{Evin-2018}, who found no clear time trend for the probability of extreme fires (GPD).

\subsection{Shared spatial effects for improved regionalized predictions}

The shared spatial effects shown in \S\ref{sec:mode:results} highlight regional differences in fire size distributions and provide quantitative interpretations of effects.
They also reveal substantial  regional variation in proportions of  moderate and extreme fires. In particular, the sharing effect with significantly negative $\beta^{\text{COX-BIN}}$ allows for interpretation with respect to different wildland-to-urban interactions. The lowland area in the western Pyr\'en\'ees-Orientales region, fairly densely populated with a large proportion of abandoned agricultural land intermixed with urban surfaces, appears to have high occurrence intensities, but its combustible area is strongly fragmented, so wildfires are mostly small. More fires than expected from weather/climate and forest area occur in densely populated landscapes or in rural landscapes with significant human activities promoting fire ignitions, while landscape fragmentation and landscape management reduce the likelihood of large fires. The COX-BETA sharing effect is highly positive in Corsica, where moderately large espaced fires become larger more often than elsewhere, perhaps due to longer arrival times of firefighters in remote Corsican forests and less frequent airborne firefighting. Moreover, extreme fires tend to be more frequent because of  large contiguous forests. 
Further regional disparities in predictions are  illustrated in the Supplement 
where the right panel of Figure~\ref{fig:application::fires}  highlights significant differences in threshold exceedance probabilities.

\section{Conclusion}
\label{sec:conclusion}

We have implemented a novel Bayesian spatiotemporal model for wildfire activity with specific components for extreme events, and with shared random effects to account for stochastic dependence among components not explained by covariates.
Due to the complex structure of wildfire activity and its drivers, the sophisticated structure of our fully Bayesian hierarchical models allows us to accurately disentangle the effects and interactions of various observed and unobserved drivers while limiting estimation uncertainty. The use of Gaussian random effects at relatively high spatial resolution provides crucial benefits over frequentist generalized additive models since fine-scale spatial variation and associated uncertainties can be identified properly.

Different sharing strategies respond to different considerations. If statistical stability is the focus, then sharing from well-identified model components towards those less informed by data is appropriate. If focus is on accurate inference of a specific component (\eg extremes), then it is sensible to share effects from this component towards others. In both cases,  component-specific effects without sharing remain important and should be included as far as data allow estimating them. 
In some applications however, introducing common components by sharing is the only way to incorporate spatial effects in certain response variables. For example, had we chosen a threshold larger than $79$ha for large wildfires, we would have had even fewer observations available for the extreme fire size component. A separate spatial effect in this component would provide wider credible intervals than those in Figure \ref{fig:result:CIshare} (top right), and be of even less practical use. 
Our findings improve decision support in wildfire management: shared spatial effects explain how wildfire numbers and extreme sizes interact by providing maps of the significant disparities between regions. Moreover, FWI maps used for fire danger rating must be interpreted with care because of the strong nonlinear and seasonal effect on wildfire risk identified by our model. Our framework allows for including more general space-varying temporal trends in fire weather relationship in future work to explore the spatial disparity in temporal trends due to changes in land-use practices and fire management. 

While our focus here is on generative and predictive modeling, the adaptation of descriptive tools from stochastic geometry \citep[K-functions, mark correlation functions, see][]{Chiu2013} would further improve the analysis of point processes with extreme marks.
Beyond wildfire modeling, our flexible and generic approach could be used to provide new insights and improved extreme-value predictions for a variety of other problems. Landslide inventories can  be represented as point processes 
with heavy-tailed magnitude marks \citep{Stark2001,Lombardo2020}. Another promising application  consists in modeling  locations, times and values of high-impact events extracted from processes indexed over space and time, such as  local extremes in gridded climate data. This would yield a parsimonious representation of extreme events in such processes. 
Models for preferentially sampled spatial data \citep{Diggle2010} can be viewed as marked point processes with shared effects, such that our approach would allow capturing preferential sampling effects specifically in extreme values.






\begin{acks}[Acknowledgments]
The authors would like to thank Anthony Davison for helpful comments and discussions, and the Swiss National Science Foundation for financial support. 

\end{acks}

\begin{funding}
The first author gratefully acknowledges the Swiss National Science Foundation (project $200021\_178824$).
\end{funding}


\bibliographystyle{imsart-nameyear} 
\bibliography{biblio}       

\begin{thebibliography}{59}

\bibitem[\protect\citeauthoryear{Baddeley, Rubak and
  Turner}{2015}]{Baddeley2015}
\begin{bbook}[author]
\bauthor{\bsnm{Baddeley},~\bfnm{Adrian}\binits{A.}},
  \bauthor{\bsnm{Rubak},~\bfnm{Ege}\binits{E.}} \AND
  \bauthor{\bsnm{Turner},~\bfnm{Rolf}\binits{R.}}
(\byear{2015}).
\btitle{Spatial Point Patterns: Methodology and Applications with {R}}.
\bpublisher{Chapman and Hall/CRC Press}.
\end{bbook}
\endbibitem

\bibitem[\protect\citeauthoryear{Baddeley and
  Turner}{2000}]{Baddeley.Turner.2000}
\begin{barticle}[author]
\bauthor{\bsnm{Baddeley},~\bfnm{Adrian}\binits{A.}} \AND
  \bauthor{\bsnm{Turner},~\bfnm{Rolf}\binits{R.}}
(\byear{2000}).
\btitle{Practical maximum pseudolikelihood for spatial point patterns}.
\bjournal{Australian \& New Zealand Journal of Statistics}
\bvolume{42}
\bpages{283--322}.
\end{barticle}
\endbibitem

\bibitem[\protect\citeauthoryear{Baddeley et~al.}{2010}]{Baddeley2010}
\begin{barticle}[author]
\bauthor{\bsnm{Baddeley},~\bfnm{Adrian}\binits{A.}},
  \bauthor{\bsnm{Berman},~\bfnm{Mark}\binits{M.}},
  \bauthor{\bsnm{Fisher},~\bfnm{Nicholas~I}\binits{N.~I.}},
  \bauthor{\bsnm{Hardegen},~\bfnm{Andrew}\binits{A.}},
  \bauthor{\bsnm{Milne},~\bfnm{Robin~K}\binits{R.~K.}},
  \bauthor{\bsnm{Schuhmacher},~\bfnm{Dominic}\binits{D.}},
  \bauthor{\bsnm{Shah},~\bfnm{Rohan}\binits{R.}},
  \bauthor{\bsnm{Turner},~\bfnm{Rolf}\binits{R.}} \betal{et~al.}
(\byear{2010}).
\btitle{Spatial logistic regression and change-of-support in {P}oisson point
  processes}.
\bjournal{Electronic Journal of Statistics}
\bvolume{4}
\bpages{1151--1201}.
\end{barticle}
\endbibitem

\bibitem[\protect\citeauthoryear{Baddeley et~al.}{2014}]{Baddeley.al.2014}
\begin{barticle}[author]
\bauthor{\bsnm{Baddeley},~\bfnm{Adrian}\binits{A.}},
  \bauthor{\bsnm{Coeurjolly},~\bfnm{Jean-Fran{\c{c}}ois}\binits{J.-F.}},
  \bauthor{\bsnm{Rubak},~\bfnm{Ege}\binits{E.}} \AND
  \bauthor{\bsnm{Waagepetersen},~\bfnm{Rasmus}\binits{R.}}
(\byear{2014}).
\btitle{Logistic regression for spatial {G}ibbs point processes}.
\bjournal{Biometrika}
\bvolume{101}
\bpages{377--392}.
\end{barticle}
\endbibitem

\bibitem[\protect\citeauthoryear{Berman and Turner}{1992}]{Berman1992}
\begin{barticle}[author]
\bauthor{\bsnm{Berman},~\bfnm{Mark}\binits{M.}} \AND
  \bauthor{\bsnm{Turner},~\bfnm{T~Rolf}\binits{T.~R.}}
(\byear{1992}).
\btitle{Approximating point process likelihoods with {GLIM}}.
\bjournal{Journal of the Royal Statistical Society: Series C (Applied
  Statistics)}
\bvolume{41}
\bpages{31--38}.
\end{barticle}
\endbibitem

\bibitem[\protect\citeauthoryear{Bolin and
  Lindgren}{2015}]{Bolin-Lindgren.2015}
\begin{barticle}[author]
\bauthor{\bsnm{Bolin},~\bfnm{David}\binits{D.}} \AND
  \bauthor{\bsnm{Lindgren},~\bfnm{Finn}\binits{F.}}
(\byear{2015}).
\btitle{Excursion and contour uncertainty regions for latent {G}aussian
  models}.
\bjournal{Journal of the Royal Statistical Society: Series B (Statistical
  Methodology)}
\bvolume{77}
\bpages{85--106}.
\bdoi{https://doi.org/10.1111/rssb.12055}
\end{barticle}
\endbibitem

\bibitem[\protect\citeauthoryear{Bolin and Wallin}{2020}]{Bolin-Wallin.2020}
\begin{barticle}[author]
\bauthor{\bsnm{Bolin},~\bfnm{David}\binits{D.}} \AND
  \bauthor{\bsnm{Wallin},~\bfnm{Jonas}\binits{J.}}
(\byear{2020}).
\btitle{Scale dependence: Why the average {CRPS} often is inappropriate for
  ranking probabilistic forecasts}.
\bjournal{arXiv preprint arXiv:1912.05642}.
\end{barticle}
\endbibitem

\bibitem[\protect\citeauthoryear{Brier}{1950}]{Brier.1950}
\begin{barticle}[author]
\bauthor{\bsnm{Brier},~\bfnm{G.~W.}\binits{G.~W.}}
(\byear{1950}).
\btitle{Verification of Forecasts Expressed in Terms of Probability}.
\bjournal{Monthly Weather Review}
\bvolume{78}
\bpages{1--3}.
\end{barticle}
\endbibitem

\bibitem[\protect\citeauthoryear{Chiu et~al.}{2013}]{Chiu2013}
\begin{bbook}[author]
\bauthor{\bsnm{Chiu},~\bfnm{Sung~Nok}\binits{S.~N.}},
  \bauthor{\bsnm{Stoyan},~\bfnm{Dietrich}\binits{D.}},
  \bauthor{\bsnm{Kendall},~\bfnm{Wilfrid~S}\binits{W.~S.}} \AND
  \bauthor{\bsnm{Mecke},~\bfnm{Joseph}\binits{J.}}
(\byear{2013}).
\btitle{{Stochastic geometry and its applications; 3rd ed.}}
\bpublisher{Wiley}, \baddress{Hoboken, New Jersey}.
\end{bbook}
\endbibitem

\bibitem[\protect\citeauthoryear{Coles}{2001}]{Coles.2001}
\begin{bbook}[author]
\bauthor{\bsnm{Coles},~\bfnm{Stuart}\binits{S.}}
(\byear{2001}).
\btitle{An introduction to statistical modeling of extreme values}.
\bpublisher{Springer}.
\end{bbook}
\endbibitem

\bibitem[\protect\citeauthoryear{Cui and Perera}{2008}]{Cui2008}
\begin{barticle}[author]
\bauthor{\bsnm{Cui},~\bfnm{Wenbin}\binits{W.}} \AND
  \bauthor{\bsnm{Perera},~\bfnm{Ajith~H}\binits{A.~H.}}
(\byear{2008}).
\btitle{What do we know about forest fire size distribution, and why is this
  knowledge useful for forest management?}
\bjournal{International Journal of Wildland Fire}
\bvolume{17}
\bpages{234--244}.
\end{barticle}
\endbibitem

\bibitem[\protect\citeauthoryear{Cumming}{2001}]{Cumming2001}
\begin{barticle}[author]
\bauthor{\bsnm{Cumming},~\bfnm{SG}\binits{S.}}
(\byear{2001}).
\btitle{A parametric model of the fire-size distribution}.
\bjournal{Canadian Journal of Forest Research}
\bvolume{31}
\bpages{1297--1303}.
\end{barticle}
\endbibitem

\bibitem[\protect\citeauthoryear{Davison and Smith}{1990}]{Davison-Smith.1990}
\begin{barticle}[author]
\bauthor{\bsnm{Davison},~\bfnm{A.~C.}\binits{A.~C.}} \AND
  \bauthor{\bsnm{Smith},~\bfnm{R.~L.}\binits{R.~L.}}
(\byear{1990}).
\btitle{Models for Exceedances over High Thresholds}.
\bjournal{Journal of the Royal Statistical Society. Series B (Methodological)}
\bvolume{52}
\bpages{393--442}.
\end{barticle}
\endbibitem

\bibitem[\protect\citeauthoryear{{De Zea Bermudez}
  et~al.}{2009}]{DeZeaBermudez2009}
\begin{barticle}[author]
\bauthor{\bsnm{{De Zea Bermudez}},~\bfnm{P.}\binits{P.}},
  \bauthor{\bsnm{Mendes},~\bfnm{J.}\binits{J.}},
  \bauthor{\bsnm{Pereira},~\bfnm{J.~M.~C.}\binits{J.~M.~C.}},
  \bauthor{\bsnm{Turkman},~\bfnm{K.~F.}\binits{K.~F.}} \AND
  \bauthor{\bsnm{Vasconcelos},~\bfnm{M.~J.~P.}\binits{M.~J.~P.}}
(\byear{2009}).
\btitle{{Spatial and temporal extremes of wildfire sizes in Portugal
  (1984–2004)}}.
\bjournal{International Journal of Wildland Fire}
\bvolume{18}
\bpages{983--991}.
\bdoi{10.1071/WF07044}
\end{barticle}
\endbibitem

\bibitem[\protect\citeauthoryear{Diggle, Menezes and Su}{2010}]{Diggle2010}
\begin{barticle}[author]
\bauthor{\bsnm{Diggle},~\bfnm{Peter~J}\binits{P.~J.}},
  \bauthor{\bsnm{Menezes},~\bfnm{Raquel}\binits{R.}} \AND
  \bauthor{\bsnm{Su},~\bfnm{Ting-li}\binits{T.-l.}}
(\byear{2010}).
\btitle{Geostatistical inference under preferential sampling}.
\bjournal{Journal of the Royal Statistical Society: Series C (Applied
  Statistics)}
\bvolume{59}
\bpages{191--232}.
\end{barticle}
\endbibitem

\bibitem[\protect\citeauthoryear{Evin, Curt and Eckert}{2018}]{Evin-2018}
\begin{barticle}[author]
\bauthor{\bsnm{Evin},~\bfnm{G.}\binits{G.}},
  \bauthor{\bsnm{Curt},~\bfnm{T.}\binits{T.}} \AND
  \bauthor{\bsnm{Eckert},~\bfnm{N.}\binits{N.}}
(\byear{2018}).
\btitle{Has fire policy decreased the return period of the largest wildfire
  events in France? A Bayesian assessment based on extreme value theory}.
\bjournal{Natural Hazards and Earth System Sciences}
\bvolume{18}
\bpages{2641--2651}.
\bdoi{10.5194/nhess-18-2641-2018}
\end{barticle}
\endbibitem

\bibitem[\protect\citeauthoryear{Fawcett}{2006}]{Fawcett.2006}
\begin{barticle}[author]
\bauthor{\bsnm{Fawcett},~\bfnm{Tom}\binits{T.}}
(\byear{2006}).
\btitle{An Introduction to {ROC} Analysis}.
\bjournal{Pattern Recognition Letters}
\bvolume{27}
\bpages{861--874}.
\bdoi{10.1016/j.patrec.2005.10.010}
\end{barticle}
\endbibitem

\bibitem[\protect\citeauthoryear{Fuglstad et~al.}{2018}]{Fuglstad.al.2018}
\begin{barticle}[author]
\bauthor{\bsnm{Fuglstad},~\bfnm{Geir-Arne}\binits{G.-A.}},
  \bauthor{\bsnm{Simpson},~\bfnm{Daniel}\binits{D.}},
  \bauthor{\bsnm{Lindgren},~\bfnm{Finn}\binits{F.}} \AND
  \bauthor{\bsnm{Rue},~\bfnm{H{\aa}vard}\binits{H.}}
(\byear{2018}).
\btitle{Constructing priors that penalize the complexity of {G}aussian random
  fields}.
\bjournal{Journal of the American Statistical Association}
\bvolume{114}
\bpages{445--452}.
\end{barticle}
\endbibitem

\bibitem[\protect\citeauthoryear{Genton et~al.}{2006}]{Genton2006}
\begin{barticle}[author]
\bauthor{\bsnm{Genton},~\bfnm{Marc~G}\binits{M.~G.}},
  \bauthor{\bsnm{Butry},~\bfnm{David~T}\binits{D.~T.}},
  \bauthor{\bsnm{Gumpertz},~\bfnm{Marcia~L}\binits{M.~L.}} \AND
  \bauthor{\bsnm{Prestemon},~\bfnm{Jeffrey~P}\binits{J.~P.}}
(\byear{2006}).
\btitle{{Spatio-temporal analysis of wildfire ignitions in the St Johns River
  water management district, Florida}}.
\bjournal{International Journal of Wildland Fire}
\bvolume{15}
\bpages{87--97}.
\end{barticle}
\endbibitem

\bibitem[\protect\citeauthoryear{Horvitz and Thompson}{1952}]{Horvitz1952a}
\begin{barticle}[author]
\bauthor{\bsnm{Horvitz},~\bfnm{D.~G.}\binits{D.~G.}} \AND
  \bauthor{\bsnm{Thompson},~\bfnm{D.~J.}\binits{D.~J.}}
(\byear{1952}).
\btitle{A Generalization of Sampling Without Replacement from a Finite
  Universe}.
\bjournal{Journal of the American Statistical Association}
\bvolume{47}
\bpages{663--685}.
\bdoi{10.1080/01621459.1952.10483446}
\end{barticle}
\endbibitem

\bibitem[\protect\citeauthoryear{Illian, S{\o}rbye and
  Rue}{2012}]{Illian.al.2012}
\begin{barticle}[author]
\bauthor{\bsnm{Illian},~\bfnm{Janine~B}\binits{J.~B.}},
  \bauthor{\bsnm{S{\o}rbye},~\bfnm{Sigrunn~H}\binits{S.~H.}} \AND
  \bauthor{\bsnm{Rue},~\bfnm{H{\aa}vard}\binits{H.}}
(\byear{2012}).
\btitle{{A toolbox for fitting complex spatial point process models using
  integrated nested Laplace approximation (INLA)}}.
\bjournal{The Annals of Applied Statistics}
\bvolume{6}
\bpages{1499--1530}.
\end{barticle}
\endbibitem

\bibitem[\protect\citeauthoryear{Jones et~al.}{2020}]{Jones.2020}
\begin{bincollection}[author]
\bauthor{\bsnm{Jones},~\bfnm{Matthew~W}\binits{M.~W.}},
  \bauthor{\bsnm{Smith},~\bfnm{Adam}\binits{A.}},
  \bauthor{\bsnm{Betts},~\bfnm{Richard}\binits{R.}},
  \bauthor{\bsnm{Canadell},~\bfnm{Josep~G}\binits{J.~G.}},
  \bauthor{\bsnm{Prentice},~\bfnm{I~Colin}\binits{I.~C.}} \AND
  \bauthor{\bsnm{Le~Qu{\'e}r{\'e}},~\bfnm{Corinne}\binits{C.}}
(\byear{2020}).
\btitle{Science{B}rief {R}eview: Climate {c}hange increases the risk of
  wildfires}.
In \bbooktitle{Critical Issues in Climate Change Science}
(\beditor{\bfnm{Corinne}\binits{C.}~\bsnm{Le~Qu{\'e}r{\'e}}},
  \beditor{\bfnm{P.}\binits{P.}~\bsnm{Liss}} \AND
  \beditor{\bfnm{P.}\binits{P.}~\bsnm{Forster}}, eds.).
\end{bincollection}
\endbibitem

\bibitem[\protect\citeauthoryear{Joseph et~al.}{2019}]{Joseph2019}
\begin{barticle}[author]
\bauthor{\bsnm{Joseph},~\bfnm{Maxwell~B}\binits{M.~B.}},
  \bauthor{\bsnm{Rossi},~\bfnm{Matthew~W}\binits{M.~W.}},
  \bauthor{\bsnm{Mietkiewicz},~\bfnm{Nathan~P}\binits{N.~P.}},
  \bauthor{\bsnm{Mahood},~\bfnm{Adam~L}\binits{A.~L.}},
  \bauthor{\bsnm{Cattau},~\bfnm{Megan~E}\binits{M.~E.}},
  \bauthor{\bsnm{St.~Denis},~\bfnm{Lise~Ann}\binits{L.~A.}},
  \bauthor{\bsnm{Nagy},~\bfnm{R~Chelsea}\binits{R.~C.}},
  \bauthor{\bsnm{Iglesias},~\bfnm{Virginia}\binits{V.}},
  \bauthor{\bsnm{Abatzoglou},~\bfnm{John~T}\binits{J.~T.}} \AND
  \bauthor{\bsnm{Balch},~\bfnm{Jennifer~K}\binits{J.~K.}}
(\byear{2019}).
\btitle{Spatiotemporal prediction of wildfire size extremes with {B}ayesian
  finite sample maxima}.
\bjournal{Ecological Applications}
\bvolume{29}
\bpages{e01898}.
\end{barticle}
\endbibitem

\bibitem[\protect\citeauthoryear{Krainski et~al.}{2018}]{Krainski.al.2018}
\begin{bbook}[author]
\bauthor{\bsnm{Krainski},~\bfnm{Elias~T}\binits{E.~T.}},
  \bauthor{\bsnm{G{\'o}mez-Rubio},~\bfnm{Virgilio}\binits{V.}},
  \bauthor{\bsnm{Bakka},~\bfnm{Haakon}\binits{H.}},
  \bauthor{\bsnm{Lenzi},~\bfnm{Amanda}\binits{A.}},
  \bauthor{\bsnm{Castro-Camilo},~\bfnm{Daniela}\binits{D.}},
  \bauthor{\bsnm{Simpson},~\bfnm{Daniel}\binits{D.}},
  \bauthor{\bsnm{Lindgren},~\bfnm{Finn}\binits{F.}} \AND
  \bauthor{\bsnm{Rue},~\bfnm{H{\aa}vard}\binits{H.}}
(\byear{2018}).
\btitle{{Advanced Spatial Modeling with Stochastic Partial Differential
  Equations Using R and INLA}}.
\bpublisher{Chapman and Hall/CRC}.
\end{bbook}
\endbibitem

\bibitem[\protect\citeauthoryear{Lindgren, Rue and
  Lindstr{\"o}m}{2011}]{Lindgren.al.2011}
\begin{barticle}[author]
\bauthor{\bsnm{Lindgren},~\bfnm{Finn}\binits{F.}},
  \bauthor{\bsnm{Rue},~\bfnm{H{\aa}vard}\binits{H.}} \AND
  \bauthor{\bsnm{Lindstr{\"o}m},~\bfnm{Johan}\binits{J.}}
(\byear{2011}).
\btitle{An explicit link between {G}aussian fields and {G}aussian {M}arkov
  random fields: the stochastic partial differential equation approach}.
\bjournal{Journal of the Royal Statistical Society: Series B (Statistical
  Methodology)}
\bvolume{73}
\bpages{423--498}.
\end{barticle}
\endbibitem

\bibitem[\protect\citeauthoryear{Lindgren and Rue}{2015}]{Lindgren.al.2015}
\begin{barticle}[author]
\bauthor{\bsnm{Lindgren},~\bfnm{Finn}\binits{F.}} \AND
  \bauthor{\bsnm{Rue},~\bfnm{H{\aa}vard}\binits{H.}}
(\byear{2015}).
\btitle{Bayesian spatial modelling with {R-INLA}}.
\bjournal{Journal of Statistical Software}
\bvolume{63}.
\end{barticle}
\endbibitem

\bibitem[\protect\citeauthoryear{Lombardo et~al.}{2020}]{Lombardo2020}
\begin{barticle}[author]
\bauthor{\bsnm{Lombardo},~\bfnm{Luigi}\binits{L.}},
  \bauthor{\bsnm{Opitz},~\bfnm{Thomas}\binits{T.}},
  \bauthor{\bsnm{Ardizzone},~\bfnm{Francesca}\binits{F.}},
  \bauthor{\bsnm{Guzzetti},~\bfnm{Fausto}\binits{F.}} \AND
  \bauthor{\bsnm{Huser},~\bfnm{Rapha{\"{e}}l}\binits{R.}}
(\byear{2020}).
\btitle{{Space-Time Landslide Predictive Modelling}}.
\bjournal{Earth Science Reviews}
\bvolume{209}
\bpages{103318}.
\end{barticle}
\endbibitem

\bibitem[\protect\citeauthoryear{Mendes et~al.}{2010}]{Mendes2010}
\begin{barticle}[author]
\bauthor{\bsnm{Mendes},~\bfnm{Jorge~M.}\binits{J.~M.}}, \bauthor{\bsnm{{de Zea
  Bermudez}},~\bfnm{Patr{\'{i}}cia~Cort{\'{e}}s}\binits{P.~C.}},
  \bauthor{\bsnm{Pereira},~\bfnm{Jos{\'{e}}}\binits{J.}},
  \bauthor{\bsnm{Turkman},~\bfnm{K.~F.}\binits{K.~F.}} \AND
  \bauthor{\bsnm{Vasconcelos},~\bfnm{M.~J.~P.}\binits{M.~J.~P.}}
(\byear{2010}).
\btitle{{Spatial extremes of wildfire sizes: Bayesian hierarchical models for
  extremes}}.
\bjournal{Environmental and Ecological Statistics}
\bvolume{17}
\bpages{1--28}.
\bdoi{10.1007/s10651-008-0099-3}
\end{barticle}
\endbibitem

\bibitem[\protect\citeauthoryear{Northrop and
  Coleman}{2014}]{Northrop-Coleman.2014}
\begin{barticle}[author]
\bauthor{\bsnm{Northrop},~\bfnm{Paul~J.}\binits{P.~J.}} \AND
  \bauthor{\bsnm{Coleman},~\bfnm{Claire~L.}\binits{C.~L.}}
(\byear{2014}).
\btitle{Improved threshold diagnostic plots for extreme value analyses}.
\bjournal{Extremes}
\bvolume{17}
\bpages{289--303}.
\end{barticle}
\endbibitem

\bibitem[\protect\citeauthoryear{Opitz}{2017}]{Opitz.2017b}
\begin{barticle}[author]
\bauthor{\bsnm{Opitz},~\bfnm{Thomas}\binits{T.}}
(\byear{2017}).
\btitle{{{L}atent {G}aussian modeling and {INLA}: A review with focus on
  space-time applications}}.
\bjournal{Journal de la Soci\'et\'e Fran\c caise de Statistique}
\bvolume{158}
\bpages{62--85}.
\end{barticle}
\endbibitem

\bibitem[\protect\citeauthoryear{Opitz, Bonneu and Gabriel}{2020}]{Opitz2020}
\begin{barticle}[author]
\bauthor{\bsnm{Opitz},~\bfnm{Thomas}\binits{T.}},
  \bauthor{\bsnm{Bonneu},~\bfnm{Florent}\binits{F.}} \AND
  \bauthor{\bsnm{Gabriel},~\bfnm{Edith}\binits{E.}}
(\byear{2020}).
\btitle{{Point-process based modeling of space-time structures of forest fire
  occurrences in Mediterranean France}}.
\bjournal{Spatial Statistics}
\bvolume{40}
\bpages{100429}.
\bdoi{10.1016/j.spasta.2020.100429}
\end{barticle}
\endbibitem

\bibitem[\protect\citeauthoryear{Opitz et~al.}{2018}]{Opitz.al.2018}
\begin{barticle}[author]
\bauthor{\bsnm{Opitz},~\bfnm{Thomas}\binits{T.}},
  \bauthor{\bsnm{Huser},~\bfnm{Rapha{\"e}l}\binits{R.}},
  \bauthor{\bsnm{Bakka},~\bfnm{Haakon}\binits{H.}} \AND
  \bauthor{\bsnm{Rue},~\bfnm{Haavard}\binits{H.}}
(\byear{2018}).
\btitle{{INLA goes extreme: Bayesian tail regression for the estimation of high
  spatio-temporal quantiles}}.
\bjournal{Extremes}
\bvolume{21}
\bpages{441--462}.
\end{barticle}
\endbibitem

\bibitem[\protect\citeauthoryear{Opitz et~al.}{2020}]{Opitz2020b}
\begin{barticle}[author]
\bauthor{\bsnm{Opitz},~\bfnm{Thomas}\binits{T.}},
  \bauthor{\bsnm{Bakka},~\bfnm{Haakon}\binits{H.}},
  \bauthor{\bsnm{Huser},~\bfnm{Rapha{\"e}l}\binits{R.}} \AND
  \bauthor{\bsnm{Lombardo},~\bfnm{Luigi}\binits{L.}}
(\byear{2020}).
\btitle{High-resolution {B}ayesian mapping of landslide hazard with unobserved
  trigger event}.
\bjournal{arXiv preprint arXiv:2006.07902}.
\end{barticle}
\endbibitem

\bibitem[\protect\citeauthoryear{Peng, Schoenberg and Woods}{2005}]{Peng2005}
\begin{barticle}[author]
\bauthor{\bsnm{Peng},~\bfnm{Roger~D}\binits{R.~D.}},
  \bauthor{\bsnm{Schoenberg},~\bfnm{Frederic~Paik}\binits{F.~P.}} \AND
  \bauthor{\bsnm{Woods},~\bfnm{James~A}\binits{J.~A.}}
(\byear{2005}).
\btitle{A space-time conditional intensity model for evaluating a wildfire
  hazard index}.
\bjournal{Journal of the American Statistical Association}
\bvolume{100}
\bpages{26--35}.
\end{barticle}
\endbibitem

\bibitem[\protect\citeauthoryear{Pereira and Turkman}{2019}]{Pereira2019}
\begin{bincollection}[author]
\bauthor{\bsnm{Pereira},~\bfnm{Jos{\'e} M.~C.}\binits{J.~M.~C.}} \AND
  \bauthor{\bsnm{Turkman},~\bfnm{Kamil~F.}\binits{K.~F.}}
(\byear{2019}).
\btitle{Statistical models of vegetation fires: Spatial and temporal patterns}.
In \bbooktitle{Handbook of Environmental and Ecological Statistics}
\bpages{401--420}.
\bpublisher{Chapman and Hall/CRC}.
\end{bincollection}
\endbibitem

\bibitem[\protect\citeauthoryear{Pimont et~al.}{2021}]{Pimont2020}
\begin{barticle}[author]
\bauthor{\bsnm{Pimont},~\bfnm{Fran{\c{c}}ois}\binits{F.}},
  \bauthor{\bsnm{Fargeon},~\bfnm{H{\'{e}}l{\`{e}}ne}\binits{H.}},
  \bauthor{\bsnm{Opitz},~\bfnm{Thomas}\binits{T.}},
  \bauthor{\bsnm{Ruffault},~\bfnm{Julien}\binits{J.}},
  \bauthor{\bsnm{Barbero},~\bfnm{Renaud}\binits{R.}},
  \bauthor{\bsnm{Martin-StPaul},~\bfnm{Nicolas}\binits{N.}},
  \bauthor{\bsnm{Rigolot},~\bfnm{Eric;~INRAE}\binits{E.~I.}},
  \bauthor{\bsnm{Rivi{\`{e}}re},~\bfnm{Miguel}\binits{M.}} \AND
  \bauthor{\bsnm{Dupuy},~\bfnm{Jean-Luc}\binits{J.-L.}}
(\byear{2021}).
\btitle{Prediction of regional wildfire activity in the probabilistic Bayesian
  framework of {F}irelihood}.
\bjournal{Ecological Applications}
\bvolume{In press.}
\bdoi{https://doi.org/10.1002/eap.2316}
\end{barticle}
\endbibitem

\bibitem[\protect\citeauthoryear{Preisler et~al.}{2004}]{Preisler.2004}
\begin{barticle}[author]
\bauthor{\bsnm{Preisler},~\bfnm{Haiganoush~K}\binits{H.~K.}},
  \bauthor{\bsnm{Brillinger},~\bfnm{David~R}\binits{D.~R.}},
  \bauthor{\bsnm{Burgan},~\bfnm{Robert~E}\binits{R.~E.}} \AND
  \bauthor{\bsnm{Benoit},~\bfnm{JW}\binits{J.}}
(\byear{2004}).
\btitle{Probability based models for estimation of wildfire risk}.
\bjournal{International Journal of wildland fire}
\bvolume{13}
\bpages{133--142}.
\end{barticle}
\endbibitem

\bibitem[\protect\citeauthoryear{Rathbun}{2013}]{Rathbun2013}
\begin{barticle}[author]
\bauthor{\bsnm{Rathbun},~\bfnm{S.~L.}\binits{S.~L.}}
(\byear{2013}).
\btitle{{Optimal estimation of Poisson intensity with partially observed
  covariates}}.
\bjournal{Biometrika}
\bvolume{100}
\bpages{277--281}.
\bdoi{10.1093/biomet/ass069}
\end{barticle}
\endbibitem

\bibitem[\protect\citeauthoryear{Rathbun, Shiffman and
  Gwaltney}{2007}]{Rathbun2007}
\begin{barticle}[author]
\bauthor{\bsnm{Rathbun},~\bfnm{Stephen~L.}\binits{S.~L.}},
  \bauthor{\bsnm{Shiffman},~\bfnm{Saul}\binits{S.}} \AND
  \bauthor{\bsnm{Gwaltney},~\bfnm{Chad~J.}\binits{C.~J.}}
(\byear{2007}).
\btitle{{Modelling the effects of partially observed covariates on Poisson
  process intensity}}.
\bjournal{Biometrika}
\bvolume{94}
\bpages{153--165}.
\bdoi{10.1093/biomet/asm009}
\end{barticle}
\endbibitem

\bibitem[\protect\citeauthoryear{R{\'\i}os-Pena et~al.}{2018}]{Rios2018}
\begin{barticle}[author]
\bauthor{\bsnm{R{\'\i}os-Pena},~\bfnm{Laura}\binits{L.}},
  \bauthor{\bsnm{Kneib},~\bfnm{Thomas}\binits{T.}},
  \bauthor{\bsnm{Cadarso-Su{\'a}rez},~\bfnm{Carmen}\binits{C.}},
  \bauthor{\bsnm{Klein},~\bfnm{Nadja}\binits{N.}} \AND
  \bauthor{\bsnm{Marey-P{\'e}rez},~\bfnm{Manuel}\binits{M.}}
(\byear{2018}).
\btitle{Studying the occurrence and burnt area of wildfires using
  zero-one-inflated structured additive beta regression}.
\bjournal{Environmental Modelling \& Software}
\bvolume{110}
\bpages{107--118}.
\end{barticle}
\endbibitem

\bibitem[\protect\citeauthoryear{Rue, Martino and Chopin}{2009}]{Rue.al.2009}
\begin{barticle}[author]
\bauthor{\bsnm{Rue},~\bfnm{H{\aa}vard}\binits{H.}},
  \bauthor{\bsnm{Martino},~\bfnm{Sara}\binits{S.}} \AND
  \bauthor{\bsnm{Chopin},~\bfnm{Nicolas}\binits{N.}}
(\byear{2009}).
\btitle{{Approximate Bayesian inference for latent Gaussian models by using
  integrated nested Laplace approximations}}.
\bjournal{Journal of the Royal Statistical Society: Series B (Statistical
  Methodology)}
\bvolume{71}
\bpages{319--392}.
\end{barticle}
\endbibitem

\bibitem[\protect\citeauthoryear{Rue et~al.}{2017}]{Rue.al.2017}
\begin{barticle}[author]
\bauthor{\bsnm{Rue},~\bfnm{H{\aa}vard}\binits{H.}},
  \bauthor{\bsnm{Riebler},~\bfnm{Andrea}\binits{A.}},
  \bauthor{\bsnm{S{\o}rbye},~\bfnm{Sigrunn~H}\binits{S.~H.}},
  \bauthor{\bsnm{Illian},~\bfnm{Janine~B}\binits{J.~B.}},
  \bauthor{\bsnm{Simpson},~\bfnm{Daniel~P}\binits{D.~P.}} \AND
  \bauthor{\bsnm{Lindgren},~\bfnm{Finn~K}\binits{F.~K.}}
(\byear{2017}).
\btitle{{Bayesian computing with INLA: a review}}.
\bjournal{Annual Review of Statistics and Its Application}
\bvolume{4}
\bpages{395--421}.
\end{barticle}
\endbibitem

\bibitem[\protect\citeauthoryear{Ruffault et~al.}{2018}]{Ruffault2018}
\begin{barticle}[author]
\bauthor{\bsnm{Ruffault},~\bfnm{Julien}\binits{J.}},
  \bauthor{\bsnm{Martin-StPaul},~\bfnm{Nicolas}\binits{N.}},
  \bauthor{\bsnm{Pimont},~\bfnm{Francois}\binits{F.}} \AND
  \bauthor{\bsnm{Dupuy},~\bfnm{Jean-Luc}\binits{J.-L.}}
(\byear{2018}).
\btitle{How well do meteorological drought indices predict live fuel moisture
  content ({LFMC})? An assessment for wildfire research and operations in
  {M}editerranean ecosystems}.
\bjournal{Agricultural and Forest Meteorology}
\bvolume{262}
\bpages{391--401}.
\end{barticle}
\endbibitem

\bibitem[\protect\citeauthoryear{Schoenberg, Peng and
  Woods}{2003}]{Schoenberg2003}
\begin{barticle}[author]
\bauthor{\bsnm{Schoenberg},~\bfnm{Frederic~Paik}\binits{F.~P.}},
  \bauthor{\bsnm{Peng},~\bfnm{Roger}\binits{R.}} \AND
  \bauthor{\bsnm{Woods},~\bfnm{James}\binits{J.}}
(\byear{2003}).
\btitle{On the distribution of wildfire sizes}.
\bjournal{Environmetrics}
\bvolume{14}
\bpages{583--592}.
\end{barticle}
\endbibitem

\bibitem[\protect\citeauthoryear{Serra et~al.}{2013}]{Serra2013}
\begin{barticle}[author]
\bauthor{\bsnm{Serra},~\bfnm{Laura}\binits{L.}},
  \bauthor{\bsnm{Juan},~\bfnm{Pablo}\binits{P.}},
  \bauthor{\bsnm{Varga},~\bfnm{Diego}\binits{D.}},
  \bauthor{\bsnm{Mateu},~\bfnm{Jorge}\binits{J.}} \AND
  \bauthor{\bsnm{Saez},~\bfnm{Marc}\binits{M.}}
(\byear{2013}).
\btitle{{Spatial pattern modelling of wildfires in Catalonia, Spain
  2004--2008}}.
\bjournal{Environmental Modelling \& Software}
\bvolume{40}
\bpages{235--244}.
\end{barticle}
\endbibitem

\bibitem[\protect\citeauthoryear{Serra et~al.}{2014}]{Serra2014}
\begin{barticle}[author]
\bauthor{\bsnm{Serra},~\bfnm{Laura}\binits{L.}},
  \bauthor{\bsnm{Saez},~\bfnm{Marc}\binits{M.}},
  \bauthor{\bsnm{Juan},~\bfnm{Pablo}\binits{P.}},
  \bauthor{\bsnm{Varga},~\bfnm{Diego}\binits{D.}} \AND
  \bauthor{\bsnm{Mateu},~\bfnm{Jorge}\binits{J.}}
(\byear{2014}).
\btitle{{A spatio-temporal Poisson hurdle point process to model wildfires}}.
\bjournal{Stochastic environmental research and risk assessment}
\bvolume{28}
\bpages{1671--1684}.
\end{barticle}
\endbibitem

\bibitem[\protect\citeauthoryear{Simpson et~al.}{2017}]{Simpson.al.2017}
\begin{barticle}[author]
\bauthor{\bsnm{Simpson},~\bfnm{Daniel}\binits{D.}},
  \bauthor{\bsnm{Rue},~\bfnm{H{\aa}vard}\binits{H.}},
  \bauthor{\bsnm{Riebler},~\bfnm{Andrea}\binits{A.}},
  \bauthor{\bsnm{Martins},~\bfnm{Thiago~G}\binits{T.~G.}},
  \bauthor{\bsnm{S{\o}rbye},~\bfnm{Sigrunn~H}\binits{S.~H.}} \betal{et~al.}
(\byear{2017}).
\btitle{Penalising model component complexity: A principled, practical approach
  to constructing priors}.
\bjournal{Statistical Science}
\bvolume{32}
\bpages{1--28}.
\end{barticle}
\endbibitem

\bibitem[\protect\citeauthoryear{Stark and Hovius}{2001}]{Stark2001}
\begin{barticle}[author]
\bauthor{\bsnm{Stark},~\bfnm{Colin~P}\binits{C.~P.}} \AND
  \bauthor{\bsnm{Hovius},~\bfnm{Niels}\binits{N.}}
(\byear{2001}).
\btitle{The characterization of landslide size distributions}.
\bjournal{Geophysical Research Letters}
\bvolume{28}
\bpages{1091--1094}.
\end{barticle}
\endbibitem

\bibitem[\protect\citeauthoryear{Stewart et~al.}{2007}]{Stewart.2007}
\begin{barticle}[author]
\bauthor{\bsnm{Stewart},~\bfnm{Susan~I.}\binits{S.~I.}},
  \bauthor{\bsnm{Radeloff},~\bfnm{Volker~C.}\binits{V.~C.}},
  \bauthor{\bsnm{Hammer},~\bfnm{Roger~B.}\binits{R.~B.}} \AND
  \bauthor{\bsnm{Hawbaker},~\bfnm{Todd~J.}\binits{T.~J.}}
(\byear{2007}).
\btitle{{Defining the Wildland–Urban Interface}}.
\bjournal{Journal of Forestry}
\bvolume{105}
\bpages{201--207}.
\bdoi{10.1093/jof/105.4.201}
\end{barticle}
\endbibitem

\bibitem[\protect\citeauthoryear{Taylor and Diggle}{2014}]{Taylor.Diggle.2014}
\begin{barticle}[author]
\bauthor{\bsnm{Taylor},~\bfnm{Benjamin~M}\binits{B.~M.}} \AND
  \bauthor{\bsnm{Diggle},~\bfnm{Peter~J}\binits{P.~J.}}
(\byear{2014}).
\btitle{{INLA or MCMC? A tutorial and comparative evaluation for spatial
  prediction in log-Gaussian Cox processes}}.
\bjournal{Journal of Statistical Computation and Simulation}
\bvolume{84}
\bpages{2266--2284}.
\end{barticle}
\endbibitem

\bibitem[\protect\citeauthoryear{Tierney and
  Kadane}{1986}]{Tierney.Kadane.1986}
\begin{barticle}[author]
\bauthor{\bsnm{Tierney},~\bfnm{Luke}\binits{L.}} \AND
  \bauthor{\bsnm{Kadane},~\bfnm{Joseph~B}\binits{J.~B.}}
(\byear{1986}).
\btitle{Accurate approximations for posterior moments and marginal densities}.
\bjournal{Journal of the American Statistical Association}
\bvolume{81}
\bpages{82--86}.
\end{barticle}
\endbibitem

\bibitem[\protect\citeauthoryear{Tokdar and Kass}{2010}]{Tokdar2010}
\begin{barticle}[author]
\bauthor{\bsnm{Tokdar},~\bfnm{Surya~T.}\binits{S.~T.}} \AND
  \bauthor{\bsnm{Kass},~\bfnm{Robert~E.}\binits{R.~E.}}
(\byear{2010}).
\btitle{Importance sampling: a review}.
\bjournal{Wiley Interdisciplinary Reviews: Computational Statistics}
\bvolume{2}
\bpages{54--60}.
\bdoi{10.1002/wics.56}
\end{barticle}
\endbibitem

\bibitem[\protect\citeauthoryear{Tonini et~al.}{2017}]{Tonini2017}
\begin{barticle}[author]
\bauthor{\bsnm{Tonini},~\bfnm{Marj}\binits{M.}},
  \bauthor{\bsnm{Pereira},~\bfnm{M{\'a}rio~Gonzalez}\binits{M.~G.}},
  \bauthor{\bsnm{Parente},~\bfnm{Joana}\binits{J.}} \AND
  \bauthor{\bsnm{Orozco},~\bfnm{Carmen~Vega}\binits{C.~V.}}
(\byear{2017}).
\btitle{Evolution of forest fires in {P}ortugal: from spatio-temporal point
  events to smoothed density maps}.
\bjournal{Natural Hazards}
\bvolume{85}
\bpages{1489--1510}.
\end{barticle}
\endbibitem

\bibitem[\protect\citeauthoryear{Turkman, {Amaral Turkman} and
  Pereira}{2010}]{Turkman2010}
\begin{barticle}[author]
\bauthor{\bsnm{Turkman},~\bfnm{Kamil~Feridun}\binits{K.~F.}},
  \bauthor{\bsnm{{Amaral Turkman}},~\bfnm{M.~A.}\binits{M.~A.}} \AND
  \bauthor{\bsnm{Pereira},~\bfnm{J.~M.}\binits{J.~M.}}
(\byear{2010}).
\btitle{{Asymptotic models and inference for extremes of spatio-temporal
  data}}.
\bjournal{Extremes}
\bvolume{13}
\bpages{375--397}.
\bdoi{10.1007/s10687-009-0092-8}
\end{barticle}
\endbibitem

\bibitem[\protect\citeauthoryear{van Niekerk et~al.}{2019}]{Niekerk2019}
\begin{barticle}[author]
\bauthor{\bparticle{van} \bsnm{Niekerk},~\bfnm{Janet}\binits{J.}},
  \bauthor{\bsnm{Bakka},~\bfnm{Haakon}\binits{H.}},
  \bauthor{\bsnm{Rue},~\bfnm{Haavard}\binits{H.}} \AND
  \bauthor{\bsnm{Schenk},~\bfnm{Loaf}\binits{L.}}
(\byear{2019}).
\btitle{{New frontiers in Bayesian modeling using the INLA package in R}}.
\bjournal{arXiv preprint arXiv:1907.10426}.
\end{barticle}
\endbibitem

\bibitem[\protect\citeauthoryear{van Wagner}{1977}]{vanWagner1977}
\begin{barticle}[author]
\bauthor{\bparticle{van} \bsnm{Wagner},~\bfnm{C.~E.}\binits{C.~E.}}
(\byear{1977}).
\btitle{Conditions for the start and spread of crown fire}.
\bjournal{Canadian Journal of Forest Research}
\bvolume{7}
\bpages{23--34}.
\end{barticle}
\endbibitem

\bibitem[\protect\citeauthoryear{Watanabe}{2010}]{Watanabe.2010}
\begin{barticle}[author]
\bauthor{\bsnm{Watanabe},~\bfnm{Sumio}\binits{S.}}
(\byear{2010}).
\btitle{Asymptotic Equivalence of {B}ayes Cross Validation and {W}idely
  {A}pplicable {I}nformation {C}riterion in Singular Learning Theory}.
\bjournal{Journal of Machine Learning Research}
\bvolume{11}
\bpages{3571--3594}.
\end{barticle}
\endbibitem

\bibitem[\protect\citeauthoryear{Xi et~al.}{2019}]{Xi2019}
\begin{barticle}[author]
\bauthor{\bsnm{Xi},~\bfnm{Dexen~DZ}\binits{D.~D.}},
  \bauthor{\bsnm{Taylor},~\bfnm{Stephen~W}\binits{S.~W.}},
  \bauthor{\bsnm{Woolford},~\bfnm{Douglas~G}\binits{D.~G.}} \AND
  \bauthor{\bsnm{Dean},~\bfnm{CB}\binits{C.}}
(\byear{2019}).
\btitle{Statistical models of key components of wildfire risk}.
\bjournal{Annual review of statistics and its application}
\bvolume{6}
\bpages{197--222}.
\end{barticle}
\endbibitem

\bibitem[\protect\citeauthoryear{Xu and Schoenberg}{2011}]{Xu2011}
\begin{barticle}[author]
\bauthor{\bsnm{Xu},~\bfnm{Haiyong}\binits{H.}} \AND
  \bauthor{\bsnm{Schoenberg},~\bfnm{Frederic~Paik}\binits{F.~P.}}
(\byear{2011}).
\btitle{{Point process modeling of wildfire hazard in Los Angeles County,
  California}}.
\bjournal{The Annals of Applied Statistics}
\bvolume{5}
\bpages{684--704}.
\end{barticle}
\endbibitem

\end{thebibliography}


\newpage

\begin{supplement}
\stitle{Plots for the inspection of posterior predictive densities}
\sdescription{Figures \ref{fig:val:exceedreg2:test}, \ref{fig:val:prediction:BA} and \ref{fig:val:reg1} show our visual assessment of the predictive behavior of our chosen model M1.}

\begin{figure}[!t]
\centering
    \includegraphics[width=.25\textwidth]{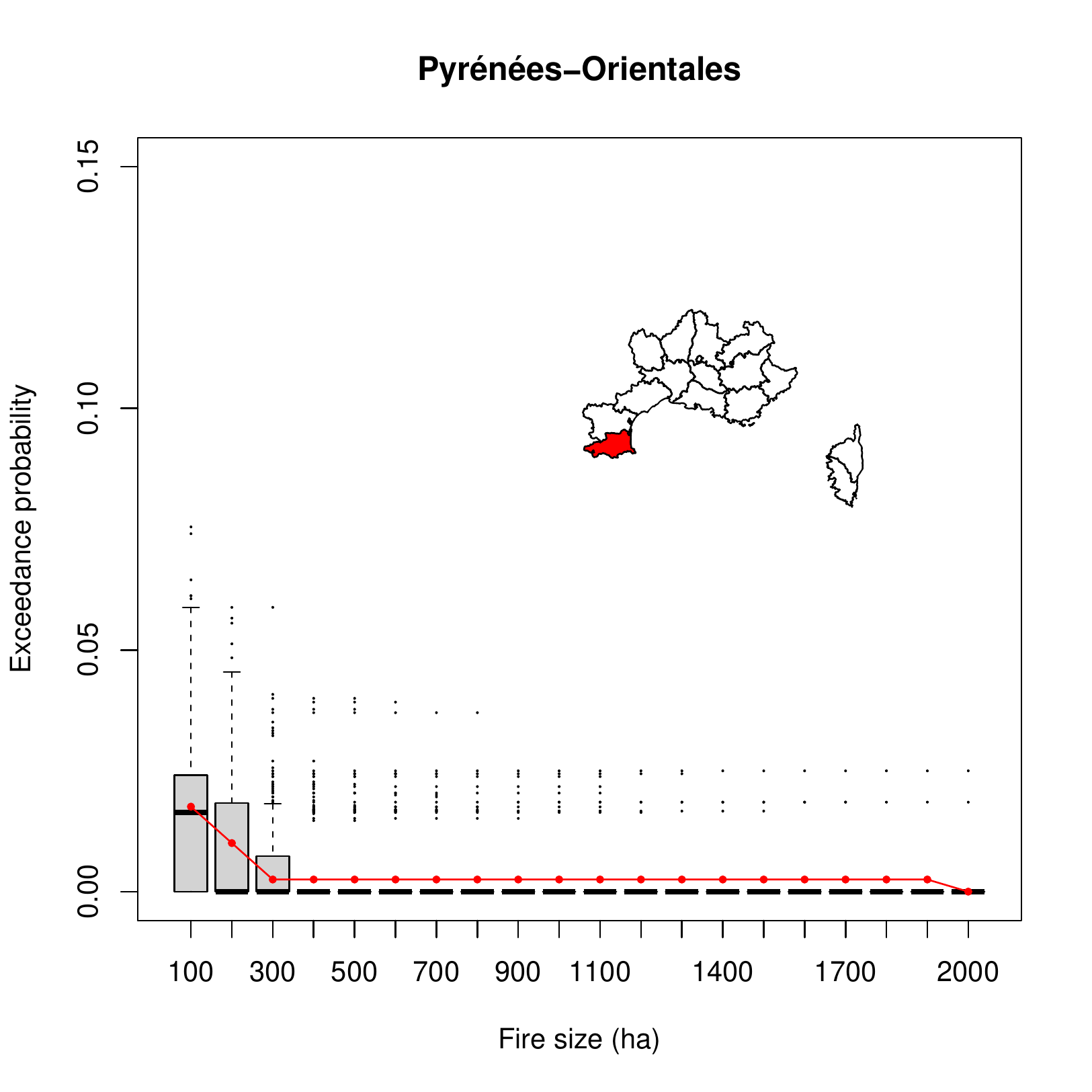} 
        \includegraphics[width=.25\textwidth]{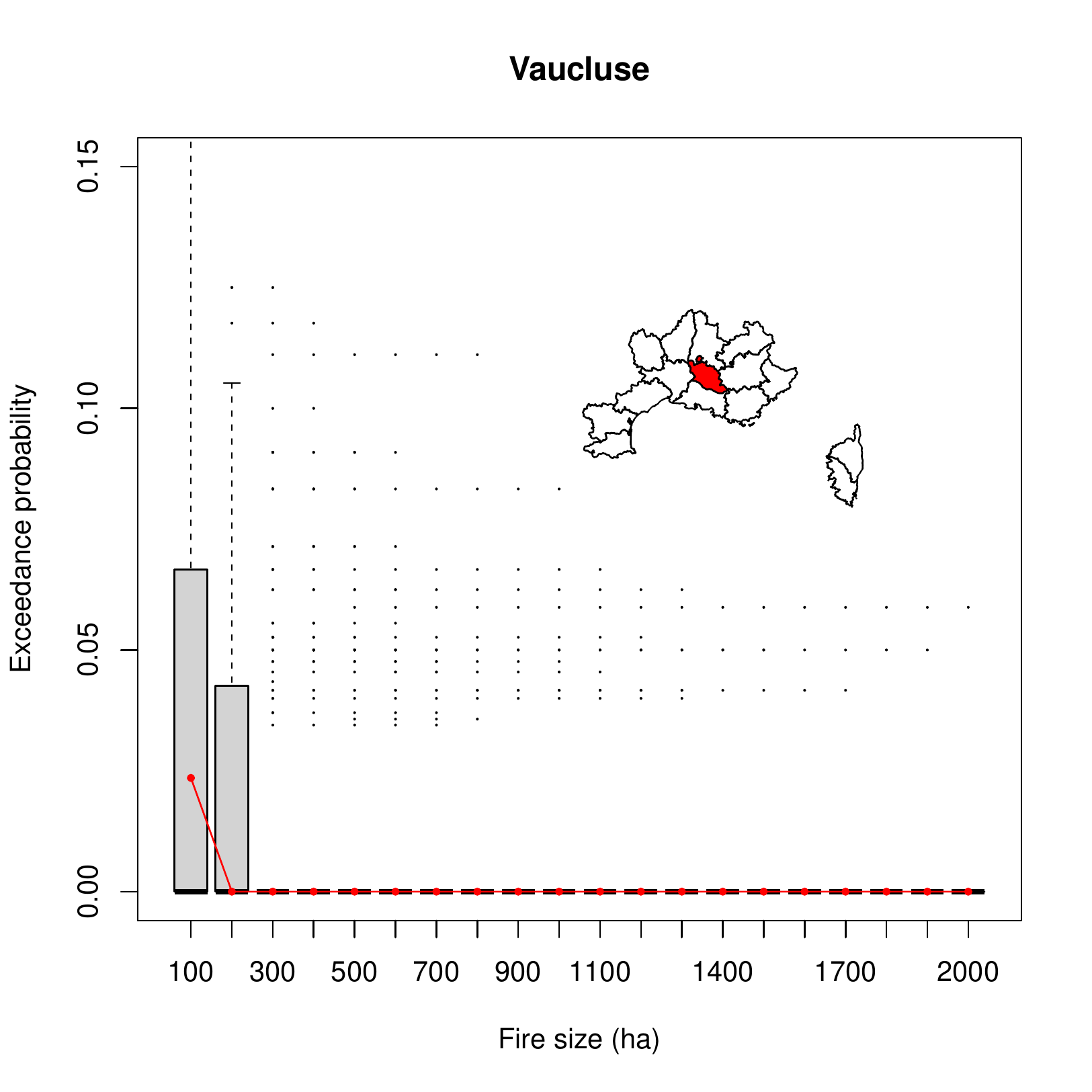} 
     \includegraphics[width=.25\textwidth]{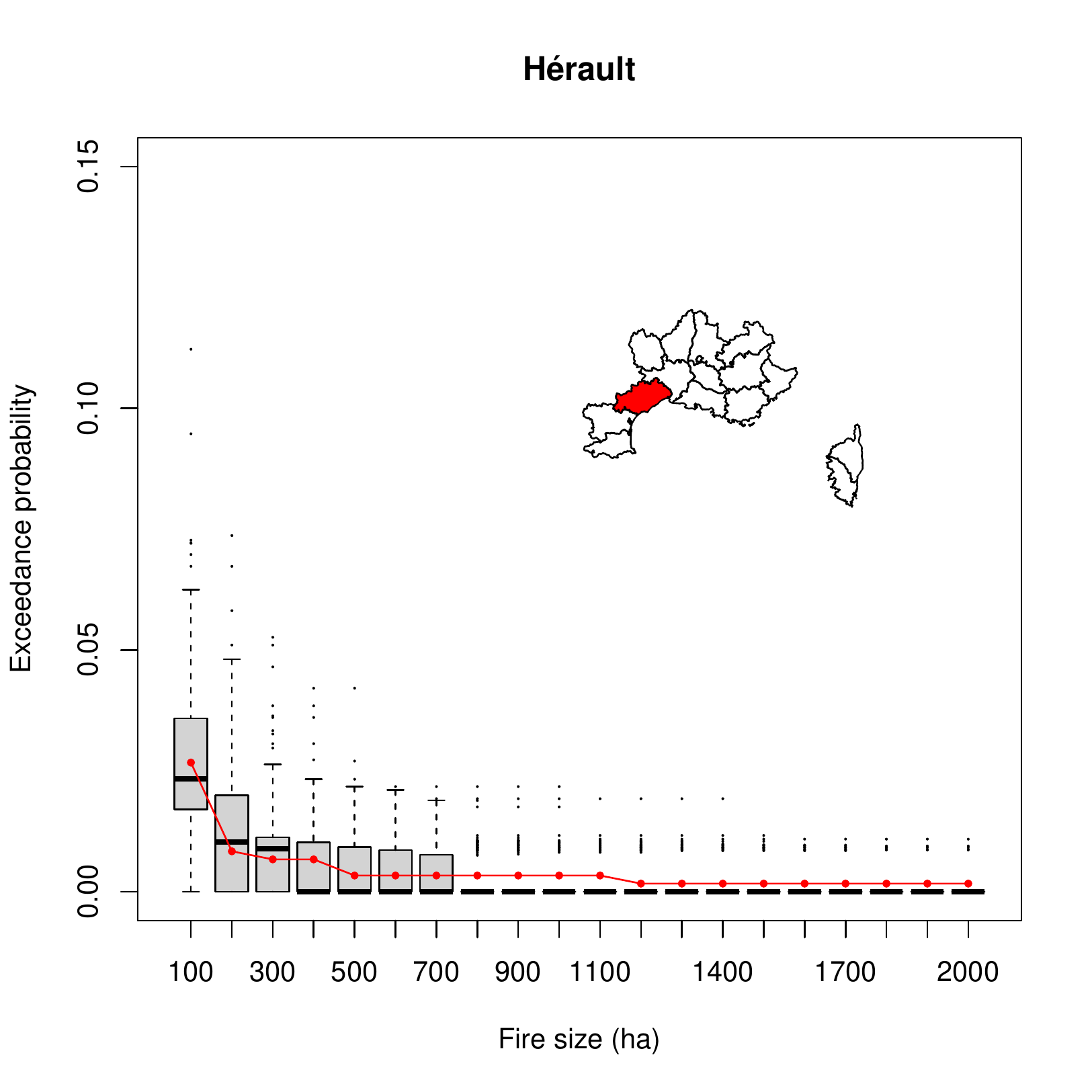} \\
     \includegraphics[width=.25\textwidth]{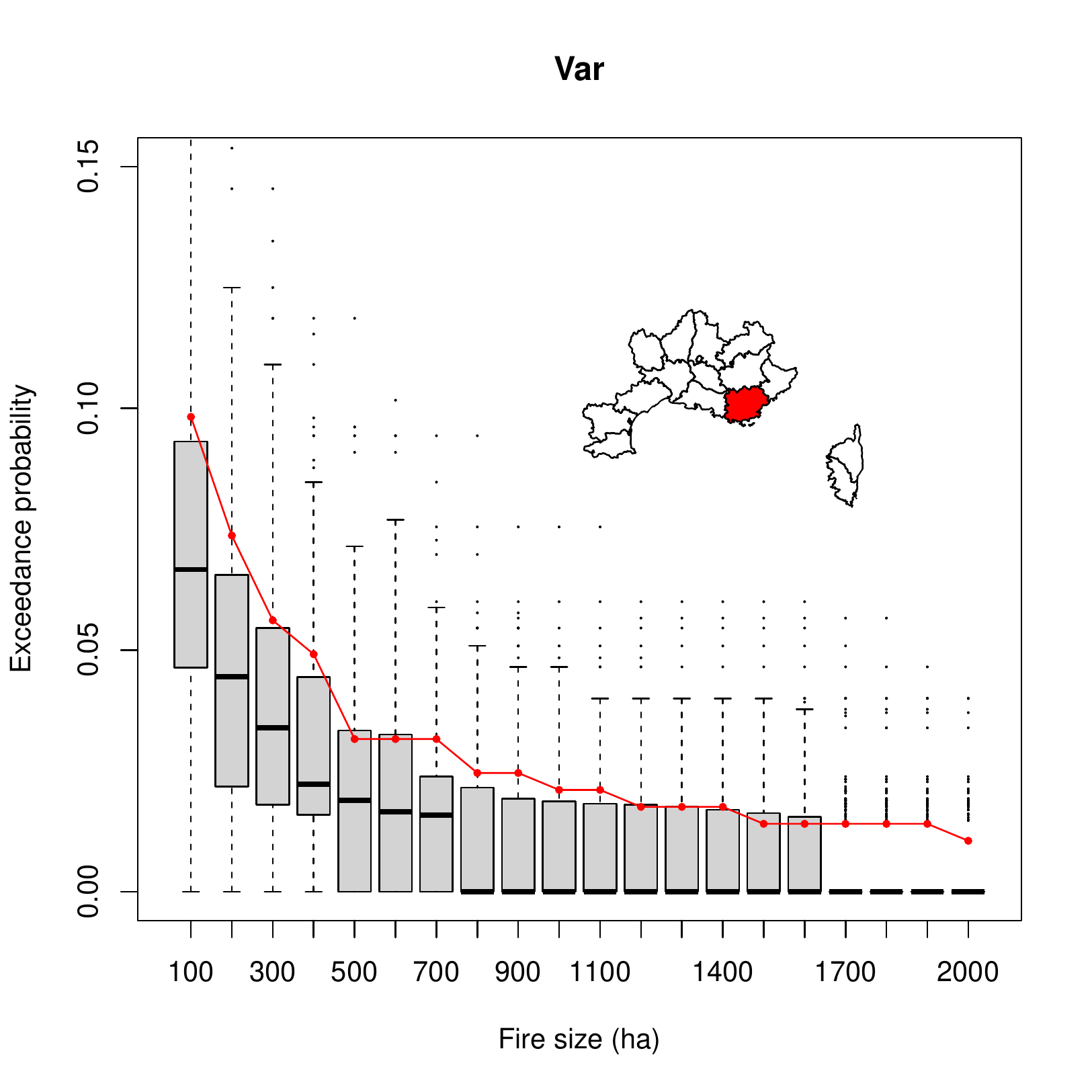} 
     \includegraphics[width=.25\textwidth]{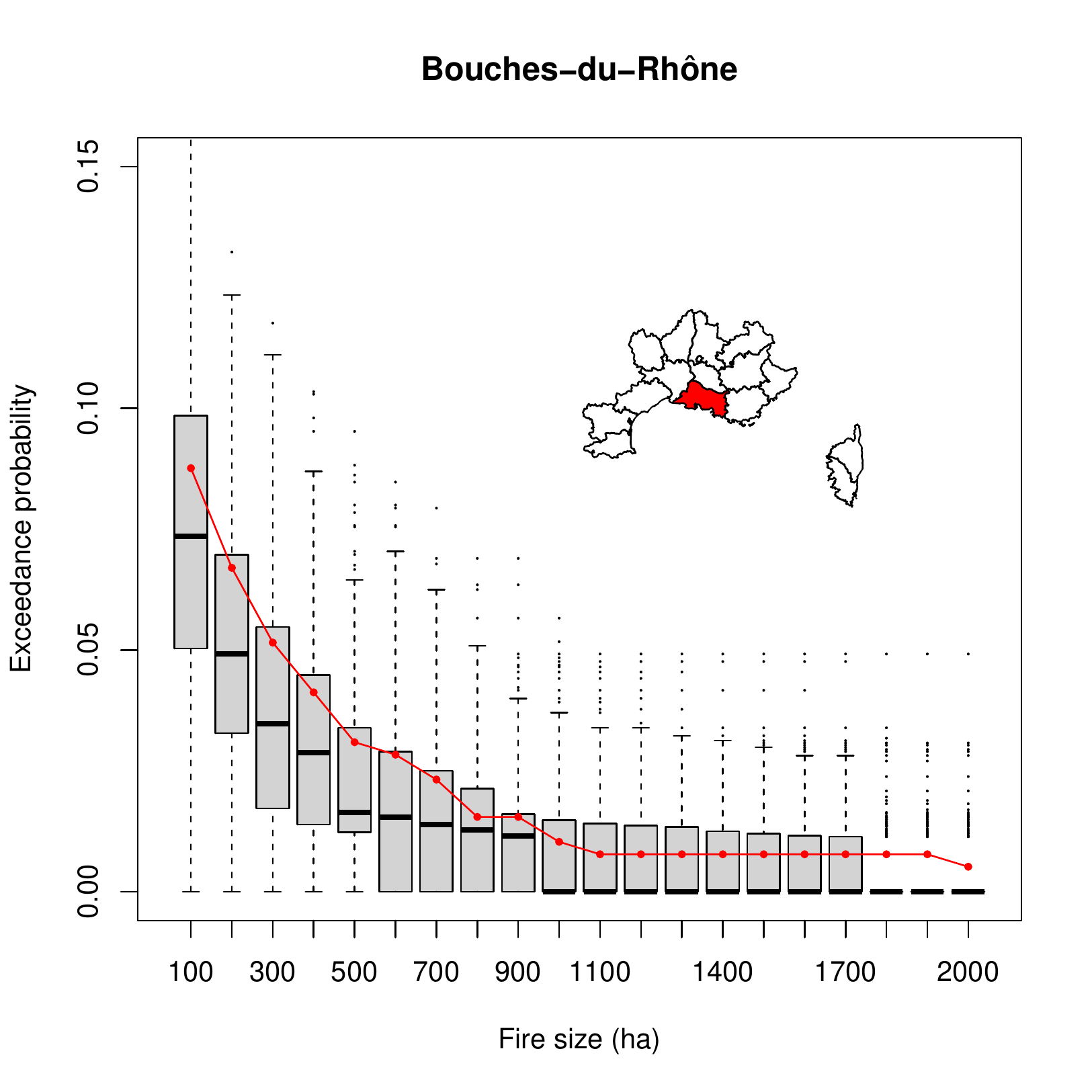}
     \includegraphics[width=.25\textwidth]{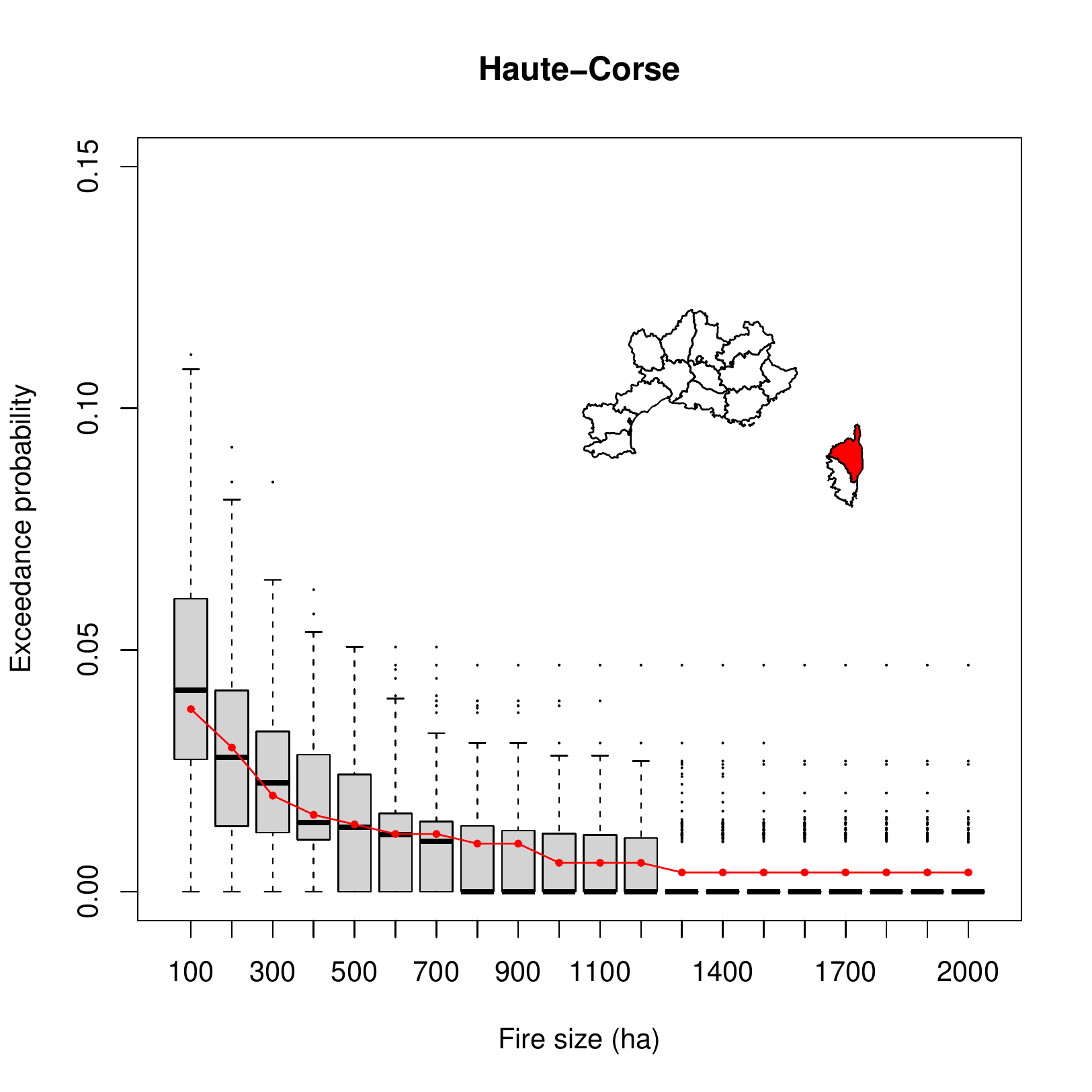} \\
  \caption{Exceedance probability plots for six  d\'epartements (in red on the maps) in the validation period (2015--2018). Boxplots are based on 200 posterior simulations. Red lines represent observed empirical exceedance probabilities.}  
  \label{fig:val:exceedreg2:test}
\end{figure}

\begin{figure}[!t]
\centering
     \includegraphics[width=.8\textwidth]{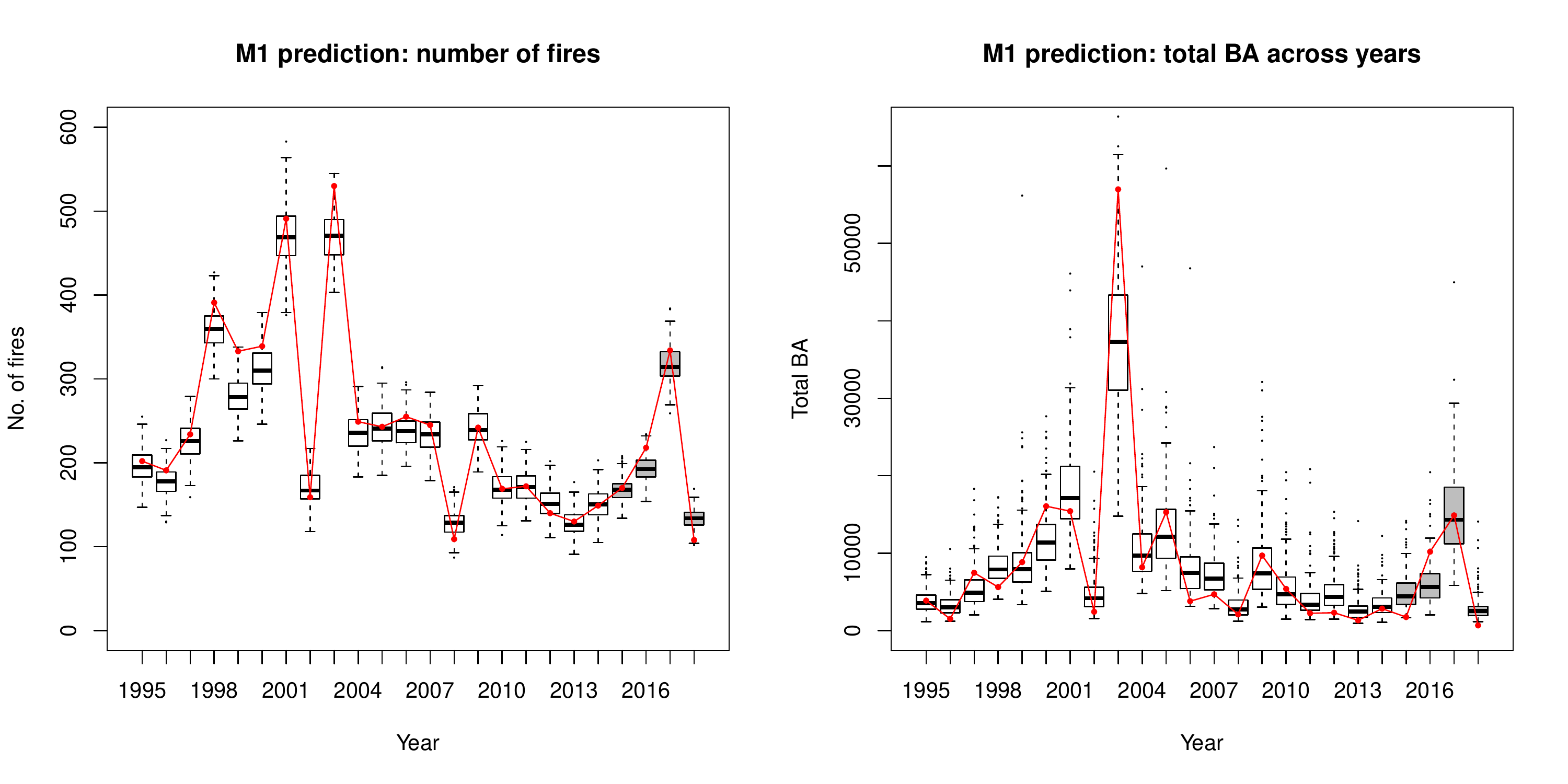}
  \caption{Boxplots by year for the predicted number of fires (left) and predicted total burnt area (right) across the whole region  from 200 simulations of the posterior model. The grey boxplots indicate the out-sample years. The red lines represent the observed annual total number of fires and total burnt area in the whole region.}  
  \label{fig:val:prediction:BA}
\end{figure}

\begin{figure}[!t]
\centering
    \includegraphics[width=.25\textwidth]{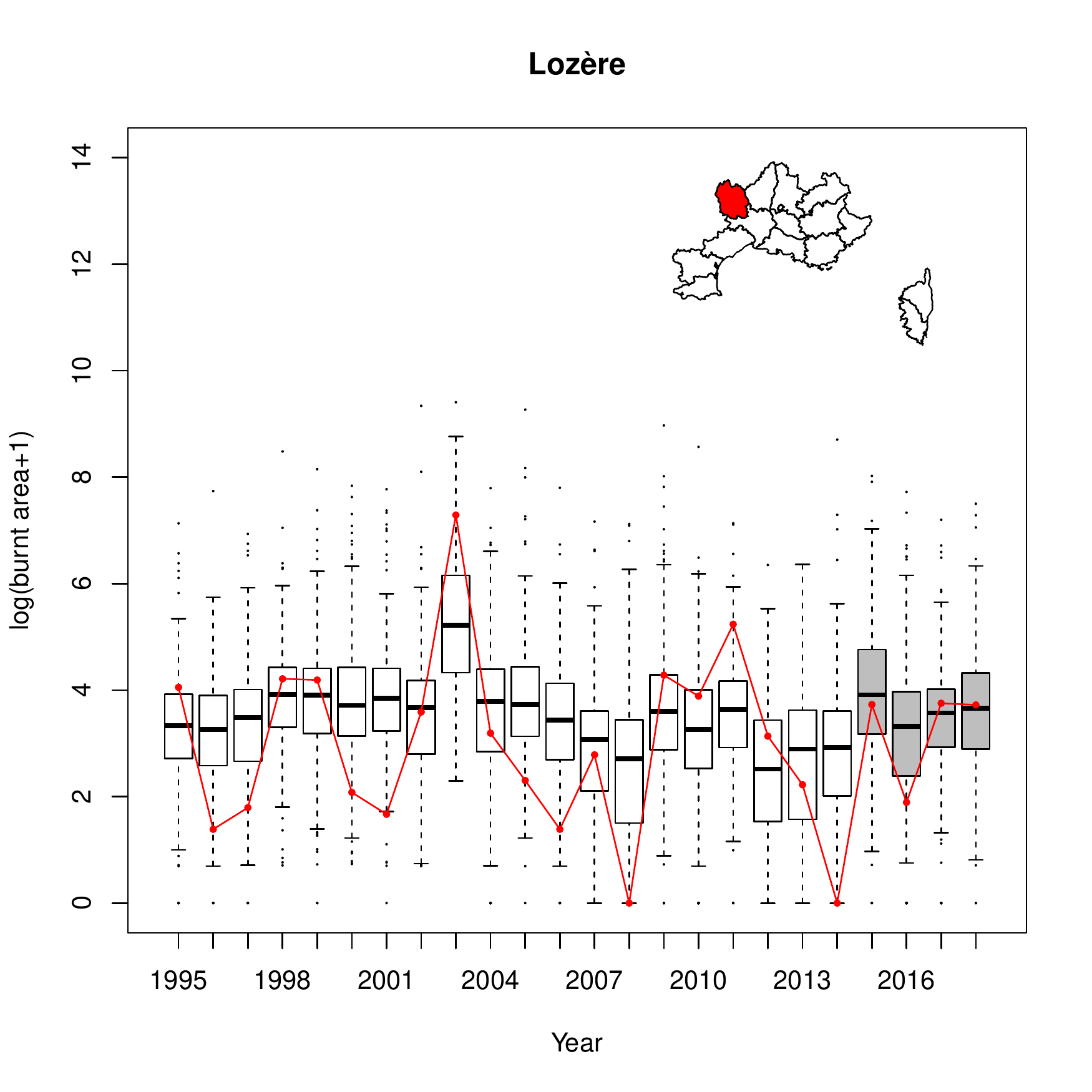} 
     \includegraphics[width=.25\textwidth]{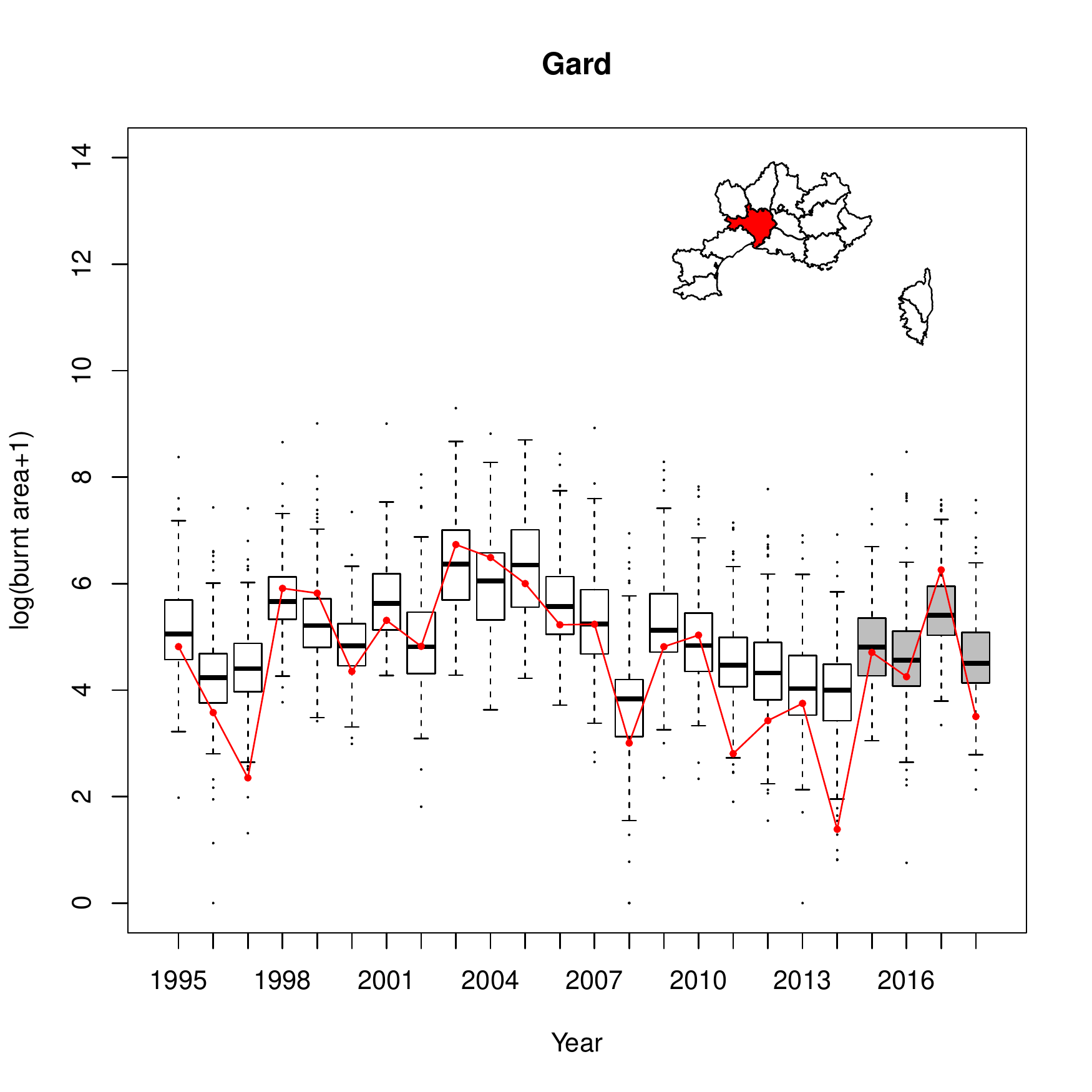} 
     \includegraphics[width=.25\textwidth]{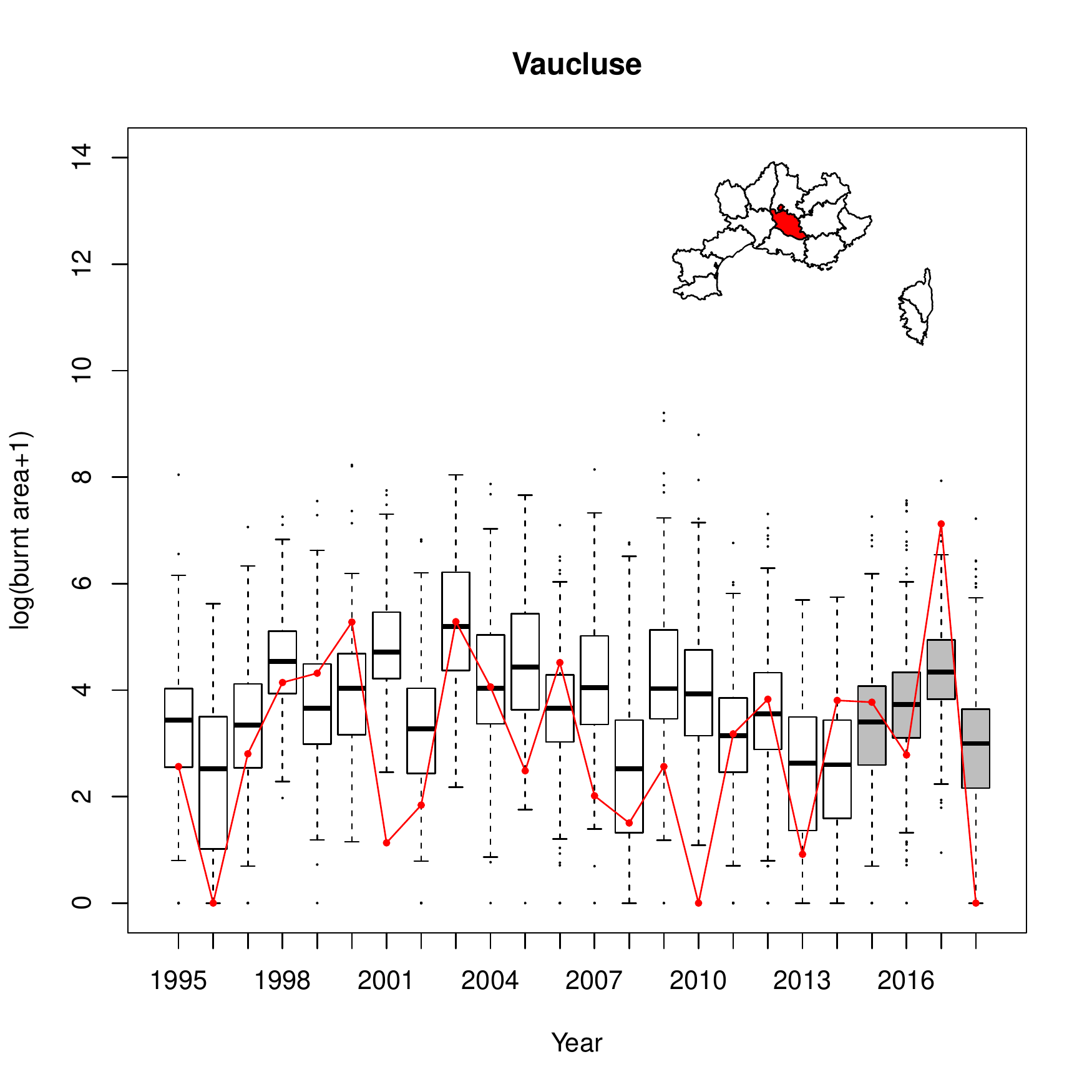}\\
     \includegraphics[width=.25\textwidth]{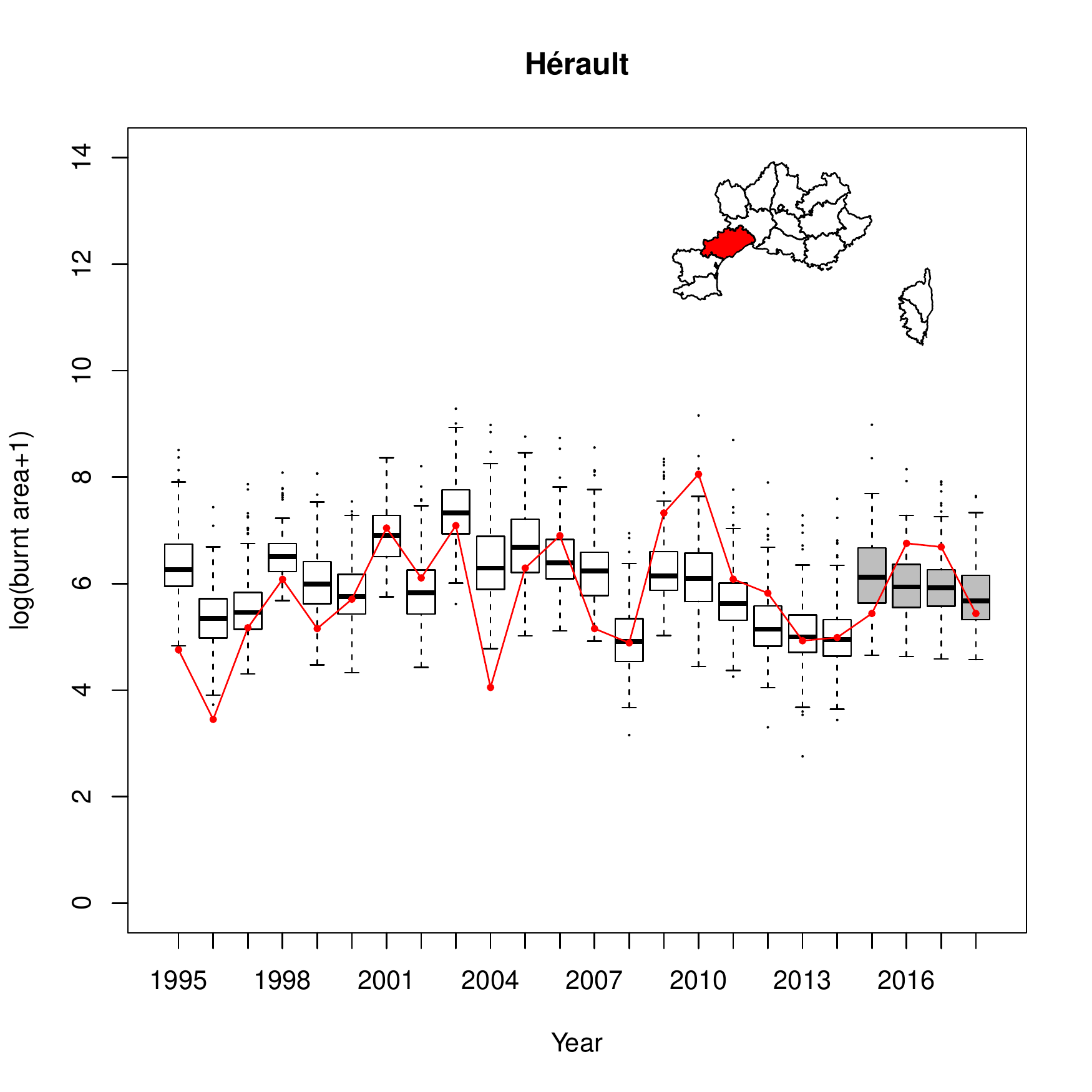} 
     \includegraphics[width=.25\textwidth]{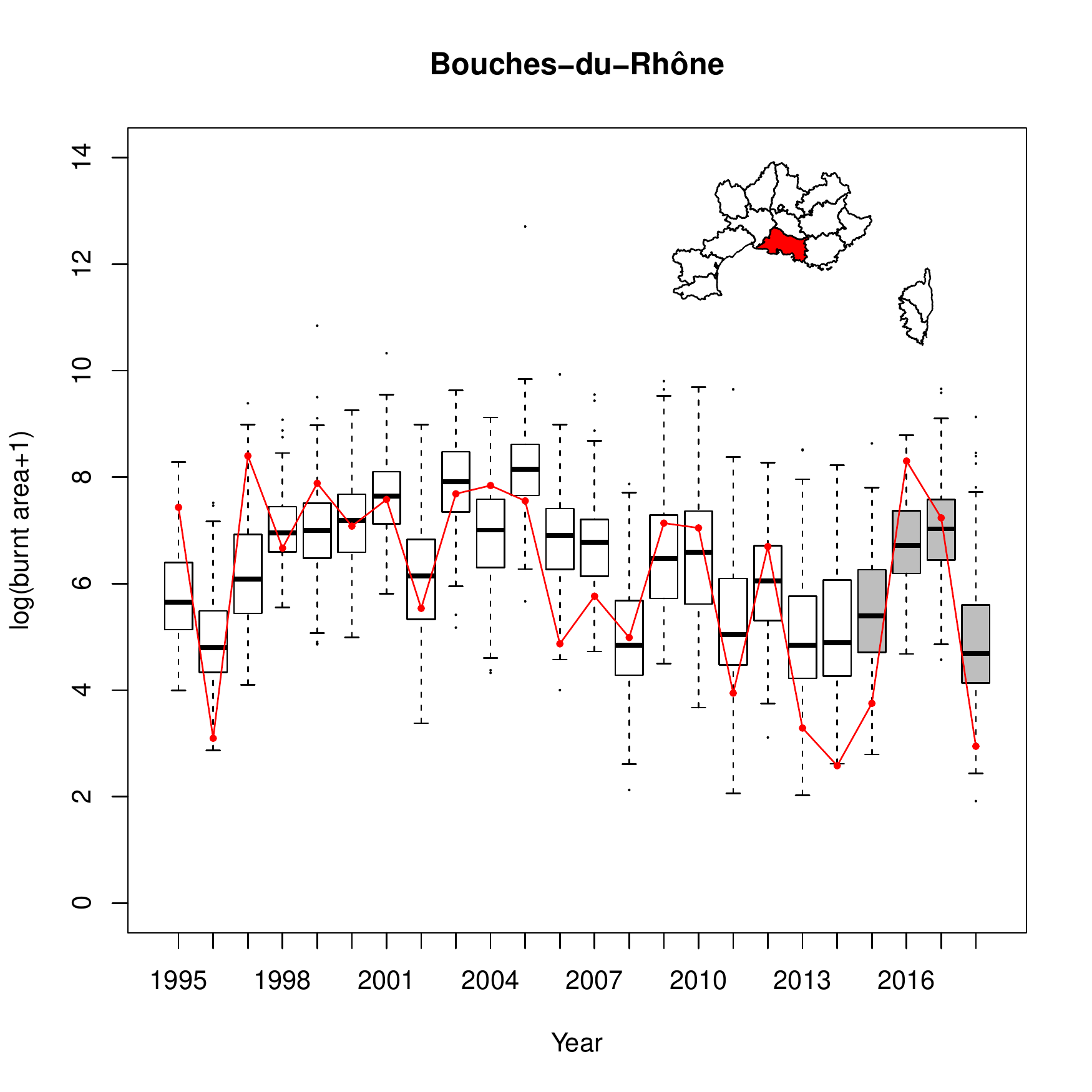} 
     \includegraphics[width=.25\textwidth]{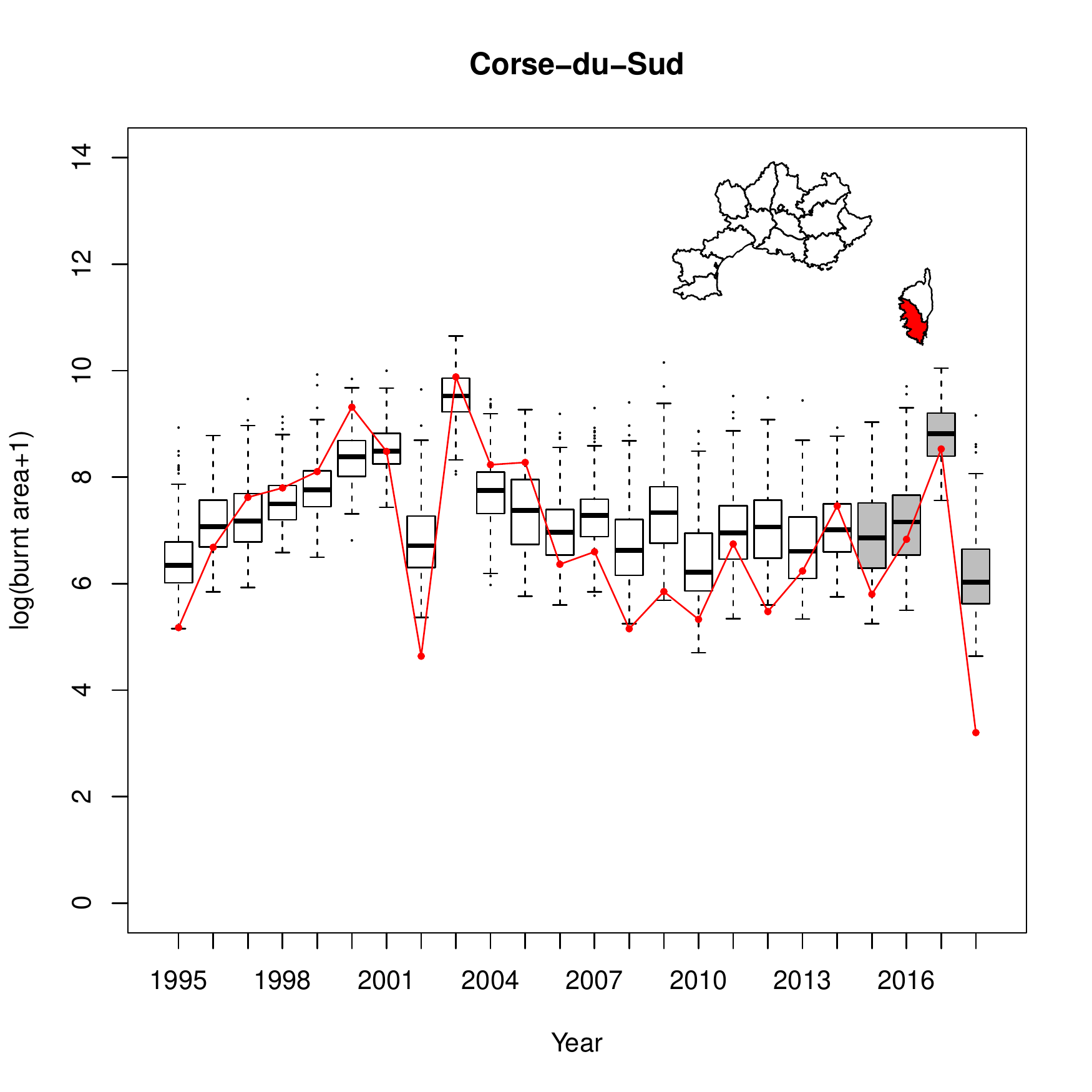} 
  \caption{Boxplots by year and d\'epartment for the predicted logarithmic total burnt area from 200 simulations of the model. Red dots represent the observed annual log total burnt area in each d\'epartment. The d\'epartment used for each panel is shown in red on the maps.}  
  \label{fig:val:reg1}
\end{figure}

\end{supplement}

\begin{supplement}

\stitle{Plots showing regionalized predictions}\label{sec:supplement:regionalized}

\sdescription{Figure \ref{fig:application::fires} shows the regionalized predictions due to the spatial effects used in our model. }

\begin{figure}[t!]
\centering
    \includegraphics[width=.40\textwidth]{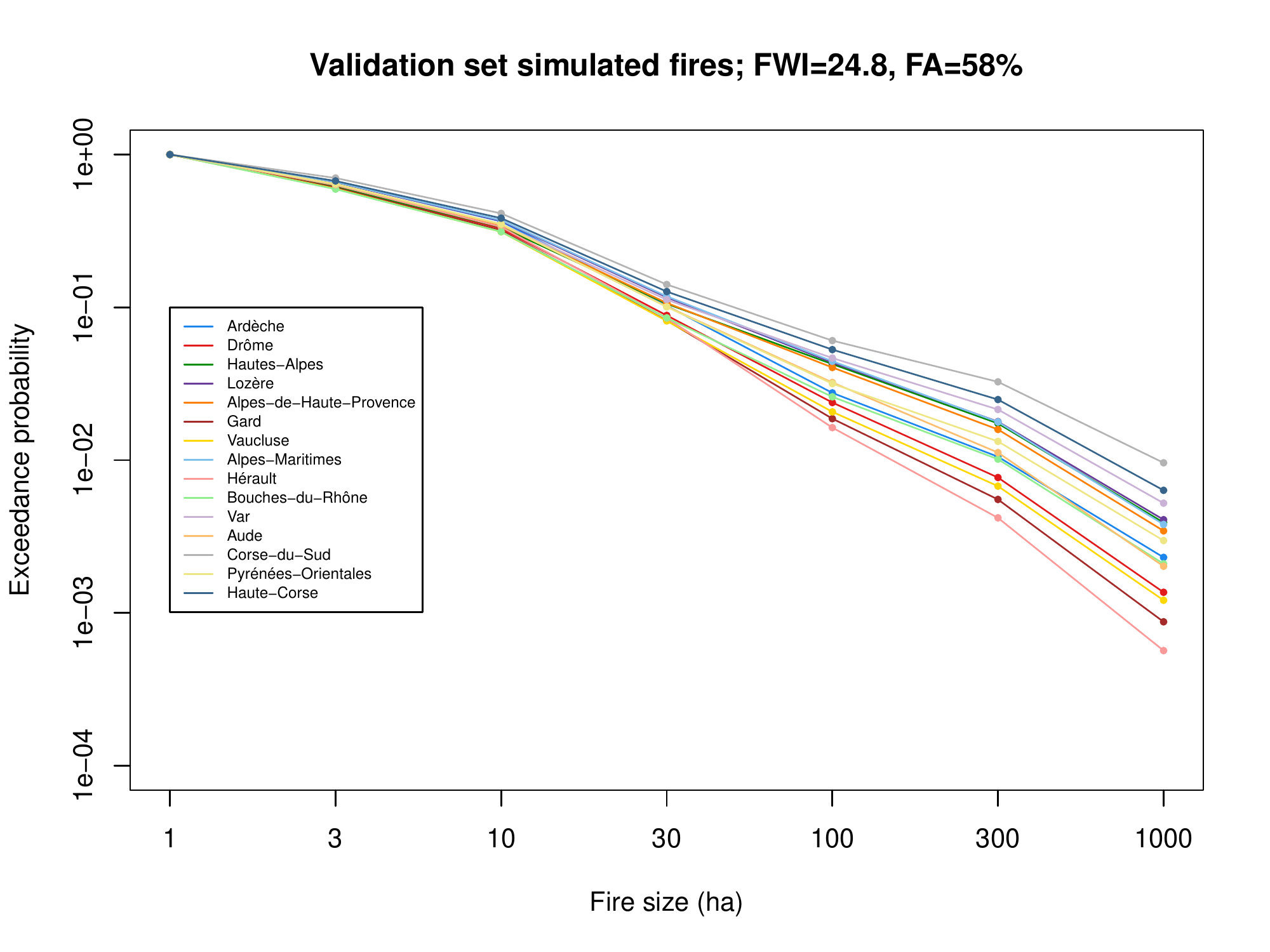} 
     \includegraphics[width=.40\textwidth]{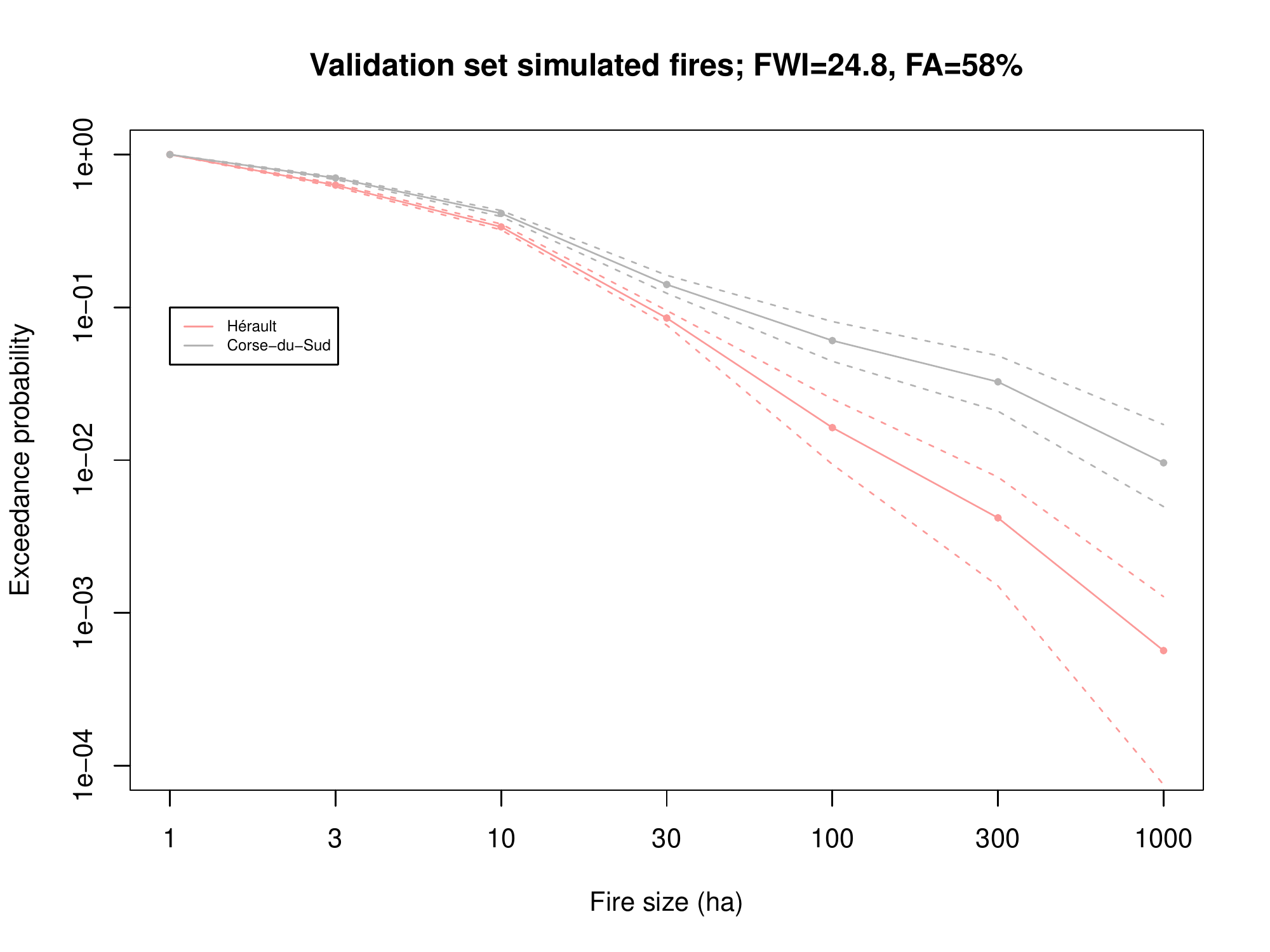} 
  \caption{Left: Exceedance probability plots by d\'epartment from 1000 posterior samples of the model M1 on the validation period given fixed FWI and FA. Right: Same as the left panel, but only for two d\'epartements with corresponding $95\%$ credible intervals.}  
  \label{fig:application::fires}
\end{figure}

\end{supplement}

\begin{supplement}

\stitle{Kernel intensity plot}\label{sec:supplement:kernel}

\sdescription{Figure~\ref{fig:map-dfci-grid} shows a map of the wildfire locations as recorded in the Prom\'eth\'ee database. The overlaid contour lines of a kernel intensity estimation highlight the strong spatial nonstationarity, with several relatively small hotspot areas characterized by  high occurrence numbers.}

\begin{figure}[t]
    \centering
    \includegraphics[width=0.7\textwidth]{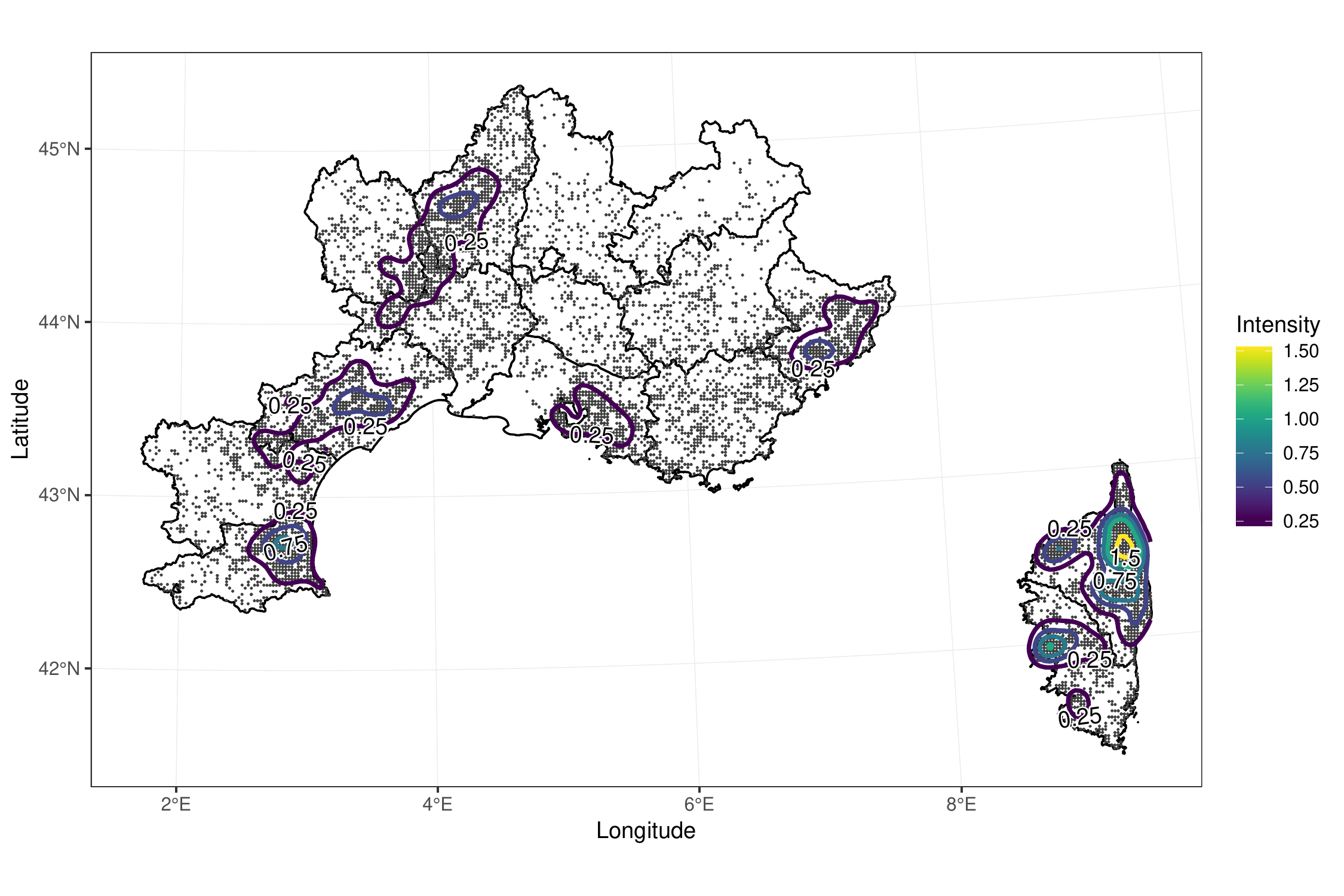}
    \caption{Map of Prom\'eth\'ee data in Southern France with the island of Corsica on the lower right, based on  the original DFCI grid used for recording wildfires.  Black lines indicate boundaries of administrative regions (``d\'epartements").  Coloured lines correspond to intensities (\ie  to average numbers of wildfires per km$^2$) and highlight areas with many wildfires. Some of the gray points correspond to multiple wildfire occurrences.}
    \label{fig:map-dfci-grid}
\end{figure}

\end{supplement}

\begin{supplement}

\stitle{Mean excess plots}\label{sec:supplement:me}

\sdescription{We consider the mean excess plots of burnt areas and log10 burnt areas in the middle and right displays of Figure~\ref{fig:mean-excess}. Given a threshold value $u$ set for a random variable $Y$, the mean excess corresponds to the conditional expectation $\mathbb{E}[Y-u\mid Y>u]$, \ie the  expectation of the positive excess above the threshold. Mean excess plots report the corresponding empirical means. In case of exponential tail decay $\text{Pr}(Y>y)=\exp\{- (y-\mu)/\lambda\}$ for $y\geq u_0$ with scale $\lambda>0$ and an arbitrary shift $\mu\in\mathbb{R}$, the mean excess would be constant $\lambda$ for thresholds $u$ above $u_0$. The  mean excess plot for log10 of BA-log10 indicates approximately exponential tail decay  for low thresholds where mean excess values are relatively stable for threshold values in $(0,1.5)$ except for rounding of burnt areas. However, the tail decay becomes faster at higher levels, starting at around $30$~ha. Exponential decay on log-scale would correspond to power-law decay at the original scale; \ie to Pareto-like behavior. By contrast, the mean excess plot of original BA values becomes relatively stable for thresholds above $500$~ha, such that the true, ultimate tail decay rate at very high quantiles could be exponential. These plots reveal the difficulty of choosing an appropriate probability distribution for burnt areas.}

\begin{figure}[t]
    \centering
    \includegraphics[width=.3\textwidth]{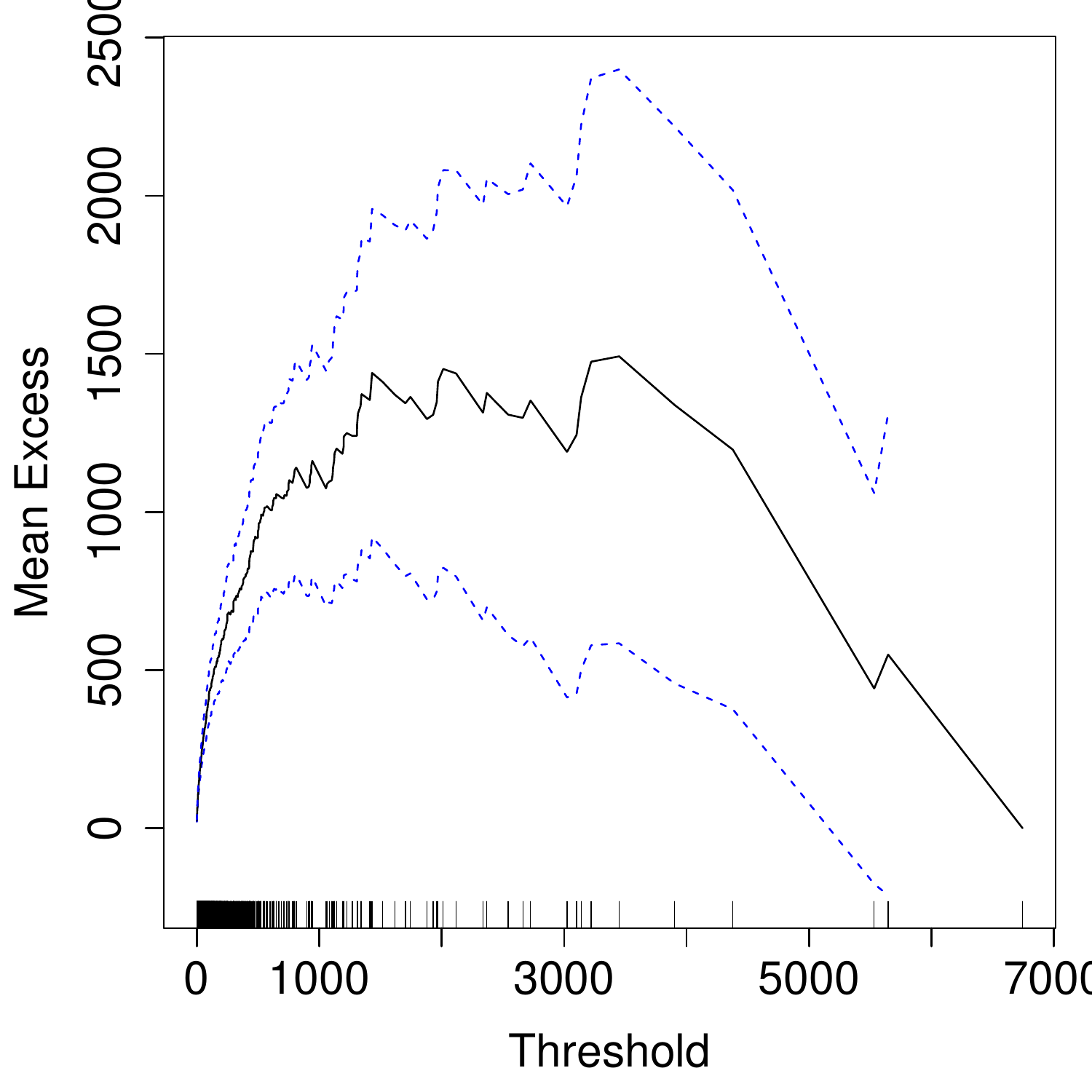}
    \includegraphics[width=.3\textwidth]{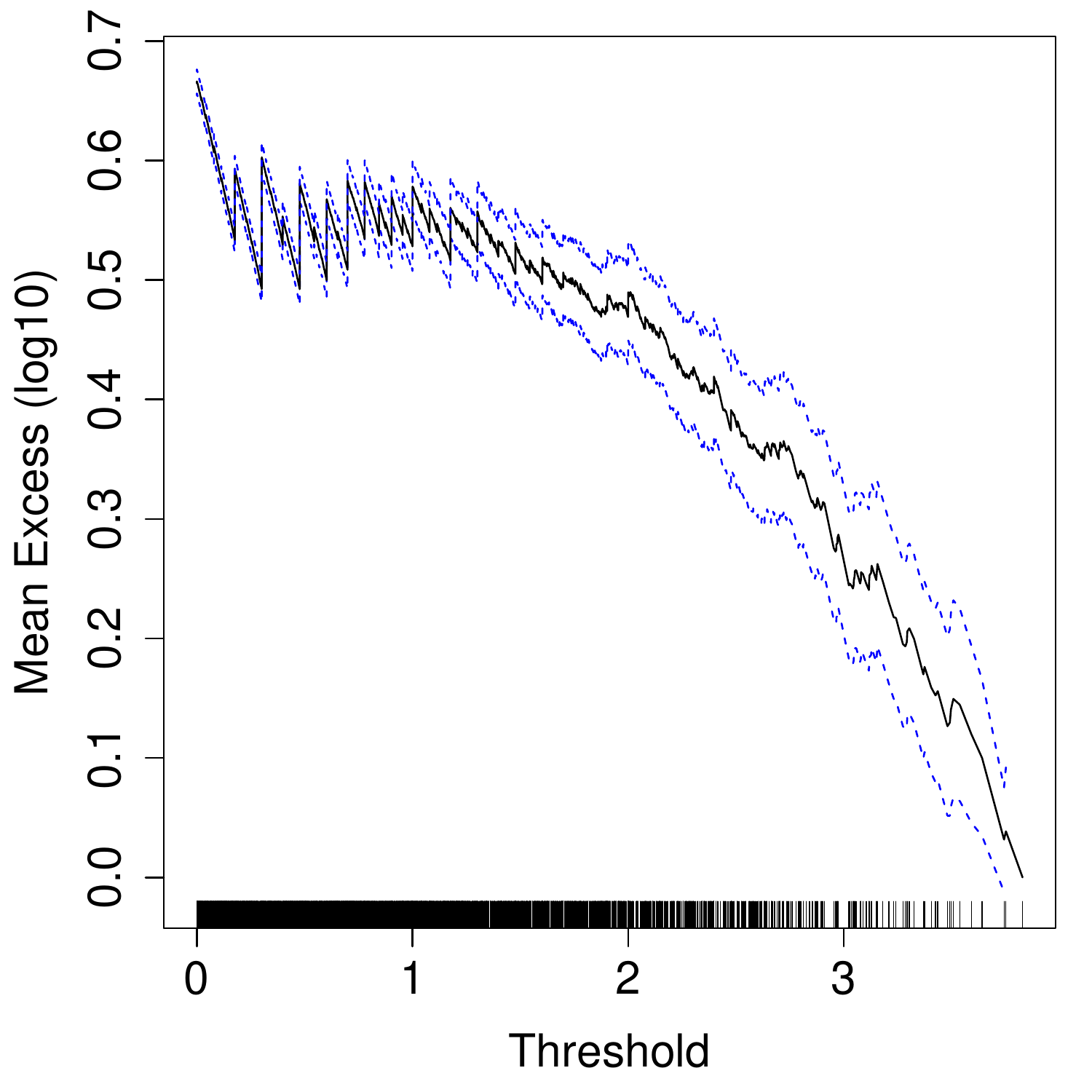}
    \caption{Mean excess plots. Left: for burnt areas (in ha). Right: for $\log$10 of burnt areas. Original observations are indicated at the bottom of the mean excess plots. Blue lines indicate symmetric pointwise confidence intervals at $95\%$.}
    \label{fig:mean-excess}
\end{figure}

\end{supplement}

\begin{supplement}

\stitle{Subsampling experiments}\label{sec:supplement:subsample}

\sdescription{We implement several experiments to aid the choice of the parameters of the subsampling scheme detailed in~\S\ref{ssec:ss}. We fix the sampling probability parameter to $p_\text{SS}=0.9$ but allow the empirical FWI probability $p_{\text{FWI}}$ to equal $\{0.1,0.3,0.5,0.7,0.9\}$. The case $p_\text{SS}=1-p_\text{FWI}$, \ie $p_\text{FWI}=0.1$, corresponds to uniform subsampling, whereas higher values of $p_\text{FWI}$ include a relatively larger number of high FWI observations in the subsample. In a first simulation experiment, we sample from the COX model  with log-linear intensity in \S\ref{sec:regression-equation},
\begin{align*}
\mu^\mathrm{COX}_{i,t} = &\alpha + \beta_1 z_\text{FWI}(s_i,t) + \beta_2 m(t),
\end{align*}
with $\alpha=-11$, $\beta_1 = 0.15$ and $\beta_2=0.1$ to reflect intensities that could be realistic in a  wildfire application, and fit this model with INLA. Figure~\ref{fig:boxplot:subsampling} highlights the improvement in estimation quality by moving away from uniform subsampling, with lower root mean squared errors of the posterior means.  In another experiment, we estimated the COX model with the linear predictor in~\S\ref{sec:model:point} and evaluated the sCRPS scores for the annually aggregated predicted and observed number of fires over the whole spatial region in the training set with $500$ posterior simulations and 50 different subsampling seeds. The left panel of Figure \ref{fig:boxplot:subsampling:traintest} shows that subsampling scheme with ($p_\text{FWI}$, $p_\text{SS}$)$=(0.7,0.9)$ achieves the best score. Next, we repeated the experiment with a fixed $p_\text{FWI}$ and $p_\text{SS}$, but increased the number of subsamples taken within each pixel-year. The right panel of Figure \ref{fig:boxplot:subsampling:traintest} shows that there is little improvement in sCRPS score beyond two subsamples per pixel-year, while the computational time and memory requirements increase strongly non-linearly with the number of subsamples (not shown).   }

\begin{figure}[t]
\centering
    \includegraphics[width=.94\textwidth]{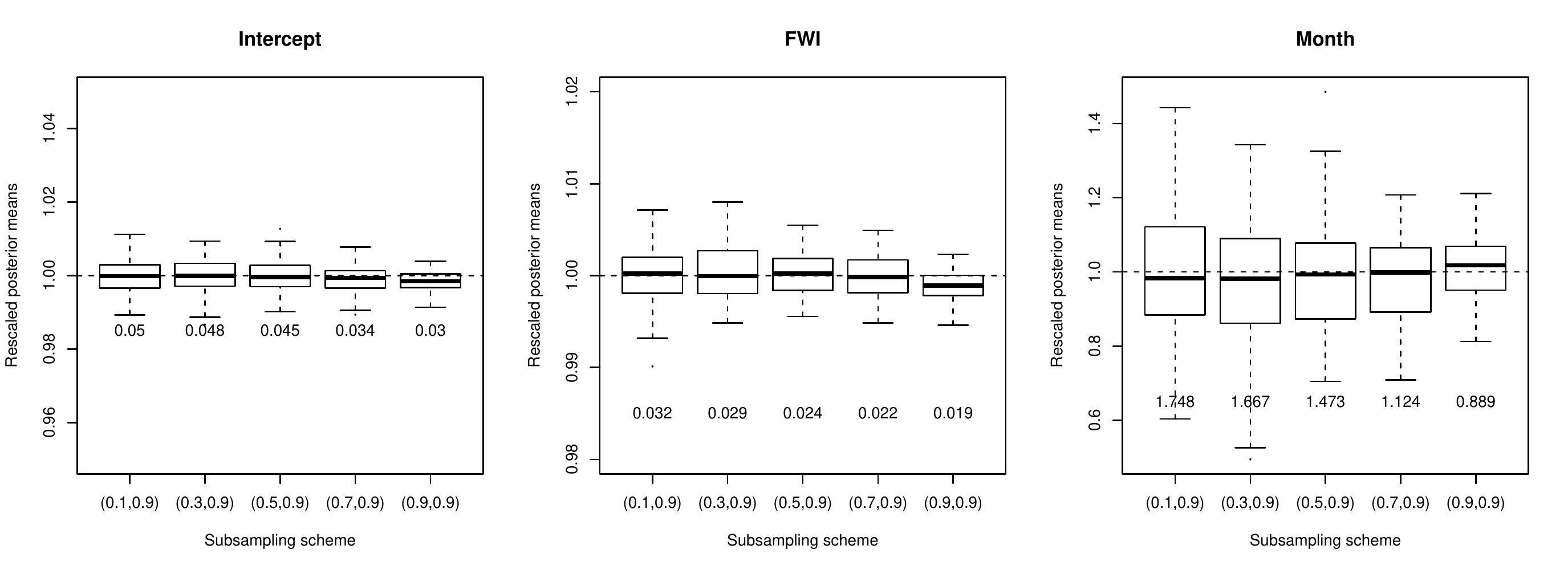}

  \caption{Boxplots of posterior means (rescaled by dividing them by the true parameter value) of fixed effect coefficients from 100 simulations with different ($p_\text{FWI}$, $p_\text{SS}$) combinations. The relative root mean square errors (rRMSE) for each subsampling scheme are displayed below the corresponding boxplots.}  
  \label{fig:boxplot:subsampling}
\end{figure}

\begin{figure}[t]
\centering
    \includegraphics[width=.45\textwidth]{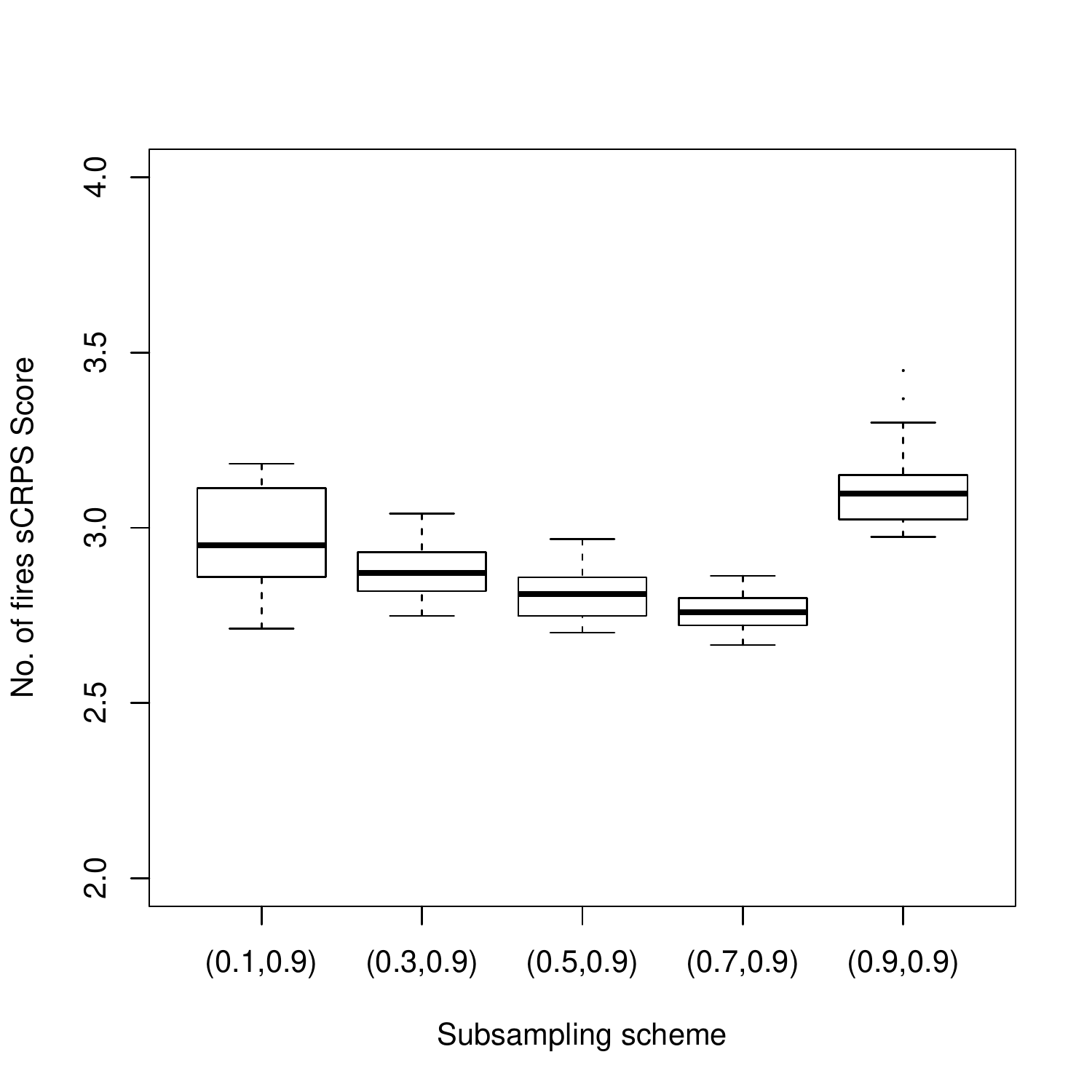}
\includegraphics[width=.45\textwidth]{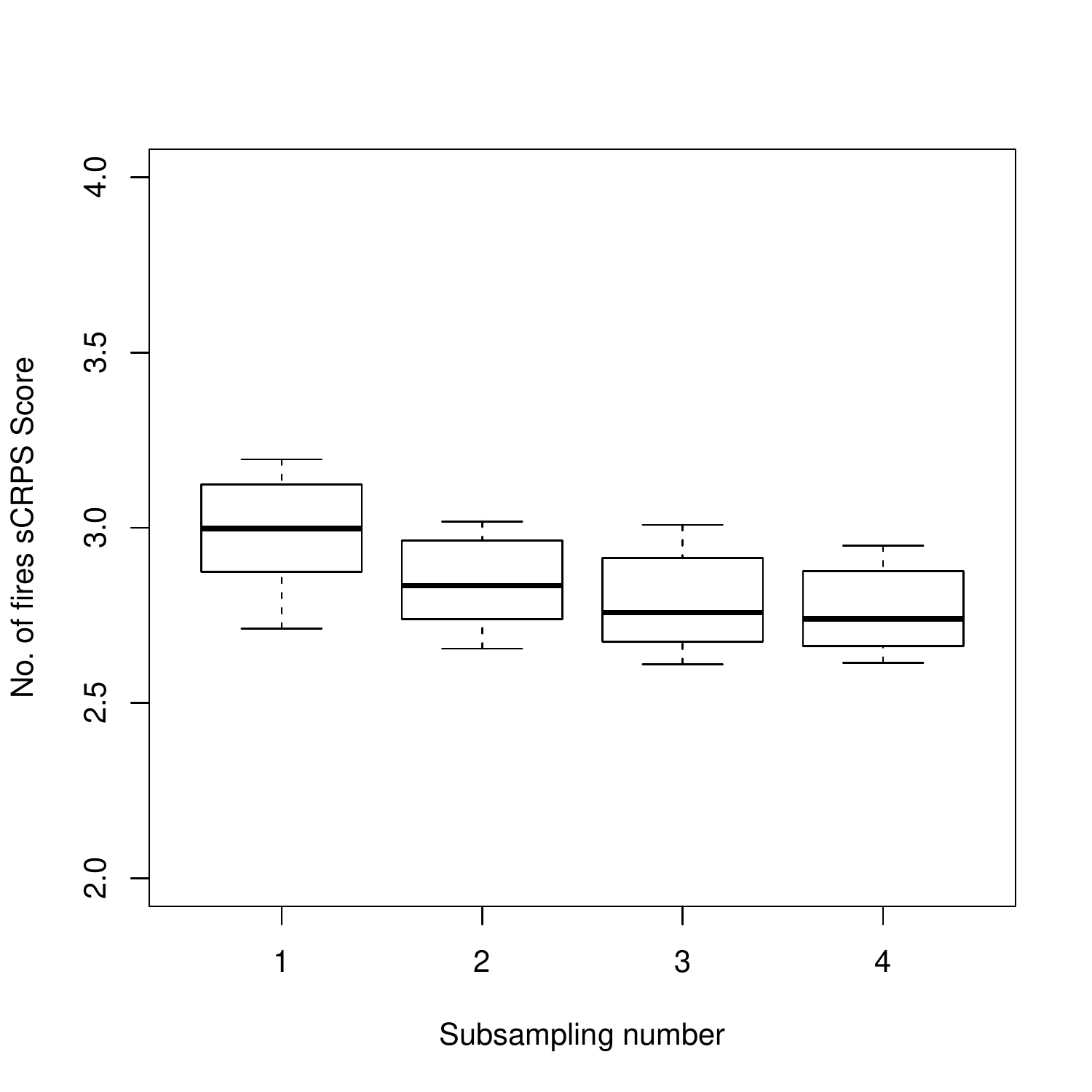}
  \caption{sCRPS for the annually aggregated predicted and observed number of fires over the whole spatial region in the training set (1994--2014), with the subsampling scheme over 50 different seeds. Right: Boxplots of sCRPS scores for the subsampling scheme with one subsample per pixel-year and different ($p_\text{FWI}$, $p_\text{SS}$) combinations. Left: Boxplots of sCRPS scores for the subsampling with ($p_\text{FWI}$, $p_\text{SS}$)=(0.1,0.9), with different number of subsamples per pixel-year.
  }  
  \label{fig:boxplot:subsampling:traintest}
\end{figure}

\end{supplement}

\begin{supplement}

\stitle{Other hyperpriors}\label{sec:supplement:priors}

\sdescription{All fixed effect coefficients in our models (\eg $\alpha^\text{COX}$, $\alpha^\text{BIN}$, $\alpha^\text{GPD}$ and $\alpha^\text{BETA}$) are assigned flat Gaussian priors with zero mean and precision $0.001$. The prior for each of the scaling parameters $\beta^\text{COX-BETA}$, $\beta^\text{COX-BIN}$ and $\beta^\text{BIN-GPD}$ is a zero-centered Gaussian distribution with precision $1/20$. To reduce the number of estimated hyperparameters, we fixed the hyperparameters associated with the priors $\mathcal{GP}_{\mathrm{1D\text{-}SPDE}}$ to values guided by prior knowledge about the relationship between FWI/FA and the relevant aspects of wildfire risk. For the tail index parameter $\xi$ in the GPD component, we assign a exponential distribution with rate unity, which corresponds to an approximate Penalized Complexity prior \citep{Opitz.al.2018} with moderate level of penalization from the base model ($\xi=0$). Lastly, we assign a log-Gamma hyperprior with mean unity and precision $0.0005$ to each of the random-walk hyperparameters $\tau_1$, $\tau_2$, $\tau_3$, $\tau_4$, $\tau_5$, $\tau_6$ and $\tau_7$.}

\end{supplement}






\begin{supplement}

\stitle{Spatial effects in model M1}\label{sec:supplement:spatial}

\sdescription{Figure \ref{fig:results:sp:spatial_add} shows the same plot as Figure \ref{fig:results:sp:spatial} but for the shared spatial random field $g^{\text{BIN-GPD}}$.
Figure \ref{fig:results:pp:spatial} shows the posterior means of all the spatial model M1, with priors detailed in \S\ref{sec:model:point}. }

\begin{figure}[t]
\centering
  \begin{subfigure}[b]{.7\linewidth}
    \centering
    \includegraphics[width=.99\textwidth]{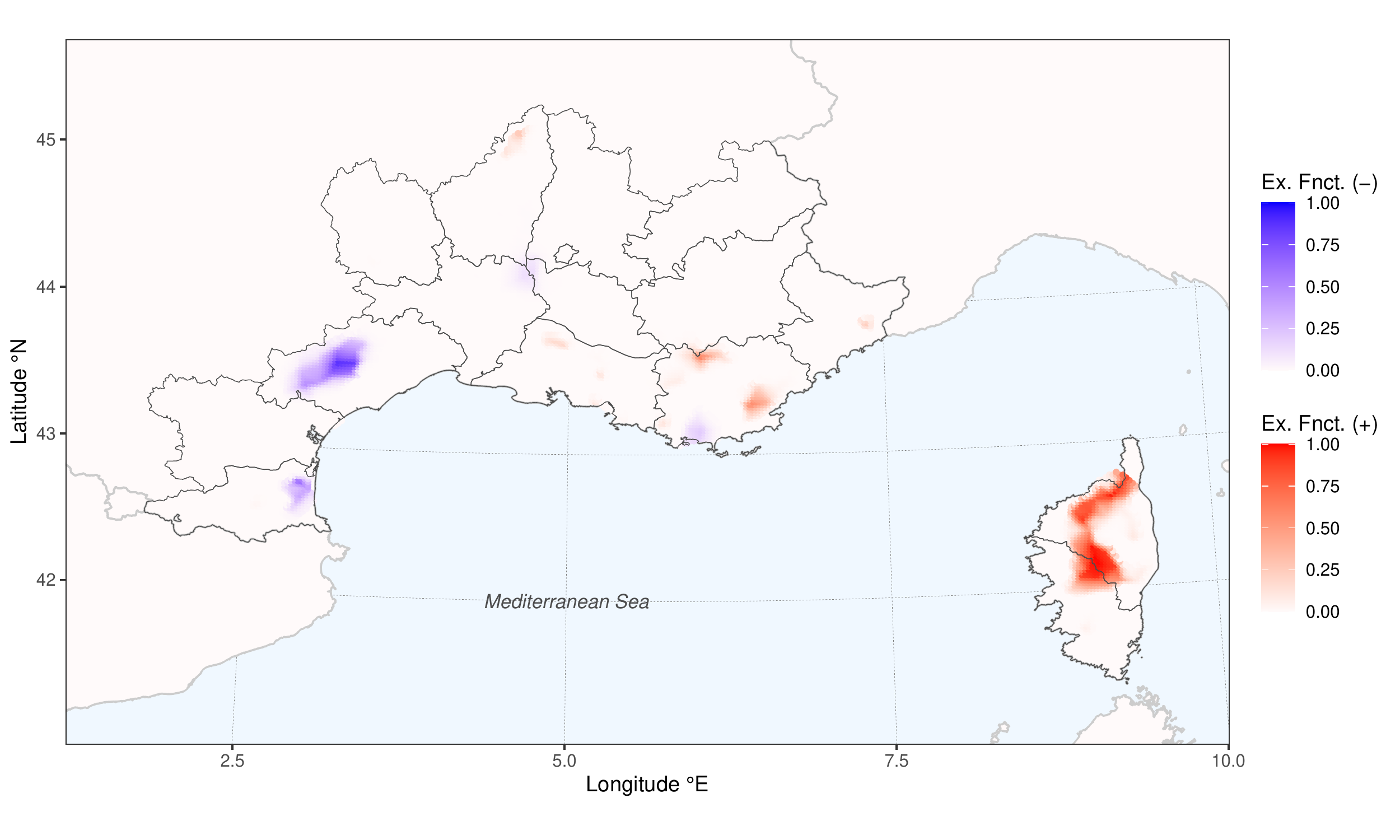}
  \end{subfigure}
  \caption{Excursion functions of posterior latent fields above $u=0.1$ and below $-u$. Plots show $\max\{ F^{+}_{0.1}(\bigcdot), F^{-}_{0.1}(\bigcdot)\}$ for the shared spatial random field $g^{\text{BIN-GPD}}$. }  
      \label{fig:results:sp:spatial_add}
\end{figure} 

\begin{figure}[t]
\centering
\begin{subfigure}[b]{.31\linewidth}
    \centering
    \includegraphics[width=.99\textwidth]{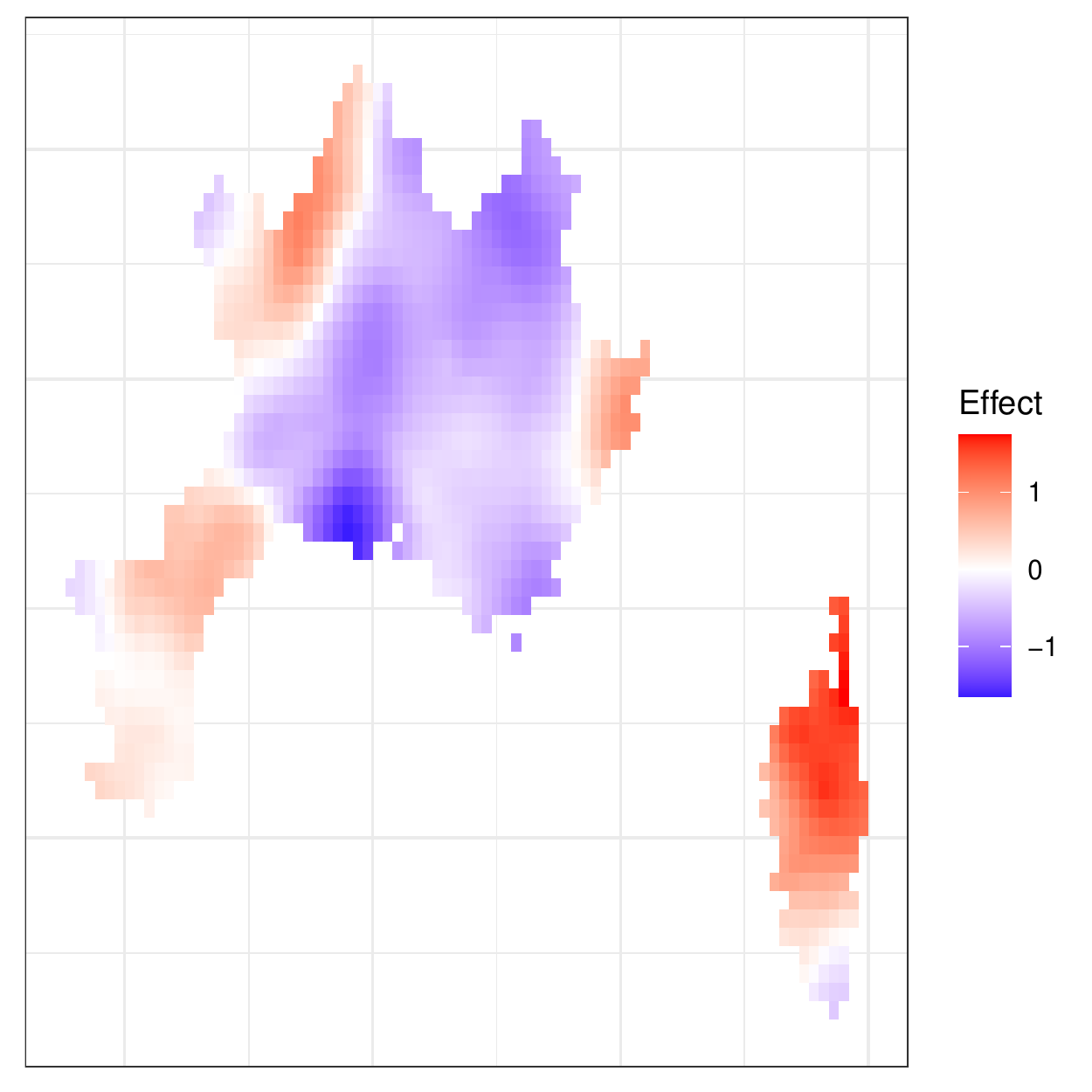}
  \end{subfigure}%
  \begin{subfigure}[b]{.31\linewidth}
    \centering
    \includegraphics[width=.99\textwidth]{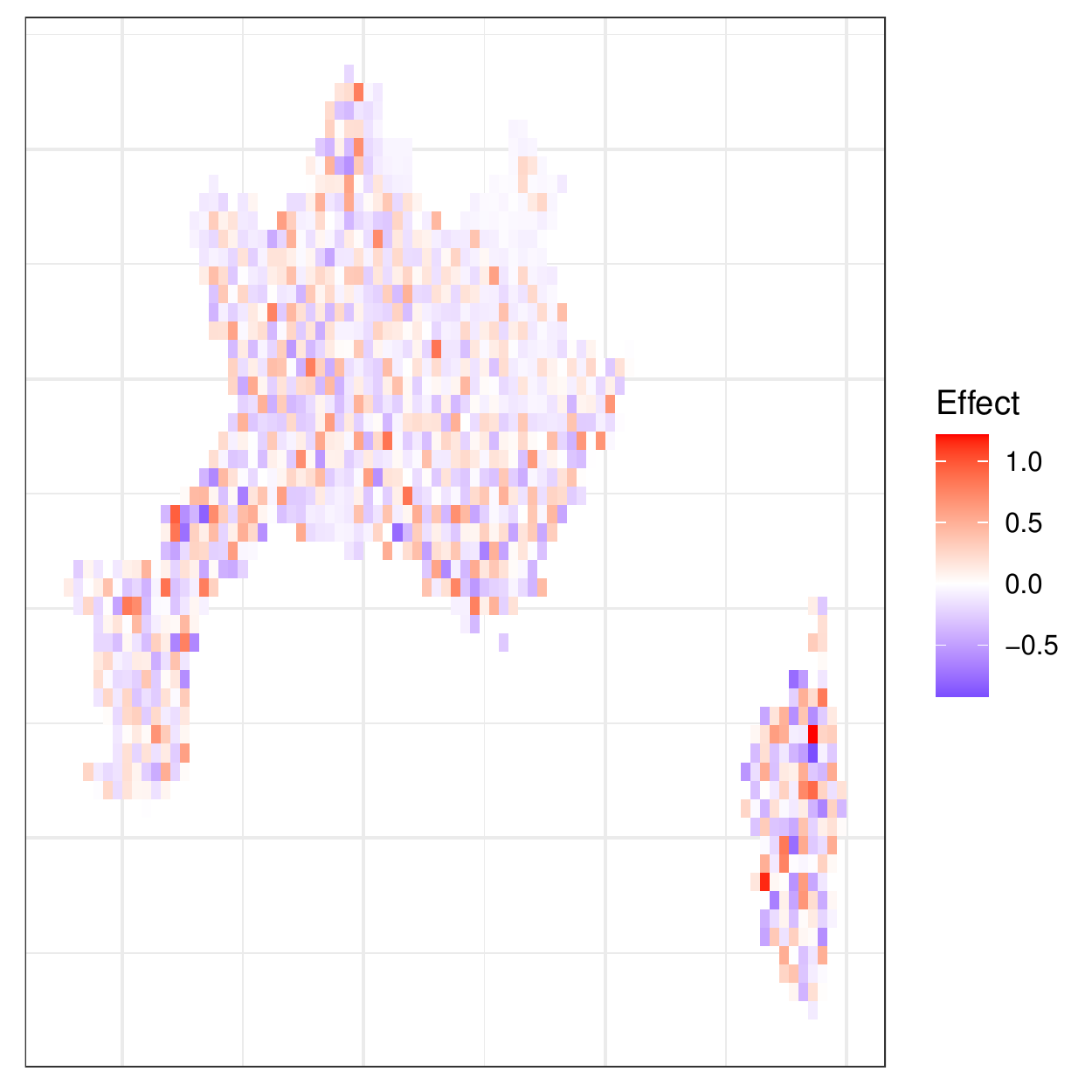}
  \end{subfigure} \\
  \begin{subfigure}[b]{.31\linewidth}
    \centering
    \includegraphics[width=.99\textwidth]{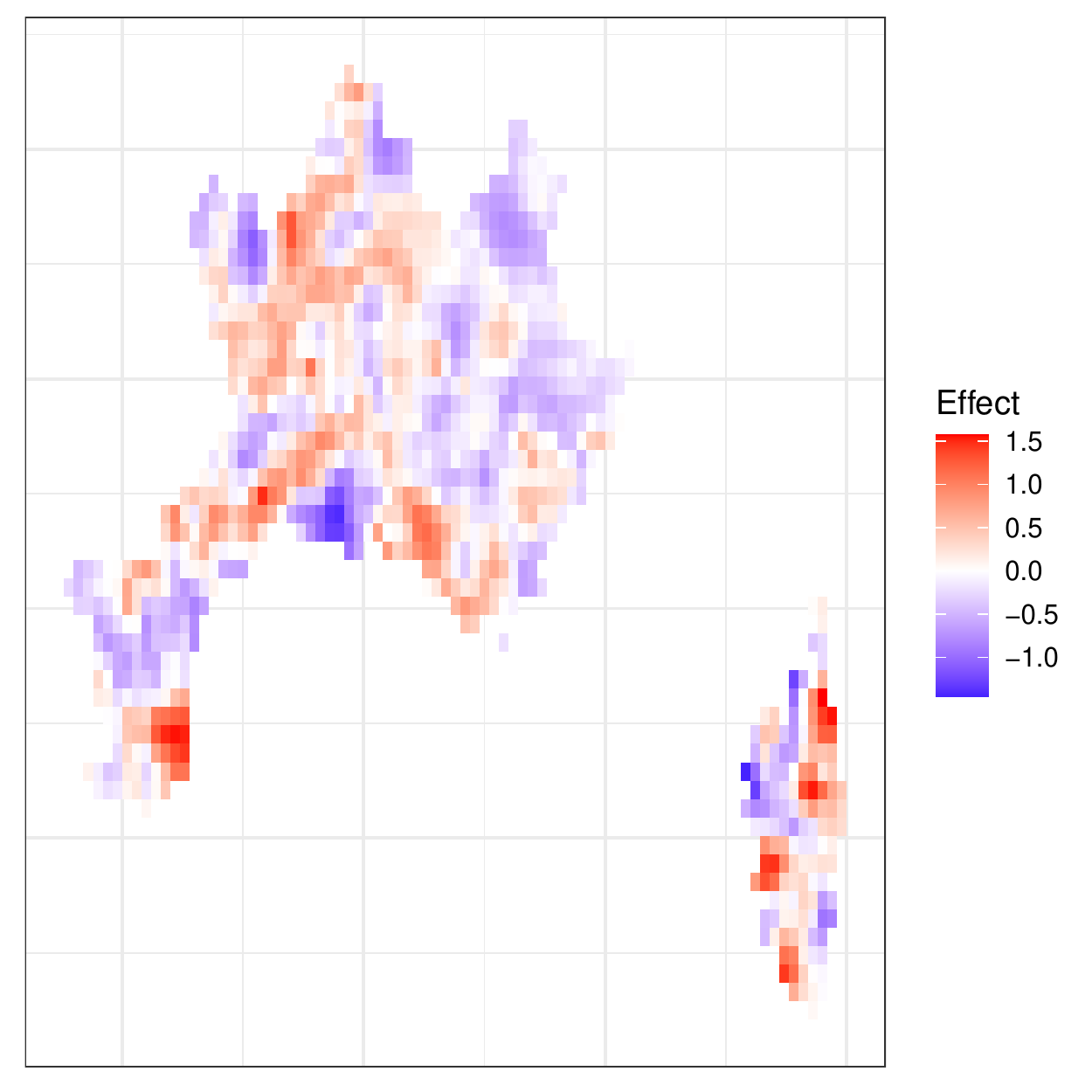}
  \end{subfigure}
  \begin{subfigure}[b]{.31\linewidth}
    \centering
    \includegraphics[width=.99\textwidth]{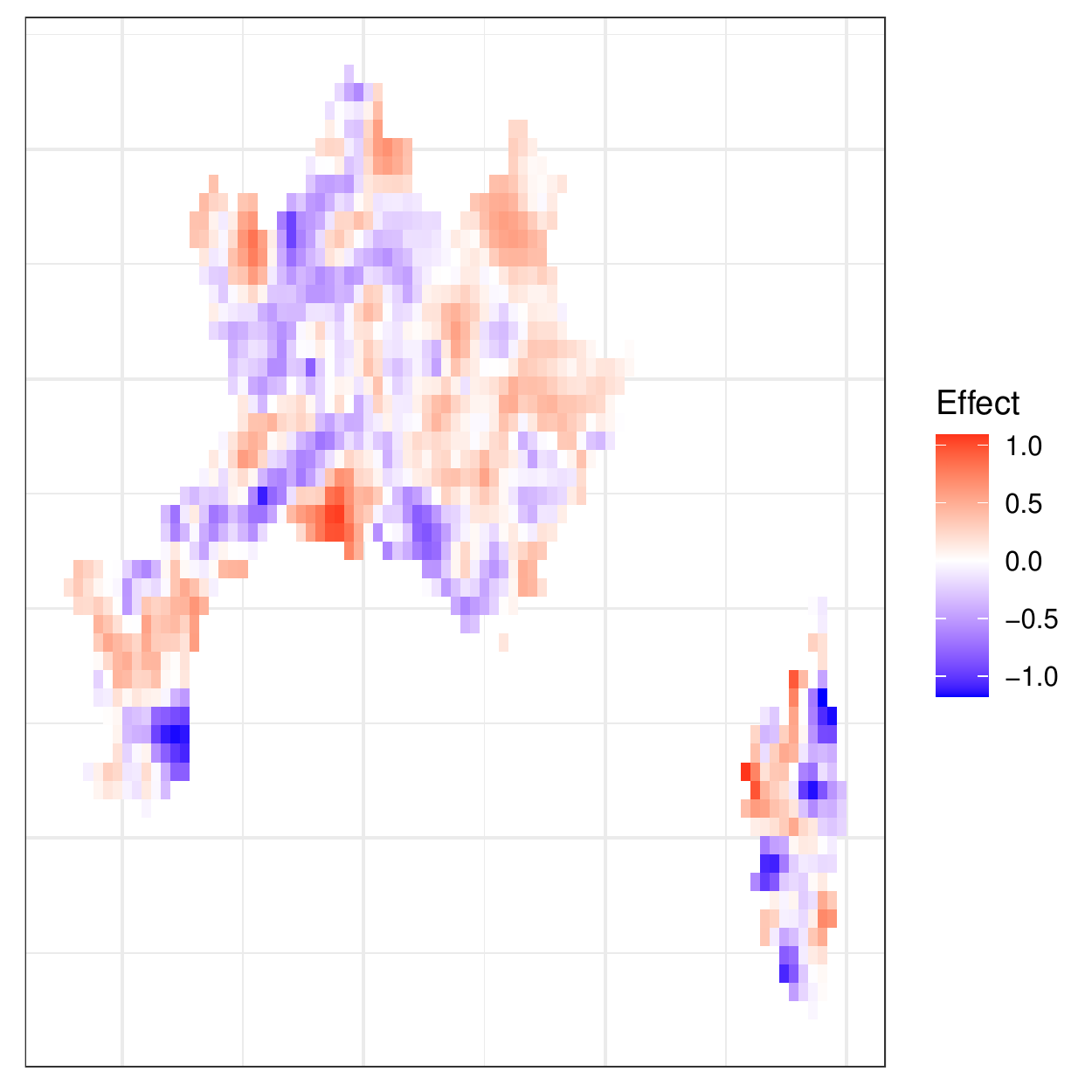}
  \end{subfigure}
  \begin{subfigure}[b]{.31\linewidth}
    \centering
    \includegraphics[width=.99\textwidth]{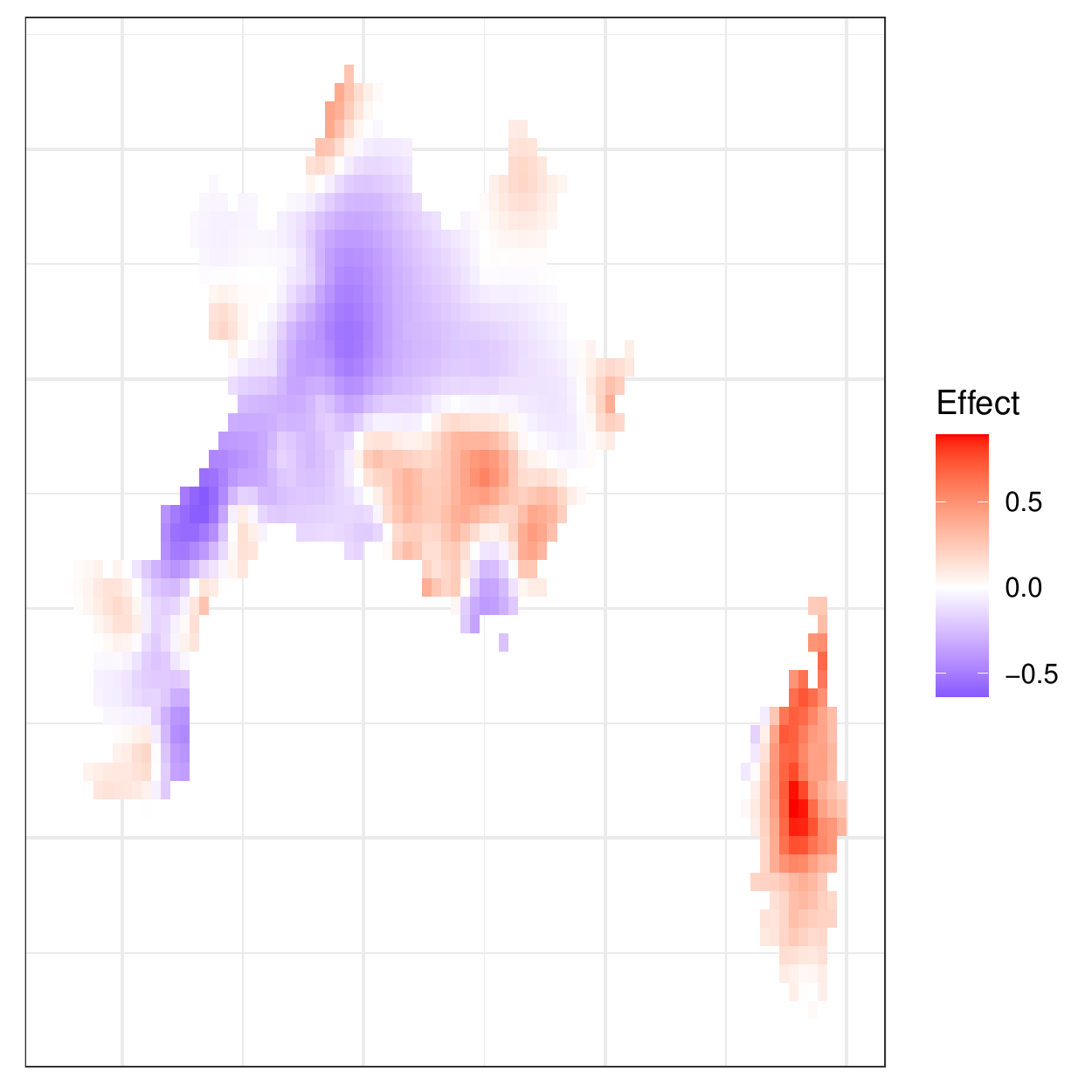}
  \end{subfigure}
  \caption{Posterior means of (top left to bottom right): $\beta^{\text{COX-BETA}} { g^{\text{COX-BETA}} ( \bigcdot )}$, $g_{1}^\text{COX} ( \bigcdot )$ , $\beta^{\text{COX-BIN}} { g^\text{COX-BIN} ( \bigcdot )}$, ${ g^\text{COX-BIN} ( \bigcdot )}$ and $ \beta^{\text{BIN-GPD}} { g^{\text{BIN-GPD}} ( \bigcdot )} $
  effects in the model.}  
    \label{fig:results:pp:spatial}
\end{figure} 

\end{supplement}

\end{document}